\documentclass[a4paper,11pt]{article}
\usepackage{bm,mathrsfs,amsmath,amsthm,amssymb,amscd,longtable,array,graphicx,bm}
\newcommand{\nc}{\newcommand}
\nc{\rnc}{\renewcommand}
\nc{\nn}{\nonumber}
\nc{\der}{{\partial}}
\rnc{\Im}{{\rm{Im}\,}}
\rnc{\Re}{{\rm{Re}\,}}
\nc{\db}{\displaybreak[0]\\}
\nc{\bra}{\langle}
\nc{\ket}{\rangle}
\nc{\bs}{\boldsymbol}

\newtheorem{theorem}{Theorem}[section]
\newtheorem{lemma}[theorem]{Lemma}

\newtheorem{proposition}[theorem]{Proposition}
\newtheorem{corollary}[theorem]{Corollary}

\theoremstyle{definition}

\numberwithin{equation}{section}

\numberwithin{equation}{section}

\begin{document}%
%%%%%%%%%%%%%%%%%%%%%%%%%%%%%%%%%%%%%%%%%%%%%%%%%%%%%%%%%
%TITLE
%%%%%%%%%%%%%%%%%%%%%%%%%%%%%%%%%%%%%%%%%%%%%%%%%%%%%%%%%
%
\title{
Integrable models and $K$-theoretic pushforward \\
of Grothendieck classes
}

\author{
Kohei Motegi \thanks{E-mail: kmoteg0@kaiyodai.ac.jp}
\\\\
{\it Faculty of Marine Technology, Tokyo University of Marine Science and Technology,}\\
 {\it Etchujima 2-1-6, Koto-Ku, Tokyo, 135-8533, Japan} \\
\\\\
\\
}

\date{\today}

\maketitle

\begin{abstract}
We show that a multiple commutation relation
of the Yang-Baxter algebra of integrable lattice models
derived by Shigechi and Uchiyama can be used to connect two types of
Grothendieck classes by the $K$-theoretic pushforward
from the Grothendieck group of Grassmann bundles
to the Grothendieck group of a nonsingular variety.
Using the commutation relation,
we show that two types of partition functions
of an integrable five-vertex model,
which can be explicitly described using skew Grothendieck polynomials, and can be viewed as Grothendieck classes,
are directly connected by the $K$-theoretic pushforward.
We show that special cases of the pushforward formula
which correspond to the nonskew version
are also special cases of the formulas derived by Buch.
We also present a skew generalization of an identity 
for the Grothendieck polynomials by Guo and Sun,
which is an extension of the one for Schur polynomials by
Feh\'er, N\'emethi and Rim\'anyi.
We also show an application of the pushforward formula
and derive an integration formula for the Grothendieck polynomials.
\end{abstract}

\section{Introduction}
Recently, various connections between integrable systems
\cite{Bethe,Baxter,KBI} and
$K$-theoretic objects in algebraic geometry have been explored
\cite{MS,MS2,RTV,Okounkov,KirillovSigma,GK,WZ,BSW,ZJ,Iwao,Iwaoskew}.
A typical object which appears in both areas
is the Grothendieck polynomials
\cite{LS,FK,Buch1,Buch2,Buch3,Buch4,Lenart}
which are polynomial representatives for $K$-theoretic Schubert classes
in algebraic geometry.
They appear in integrable models as explicit forms
of partition functions in statistical physics.
The study of partition functions of integrable lattice models 
based on the quantum inverse scattering method \cite{KBI,FST,KRS}
was initiated by Korepin \cite{Ko} and Izergin \cite{Iz}. See
\cite{Ku,Ts,PRS,Ros,Wheeler,Motegi} for examples on the developments
of the method. The partition functions are found to be connected with
famous symmetric functions and their generalizations,
and based on the correspondence between partition functions
and symmetric functions, various algebraic identities have been found.
See the aforementioned papers as well as
\cite{HK1,HK2,Bogo,ShigechiUchiyama,BBF,BMN,Lascoux,Mcnamara,KS,Korff,BW,BWZ,WZ2,Borodin,BP1,Takeyama,vDE,BBB,BBBGduality} for examples
on the developments on this subject.
As for the Grothendieck polynomials,
the connection with quantum integrable models
gave new perspectives on algebraic identities
and also on the Littlewood-Richardson coefficients,
Gromov-Witten invariants and so on \cite{MS,MS2,GK,WZ}.

In this paper, we discuss another aspect of the connection
between integrable models and algebraic geometry.
We investigate the $K$-theoretic pushforward from the Grothendieck group
of Grassmann bundles to the Grothendieck group of a nonsingular variety
from the perspective of quantum integrability.

Let us first recall the algebraic geometry side.
See \cite{Buch1,Buch2,AllPhD,All} for examples for more details.
Let $X$ be a smooth projective complex variety
and $K(X)$ the Grothendieck ring of isomorphism classes
of algebraic vector bundles over $X$.
For a subvariety $Y$ of $X$, the Grothendieck class of the
structure sheaf of $Y$ is denoted $[\mathcal{O}_Y]$, or $[Y]$ for short.
Let $\mathcal{A} \rightarrow X$ be a rank $n$ vector bundle,
and $\pi: \mathrm{Gr}_k(\mathcal{A}) \rightarrow X$ the Grassmann bundle
of $k$-planes in $\mathcal{A}$ with universal tautological exact sequence
$0 \rightarrow \mathcal{S} \rightarrow \pi^* \mathcal{A} \rightarrow \mathcal{Q} \rightarrow 0$ of vector bundles over $\mathrm{Gr}_k(\mathcal{A})$.
Let $\sigma=\{ \sigma_1, \dots, \sigma_k \}$ and $\omega=\{ \omega_1,\dots,\omega_{n-k} \}$
be the set of Grothendieck roots
for the subbundle $\mathcal{S}$ and the 
quotient bundle $\mathcal{Q}$ respectively, i.e., 
defined by the exterior powers of
$\mathcal{S}$ and $\mathcal{Q}$ by the relation
${[} \bigwedge^i \mathcal{S} {]}=e_i(\sigma)$
and ${[} \bigwedge^j \mathcal{Q} {]}=e_j(\omega)$,
where $e_i(\sigma)$ and $e_j(\omega)$ are the $i$th and $j$th
elementary symmetric functions with the set of variables $\sigma$ and $\omega$,
respectively.

Let $f(\sigma;\omega)$ be a Laurent polynomial
in $K(X)[\sigma_i^{\pm};\omega_j^{\pm}]$ which are symmetric in
$\sigma$ variables and also symmetric in $\omega$ variables,
which makes it possible for the polynomial to be viewed as 
a certain Grothendieck class in $K(\mathrm{Gr}_k(\mathcal{A}))$.

We consider the following $K$-theoretic pushforward
$\pi_*:K(\mathrm{Gr}_k(\mathcal{A})) \rightarrow K(X)$
induced from $\pi$.
The following localization type formula holds for $\pi_*(f)$
\cite{AllPhD,All,AB,BV,CG,Nie,Gr,BFQ,Tu,WebZie}:

\begin{proposition}
Let $S_k^n$ be a $k$-subset of $[n]=\{1,2,\dots,n \}$
and denote the set of $k$-subsets of $n$ as $\binom{[n]}{k}$,
and $\overline{S_k^n}=\{1,2,\dots,n \} \backslash S_k^n$.
Let $\{\alpha_1,\dots,\alpha_n \}$ be the Grothendieck roots for $\mathcal{A}$
and $\alpha_J=\{\alpha_{j_1},\dots,\alpha_{j_r} \}$ for any subset
$J=\{j_1,\dots,j_r \}$ of $[n]$. For a Grothendieck class
$f(\sigma;\omega) \in K(\mathrm{Gr}_k(\mathcal{A}))$, we have
\begin{align}
\pi_*(f(\sigma;\omega))=\sum_{S_k^n \in \binom{[n]}{k}}
\prod_{i \in S_k^n, j \in \overline{S_k^n}}
\frac{1}{1-\alpha_i/\alpha_j}
f(\alpha_{S_k^n};\alpha_{\overline{{S_k^n}}}). \label{KGysin}
\end{align}
\end{proposition}

The explicit computations of the pushforward
have been explored through various ways.
One of the approaches which have been studied extensively
in recent years is to
reexpress the right hand side of \eqref{KGysin}
as iterated residues
and find various pushforward formulas for
polynomials which can be viewed as generalized cohomology classes
\cite{AllPhD,All,Rim,AR1,AR2,RS,Zi1,Z2,Z3,Pr,DP1,DP2,HIMN,NN1,NN2,NN3}.

In this paper, we study the map $\pi_*$ of Grothendieck classes
from the point of view of quantum integrability.
On the integrable model side, the key observation is that
the right hand side of the following commutation relation
between the monodromy matrix elements of certain integrable models
\begin{align}
&\prod_{j=k+1}^{n} D(u_j) \prod_{j=1}^k B(u_j)
=\sum_{S_k^n \in \binom{[n]}{k}} \prod_{i \in S_k^n, j \in \overline{S_k^n}}
\frac{1}{1-u_i/u_j} \prod_{i \in S_k^n} B(u_i)
\prod_{j \in \overline{S_k^n}} D(u_j),
\label{SUcommutation}
\end{align}
whose details of the notations will be explained later,
resembles the right hand side of the $K$-theoretic pushforward formula
\eqref{KGysin}.
This type of commutation relation is one of the commutation relations
derived essentially  by Shigechi and Uchiyama \cite{ShigechiUchiyama}
when they investigated the integrable boson model \cite{Bogo},
and also holds for an integrable five vertex model used in this paper.
One can observe that the the expressions in the
right hand side of \eqref{KGysin} and \eqref{SUcommutation} resemble.
However, in \eqref{SUcommutation},
what appears are the so-called $B$-operators and $D$-operators which are
noncommutative objects. Noncommutative objects do not appear
in the $K$-theoretic pushforward formula \eqref{KGysin}.
To go from the noncommutative world to the commutative world,
we consider partition functions of an integrable five-vertex model,
which are fundamental objects
in statistical physics. We introduce two classes of partition functions
which can be regarded as some Grothendieck classes of
the Grassmann bundle $\mathrm{Gr}_k(\mathcal{A})$ and a
nonsingular variety $X$. Using the Shigechi-Uchiyama commutation relation
\eqref{SUcommutation},
we show that they are directly connected by the
$K$-theoretic pushforward $\pi_*$.
We also show that the partition functions introduced in this paper
can be explicitly described using the skew Grothendieck polynomials,
i.e., we show that two polynomials described by the
skew Grothendieck polynomials are connected by the
$K$-theoretic pushforward.
It is also shown that special cases of the formula which correspond
to the nonskew version are also special cases of the formulas derived 
by Buch \cite{Buch1}.
We also present an identity for the skew Grothendieck polynomials
by taking matrix elements of the commutation relation
\eqref{SUcommutation},
which can be regarded as a skew extension of the
identity for the Grothendieck polynomials by Guo and Sun \cite{GS},
which is an extension of an identity for the Schur polynomials
by Feh\'er, N\'emethi and Rim\'anyi \cite{FNR}.
By combining the pushforward formula with another approach,
we also get an integration formula for the Grothendieck polynomials.
It would be interesting to investigate other kinds of maps
which appear in algebraic geometry
from the point of view of quantum integrability by
observing similarities between formulas appearing in both areas.

This paper is organized as follows.
In the next section, we introduce an integrable five-vertex model
and the Yang-Baxter algebra associated with the model,
and review a commutation relation by Shigechi and Uchiyama.
In section 3,
we introduce partition functions
constructed from the $R$-matrix of the integrable five-vertex model,
and write their explicit forms using skew Grothendieck polynomials.
In section 4, we prove a correspondence formula between
two types of Grothendieck classes of
the Grassmann bundle $\mathrm{Gr}_k(\mathcal{A})$ and a nonsingular variety
$X$ by the $K$-theoretic pushforward.
The proof is given by using the partition function represenations
of the Grothendieck classes and the Shigechi-Uchiyama commutation relation.
We also present an identity for the skew Grothendieck polynomials
obtained as the matrix elements of the commutation relation.
We show in section 5 that special cases of the formula and identity
are also special cases of the formulas derived by Buch and Guo-Sun.
We also show an application of the pushforward formula
and derive an integration formula for the Grothendieck polynomials.

\section{Integrable five-vertex model}

In this section, we introduce an integrable five-vertex model
and the Yang-Baxter algebra associated with the model.

We first introduce a two-dimensional vector space $W$,
and denote its standard basis as $\{ | 0 \rangle, |1 \rangle  \}$,
where $|0 \rangle :=\left( 
\begin{array}{c}
1 \\
0 \\
\end{array}
\right),
|1 \rangle :=\left( 
\begin{array}{c}
0 \\
1 \\
\end{array}
\right)
$.
We denote the dual of $|0 \rangle$, $|1 \rangle$ as
$\langle 0|:=(1,0)$, $\langle 1|:=(0,1)$,
and the dual vector space as $W^*$.
We use the bra-ket notation (without taking complex conjugation) in this paper.
We define
$\langle A|B \rangle$ for two vectors
$|A \rangle =\left( 
\begin{array}{c}
A_1 \\
A_2 \\
\end{array}
\right),
|B \rangle =\left( 
\begin{array}{c}
B_1 \\
B_2 \\
\end{array}
\right)
$, $A_1,A_2,B_1,B_2 \in \mathbb{C}$
as the standard inner product between $|A \rangle \in W$
and $|B \rangle \in W$:
$\langle A|B \rangle:=A \cdot B=A_1 B_1+A_2 B_2$,
and view this as the product between $\langle A|$ and $|B \rangle$
where $\langle A|:=(A_1,A_2) \in W^*$.
The inner product between basis vectors (product between basis and dual basis vectors) are
$\langle 0|0 \rangle=\langle 1|1 \rangle=1$, $\langle 1|0 \rangle=\langle 0|1 \rangle=0$.
Using the bra-ket notation, the identity operator $\mathrm{Id}$ acting on $W$ can be written
as $\mathrm{Id}=|0 \rangle \langle 0|+|1 \rangle \langle 1|$.
This can be checked by noting that the action of
$|0 \rangle \langle 0|+|1 \rangle \langle 1|$ on $|0 \rangle$
and $|1 \rangle$ are
$(|0 \rangle \langle 0|+|1 \rangle \langle 1|)|0 \rangle
=
|0 \rangle \langle 0|0 \rangle+|1 \rangle \langle 1|0 \rangle
=|0 \rangle
$ and
$(|0 \rangle \langle 0|+|1 \rangle \langle 1|)|1 \rangle
=
|0 \rangle \langle 0|1 \rangle+|1 \rangle \langle 1|1 \rangle
=|1 \rangle$, and any vector in $W$ can be written as a linear combination
of $|0 \rangle$ and $|1 \rangle$.

Next, we consider the tensor product of
the two-dimensional vector spaces $W_i \otimes W_j$.
Here we introduce subscripts to
distinguish the two-dimensional vector spaces.
We can take $\{ |0 \rangle_i \otimes |0 \rangle_j,
|0 \rangle_i \otimes |1 \rangle_j,
|1 \rangle_i \otimes |0 \rangle_j,
|1 \rangle_i \otimes |1 \rangle_j \}
$ as a basis of $W_i \otimes W_j$.
The $R$-matrix $R_{ij}(u,w)$ acting on $W_i \otimes W_j$
is defined by acting on this basis as
\begin{align}
R_{ij}(u,w)|0 \rangle_i \otimes |0 \rangle_j
&=|0 \rangle_i \otimes |0 \rangle_j, \label{rdefone} \\
R_{ij}(u,w)|0 \rangle_i \otimes |1 \rangle_j
&=w/u|1 \rangle_i \otimes |0 \rangle_j, \label{rdeftwo} \\
R_{ij}(u,w)|1 \rangle_i \otimes |0 \rangle_j
&=|0 \rangle_i \otimes |1 \rangle_j
+(1-w/u)|1 \rangle_i \otimes |0 \rangle_j, \label{rdefthree} \\
R_{ij}(u,w)|1 \rangle_i \otimes |1 \rangle_j
&=|1 \rangle_i \otimes |1 \rangle_j \label{rdeffour}.
\end{align}
Here, $u$ and $w$ are complex numbers.
To indicate the vector spaces which the $R$-matrix act,
we use subscripts and denote the $R$-matrix as $R_{ij}(u,w)$.

We also introduce the following notation
for the matrix elements: we define $[R(u,w)]_{\epsilon_1 \epsilon_2}^{\delta_1 \delta_2}$ for $\epsilon_1,\epsilon_2,\delta_1,\delta_2=0,1$ as
$\displaystyle R_{ij}(u,w) |\epsilon_1 \rangle_i \otimes | \epsilon_2 \rangle_j 
=\sum_{\delta_1,\delta_2=0,1} [R(u,w)]_{\epsilon_1 \epsilon_2}^{\delta_1 \delta_2}
|\delta_1 \rangle_i \otimes | \delta_2 \rangle_j.
$
By defintion of the $R$-matrix
\eqref{rdefone}, \eqref{rdeftwo}, \eqref{rdefthree} and
\eqref{rdeffour}, we have
$[R(u,w)]_{00}^{00}=[R(u,w)]_{11}^{11}=[R(u,w)]_{10}^{01}=1, \
[R(u,w)]_{01}^{10}=w/u, \ [R(u,w)]_{10}^{10}=1-w/u
$ and $[R(u,w)]_{\epsilon_1 \epsilon_2}^{\delta_1 \delta_2}=0$ otherwise.
An important property of the matrix elements for this $R$-matrix
is $[R(u,w)]_{\epsilon_1 \epsilon_2}^{\delta_1 \delta_2}=0$ if 
$\epsilon_1+\epsilon_2 \neq \delta_1+\delta_2$.
This property is called as the ice-rule.
We also remark that
$[R(u,w)]_{\epsilon_1 \epsilon_2}^{\delta_1 \delta_2}$
is the same with
${}_i \langle \delta_1| \otimes {}_j \langle
\delta_2|R_{ij}(u,w)|\epsilon_1 \rangle_i \otimes |\epsilon_2 \rangle_j$
using the bra-ket notation.

\begin{figure}[ht]
\includegraphics[width=11cm]{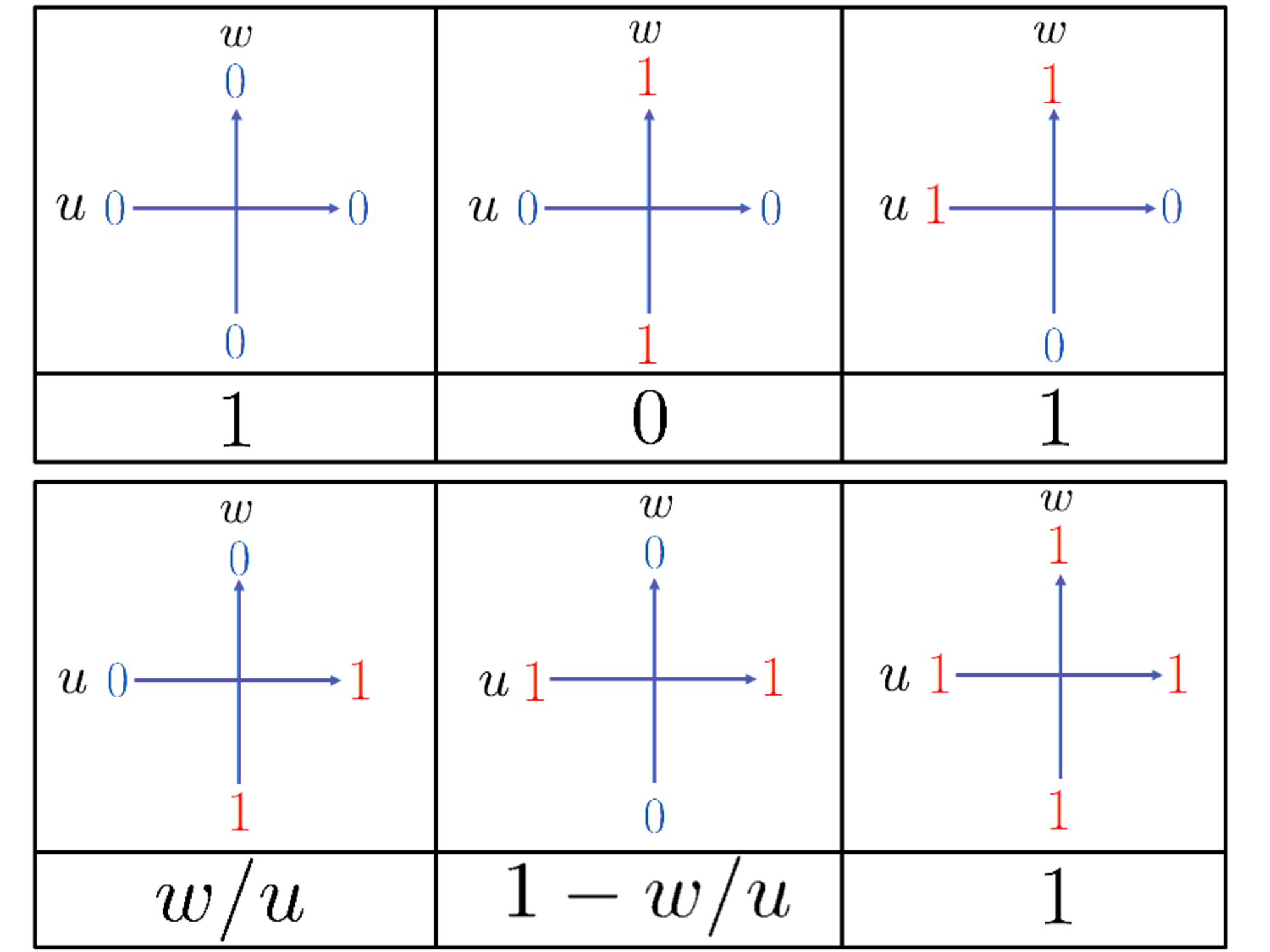}
\caption{A graphical description of the $R$-matrix $R_{ij}(u,w)$
of the integrable five-vertex model \eqref{fivevertexrmatrix}.
The elements of the $R$-matrix
are displayed. The $R$-matrix is represented as two crossing arrows.
The left and the up arrow represents the space $W_i$
and $W_j$, respectively. 0 or 1 on
the left and the right around a vertex denote the input
(basis vector) and the 
output (dual basis vector) of $W_i$, and 0 or 1
on the bottom and the top denote the input (basis vector) and the output (dual basis vector) of $W_j$, respectively.
For example, $[R_{ij}(u,w)]_{01}^{10}={}_i \langle 1 |
\otimes {}_j \langle 0| R_{ij}(u,w)|0 \rangle_i \otimes |1 \rangle_j=w/u$.
There are $2^2 \times 2^2=16$ matrix elements, and 6 of them
$[R_{ij}(u,w)]_{\epsilon_1 \epsilon_2}^{\delta_1 \delta_2}$ satisfying
$\epsilon_1+\epsilon_2=\delta_1+\delta_2$ are displayed.
The other matrix elements are identically zero.
}
\label{picturermatrix}
\end{figure}

Identifying
$|0 \rangle_i \otimes |0 \rangle_j,
|0 \rangle_i \otimes |1 \rangle_j,
|1 \rangle_i \otimes |0 \rangle_j,
|1 \rangle_i \otimes |1 \rangle_j,
$ with
 $\left( 
\begin{array}{c}
1 \\
0 \\
0 \\
0
\end{array}
\right),
\left( 
\begin{array}{c}
0 \\
1 \\
0 \\
0
\end{array}
\right),
\left( 
\begin{array}{c}
0 \\
0 \\
1 \\
0
\end{array}
\right),
\left( 
\begin{array}{c}
0 \\
0 \\
0 \\
1
\end{array}
\right),
$ respectively,
the $R$-matrix is written as the $4 \times 4$ matrix
\begin{eqnarray}
R_{ij}(u,w)=\left( 
\begin{array}{cccc}
1 & 0 & 0 & 0 \\
0 & 0 & 1 & 0 \\
0 & w/u & 1-w/u & 0 \\
0 & 0 & 0 & 1
\end{array}
\right). \label{fivevertexrmatrix}
\end{eqnarray}
We remark that the $R$-matrix \eqref{fivevertexrmatrix} can be regarded
as a certain limit of the $U_q(\widehat{sl_2})$ $R$-matrix
\cite{Dr,J,LW} for the six-vertex model.
See Figure \ref{picturermatrix} for a graphical
description of the $R$-matrix of the integrable five-vertex model.

More generally, for an integer $m \geq 2$, we define
$R_{ij}(u,w)$ $(1 \le i < j \le m)$
as an operator acting on $W_1 \otimes W_2 \otimes \cdots \otimes
W_m$. We can take $\{| \epsilon_1 \rangle_1 \otimes | \epsilon_2 \rangle_2
\otimes \cdots \otimes |\epsilon_m \rangle_m \ | \ \epsilon_1,\epsilon_2,\dots,\epsilon_m=0,1 \}$ as a basis of $W_1 \otimes W_2 \otimes \cdots \otimes
W_m$. The operator $R_{ij}(u,w)$
is defined by acting on this basis as

\begin{align}
&R_{ij}(u,w)| \epsilon_1 \rangle_1 \otimes \cdots \otimes |0 \rangle_i \otimes
\cdots \otimes |0 \rangle_j \otimes \cdots \otimes |\epsilon_m \rangle_m \nonumber \\
&=
| \epsilon_1 \rangle_1 \otimes \cdots \otimes |0 \rangle_i \otimes
\cdots \otimes |0 \rangle_j \otimes \cdots \otimes |\epsilon_m \rangle_m,
\label{defrone}
\\
&R_{ij}(u,w)| \epsilon_1 \rangle_1 \otimes \cdots \otimes |0 \rangle_i \otimes
\cdots \otimes |1 \rangle_j \otimes \cdots \otimes |\epsilon_m \rangle_m \nonumber \\
&=w/u
| \epsilon_1 \rangle_1 \otimes \cdots \otimes |1 \rangle_i \otimes
\cdots \otimes |0 \rangle_j \otimes \cdots \otimes |\epsilon_m \rangle_m,
\label{defrtwo}
\\
&R_{ij}(u,w)| \epsilon_1 \rangle_1 \otimes \cdots \otimes |1 \rangle_i \otimes
\cdots \otimes |0 \rangle_j \otimes \cdots \otimes |\epsilon_m \rangle_m \nonumber \\
&=| \epsilon_1 \rangle_1 \otimes \cdots \otimes |0 \rangle_i \otimes
\cdots \otimes |1 \rangle_j \otimes \cdots \otimes |\epsilon_m \rangle_m
\nonumber \\
&+(1-w/u)
| \epsilon_1 \rangle_1 \otimes \cdots \otimes |1 \rangle_i \otimes
\cdots \otimes |0 \rangle_j \otimes \cdots \otimes |\epsilon_m \rangle_m,
\label{defrthree}
\\
&R_{ij}(u,w)| \epsilon_1 \rangle_1 \otimes \cdots \otimes |1 \rangle_i \otimes
\cdots \otimes |1 \rangle_j \otimes \cdots \otimes |\epsilon_m \rangle_m \nonumber \\
&=
| \epsilon_1 \rangle_1 \otimes \cdots \otimes |1 \rangle_i \otimes
\cdots \otimes |1 \rangle_j \otimes \cdots \otimes |\epsilon_m \rangle_m,
\label{defrfour}
\end{align}
for any $\epsilon_k=0,1$  $(k=1,\dots,m, \ k \neq i,j)$.
The subscripts $i$ and $j$ of $R_{ij}(u,w)$
indicate the vector spaces which the operator
acts nontrivially.
Note that $R_{ij}(u,w)$ acts as identity
on vector spaces
except for the spaces $W_i$ and $W_j$,
and the action \eqref{defrone}, \eqref{defrtwo}, \eqref{defrthree} and
\eqref{defrfour} can be regarded as an extension
of the action
\eqref{rdefone}, \eqref{rdeftwo}, \eqref{rdefthree} and
\eqref{rdeffour} to the space $W_1 \otimes \cdots \otimes W_m$.
$R_{ij}(u,w)$ acts on the $2^m$-dimensional vector space
$W_1 \otimes \cdots \otimes W_m$, hence
can be regarded as a $2^m \times 2^m$ matrix.
The matrix elements
$[R_{ij}(u,w)]_{\epsilon_1 \cdots \epsilon_m}^{\delta_1 \cdots \delta_m}$ for $\epsilon_1,\dots,\epsilon_m,\delta_1,\cdots,\delta_m=0,1$ defined as
$\displaystyle R_{ij}(u,w) |\epsilon_1 \rangle_1 \otimes
\cdots \otimes | \epsilon_m \rangle_m
=\sum_{\delta_1,\dots,\delta_m=0,1} [R_{ij}(u,w)]_{\epsilon_1 \cdots \epsilon_m}^{\delta_1 \cdots \delta_m}
|\delta_1 \rangle_1 \otimes \cdots \otimes | \delta_m \rangle_m
$
can be expressed using $[R(u,w)]_{\epsilon_i \epsilon_j}^{\delta_i \delta_j}$
as
\begin{align}
\displaystyle
[R_{ij}(u,w)]_{\epsilon_1 \cdots \epsilon_m}^{\delta_1 \cdots \delta_m}
=[R(u,w)]_{\epsilon_i \epsilon_j}^{\delta_i \delta_j} 
\prod_{\substack{k=1 \\ k \neq i,j}}^m \delta_{\epsilon_k \delta_k},
\label{embeddingrelation}
\end{align}
where $\delta_{\epsilon_k \delta_k}=1$ if $\epsilon_k=\delta_k$,
and $\delta_{\epsilon_k \delta_k}=0$ otherwise.

\begin{figure}[ht]
\includegraphics[width=12cm]{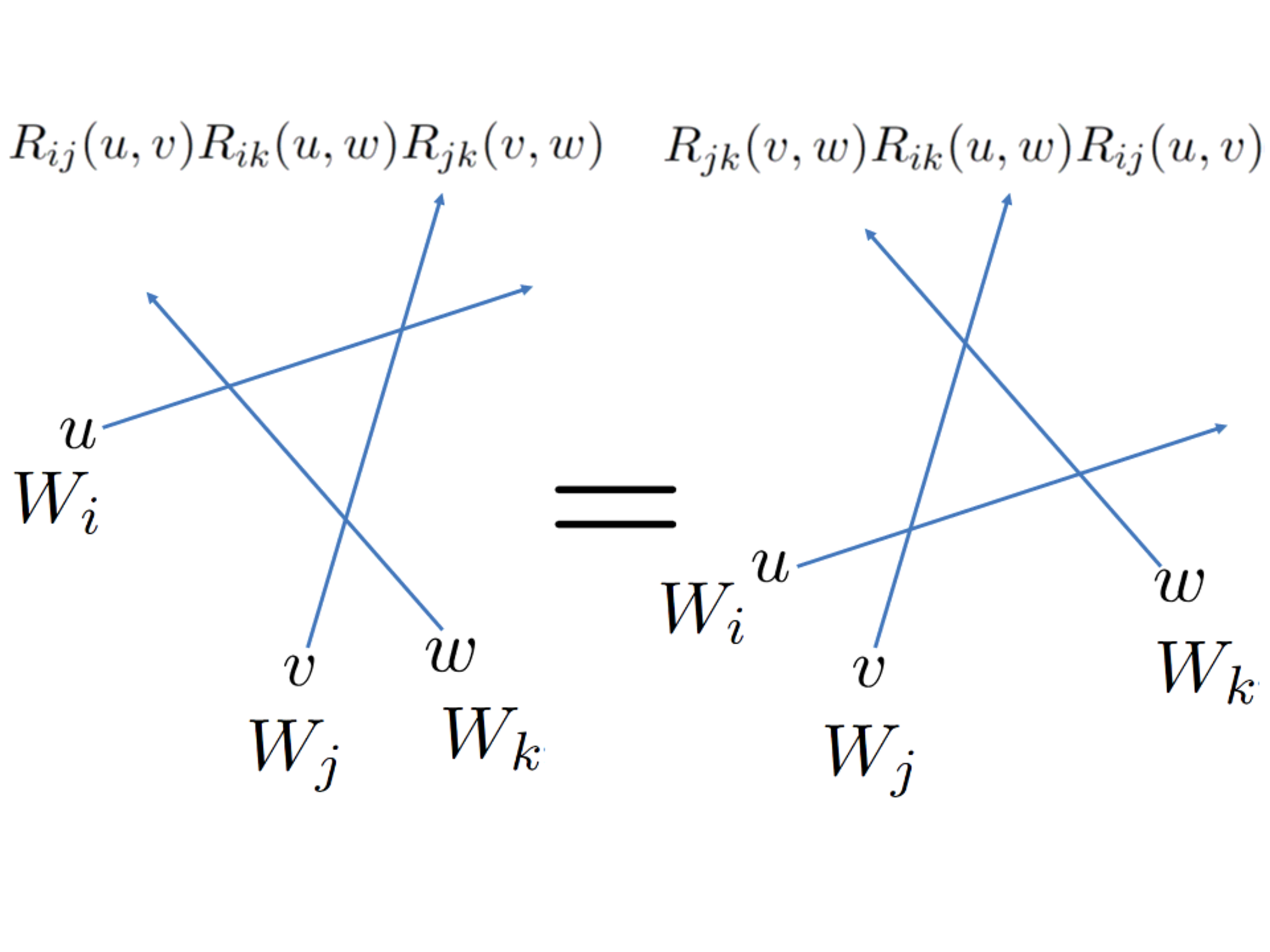}
\caption{The Yang-Baxter relation \eqref{yangbaxter}.
The left and right figure represents $R_{ij}(u,v)R_{ik}(u,w)R_{jk}(v,w)$
and $R_{jk}(v,w)R_{ik}(u,w)R_{ij}(u,v)$ respectively.
}
\label{pictureyangbaxter}
\end{figure}

The operator $R_{ij}(u,w)$ defined as
\eqref{defrone}, \eqref{defrtwo}, \eqref{defrthree}, \eqref{defrfour}
satisifies the Yang-Baxter relation
(Figure \ref{pictureyangbaxter})
\begin{align}
R_{ij}(u,v)R_{ik}(u,w)R_{jk}(v,w)
=R_{jk}(v,w)R_{ik}(u,w)R_{ij}(u,v), \label{yangbaxter}
\end{align}
which acts on $W_i \otimes W_j \otimes W_k$.
Note that
$R_{ij}(u,v)$, $R_{ik}(u,w)$ and $R_{jk}(v,w)$ are operators
acting on $W_i \otimes W_j \otimes W_k$,
and acts nontrivially on $W_i \otimes W_j$,
$W_i \otimes W_k$ and $W_j \otimes W_k$ respectively,
and acts as identity on the remaining space.
In terms of matrix entries,
the Yang-Baxter relation \eqref{yangbaxter} means that
\begin{align}
&\sum_{\gamma_1,\gamma_2,\gamma_3=0,1}
[R(u,v)]_{\gamma_3 \gamma_1}^{\delta_1 \delta_2}
[R(u,w)]_{\epsilon_1 \gamma_2}^{\gamma_3 \delta_3}
[R(v,w)]_{\epsilon_2 \epsilon_3}^{\gamma_1 \gamma_2} \nonumber \\
=&\sum_{\gamma_1,\gamma_2,\gamma_3=0,1}
[R(v,w)]_{\gamma_2 \gamma_3}^{\delta_2 \delta_3}
[R(u,w)]_{\gamma_1 \epsilon_3}^{\delta_1 \gamma_3}
[R(u,v)]_{\epsilon_1 \epsilon_2}^{\gamma_1 \gamma_2},
\label{yangbaxtermatrixelements}
\end{align}
holds for $\epsilon_1,\epsilon_2,\epsilon_3,\delta_1,\delta_2,\delta_3=0,1$.
Acting both hand sides of \eqref{yangbaxter} on $| \epsilon_1 \rangle_i \otimes |\epsilon_2 \rangle_j \otimes |\epsilon_3 \rangle_k$ and taking coefficients
of $| \delta_1 \rangle_i \otimes |\delta_2 \rangle_j \otimes |\delta_3 \rangle_k$ give \eqref{yangbaxtermatrixelements}.
Note that
$R_{ij}(u,v)$, $R_{ik}(u,w)$ and $R_{jk}(v,w)$ act on
$W_i \otimes W_j \otimes W_k$, hence are $2^3 \times 2^3$ matrices, and
we used relations \eqref{embeddingrelation}
to express the matrix elements in terms of
$[R(u,v)]_{\epsilon_1 \epsilon_2}^{\gamma_1 \gamma_2}$ etc.
For example,
$[R(u,v)]_{\epsilon_1 \epsilon_2}^{\gamma_1 \gamma_2}$
in the right hand side of \eqref{yangbaxtermatrixelements}
comes from the following computation:
$\displaystyle R_{ij}(u,v)
| \epsilon_1 \rangle_i \otimes |\epsilon_2 \rangle_j \otimes |\epsilon_3 \rangle_k=
\sum_{\gamma_1,\gamma_2,\gamma=0,1}
[R_{ij}(u,v)]_{\epsilon_1 \epsilon_2 \epsilon_3}^{\gamma_1 \gamma_2 \gamma}
| \gamma_1 \rangle_i \otimes |\gamma_2 \rangle_j \otimes |\gamma \rangle_k
=
\sum_{\gamma_1,\gamma_2,\gamma=0,1}
[R(u,v)]_{\epsilon_1 \epsilon_2}^{\gamma_1 \gamma_2} \delta_{\epsilon_3 \gamma}
| \gamma_1 \rangle_i \otimes |\gamma_2 \rangle_j \otimes |\gamma \rangle_k
=
\sum_{\gamma_1,\gamma_2=0,1}
[R(u,v)]_{\epsilon_1 \epsilon_2}^{\gamma_1 \gamma_2}
| \gamma_1 \rangle_i \otimes |\gamma_2 \rangle_j \otimes |\epsilon_3 \rangle_k
$.

We remark that
more generally, \eqref{yangbaxter} can be viewed
as a relation in $W_1 \otimes \cdots \otimes W_m$ for an integer $m \geq 3$
where
$R_{ij}(u,v)$ acts nontrivially on $W_i$ and $W_j$,
$R_{ik}(u,w)$ on $W_i$ and $W_k$,
$R_{jk}(v,w)$ on $W_j$ and $W_k$,
and acts as identity on the remaining space.
The relation \eqref{yangbaxter} also holds
as operators acting on $W_1 \otimes \cdots \otimes W_m$
due to the identities \eqref{yangbaxtermatrixelements}.

We next construct the monodromy matrix $T_a(u)$
\begin{align}
T_{a}(u)&=R_{a, m}(u,1) \cdots R_{a 1}(u,1),
\label{monodromy}
\end{align}
which acts  on
$W_a \otimes W_1 \otimes \cdots \otimes W_{m}$.
The operator $R_{a, j}(u,1)$ $(1 \le j \le m)$ acts on $W_a$ and $W_j$
nontrivially and as identity on the other spaces,
and the space $W_a$
which every $R$-matrix used to construct the monodromy matrix
acts nontrivially is called as the auxiliary space.
To distinguish the auxiliary space from the other spaces $W_1, \dots, W_m$,
we use alphabets, $a$ for example,
to label this space and denote the vector space as $W_a$.

The monodromy matrix $T_a(u)$ can be regarded as
a $2^{m+1} \times 2^{m+1}$ matrix
since it acts on the $2^{m+1}$-dimensional space $W_a \otimes W_1 \otimes \cdots \otimes W_{m}$.
The matrix elements of the monodromy matrix
$[T(u)]_{\epsilon_a,\epsilon_1,\dots, \epsilon_m}^{\delta_a,\delta_1, \dots,
\delta_m}$
for $\epsilon_a,\epsilon_1,\dots, \epsilon_m,
\delta_a,\delta_1, \dots,
\delta_m=0,1
$
defined as
\begin{align}
T_a(u) |\epsilon_a \rangle_a \otimes |\epsilon_1 \rangle_1 \otimes \cdots \otimes
|\epsilon_m \rangle_m
&=\sum_{\delta_a,\delta_1,\dots,\delta_m=0,1}
[T(u)]_{\epsilon_a,\epsilon_1,\dots, \epsilon_m}^{\delta_a,\delta_1, \dots,
\delta_m}
|\delta_a \rangle_a \otimes |\delta_1 \rangle_1 \otimes \cdots \otimes
|\delta_m \rangle_m,
\end{align}
can be written using the $R$-matrix elements as
\begin{align}
[T(u)]_{\epsilon_a,\epsilon_1,\dots, \epsilon_m}^{\delta_a,\delta_1, \dots, 
\delta_m}
&=\sum_{\gamma_1,\dots,\gamma_{m-1}=0,1}[R(u,1)]_{\gamma_{m-1} 
\epsilon_m}^{\delta_a \delta_m}
\prod_{j=1}^{m-2}
[R(u,1)]_{\gamma_{j} \epsilon_{j+1}}^{\gamma_{j+1} \delta_{j+1}}
[R(u,1)]_{\epsilon_a \epsilon_1}^{\gamma_1 \delta_1}.
\end{align}

The matrix elements of the monodromy matrix
$T_{a}(u)$
 with respect to the
auxiliary space $W_a$ are called as the $A$- $B$- $C$- and $D$-operators.
Namely, we write the results of the action of the operator
$T_{a}(u)$ on the basis vectors $| 0 \rangle_a$, $|1 \rangle_a$
of the space $W_a$ as
\begin{align}
T_a(u)|0 \rangle_a&=|0 \rangle_a \otimes A(u)+|1 \rangle_a \otimes C(u), \label{defmonodromyelementsone} \\
T_a(u)|1 \rangle_a&=|0 \rangle_a \otimes B(u)+|1 \rangle_a \otimes D(u), \label{defmonodromyelementstwo}
\end{align}
and the coefficients $A(u)$, $B(u)$, $C(u)$ and $D(u)$
are the $ABCD$-operators. Note that $A(u)$, $B(u)$, $C(u)$ and $D(u)$
act on the $2^m$-dimensional vector space
$W_1 \otimes \cdots \otimes W_{m}$,
and can be regarded as $2^m \times 2^m$ matrices.
By definition \eqref{defmonodromyelementsone}, \eqref{defmonodromyelementstwo}, the matrix elements
$[A(u)]_{\epsilon_1,\dots, \epsilon_m}^{\delta_1, \dots, \delta_m}$,
$[B(u)]_{\epsilon_1,\dots, \epsilon_m}^{\delta_1, \dots, \delta_m}$,
$[C(u)]_{\epsilon_1,\dots, \epsilon_m}^{\delta_1, \dots, \delta_m}$,
$[D(u)]_{\epsilon_1,\dots, \epsilon_m}^{\delta_1, \dots, \delta_m}$
for $\epsilon_1,\dots, \epsilon_m,\delta_1, \dots, \delta_m=0,1$
defined as
\begin{align}
A(u) |\epsilon_1 \rangle_1 \otimes \cdots \otimes
|\epsilon_m \rangle_m
&=\sum_{\delta_1,\dots,\delta_m=0,1}
[A(u)]_{\epsilon_1,\dots, \epsilon_m}^{\delta_1, \dots, \delta_m}
|\delta_1 \rangle_1 \otimes \cdots \otimes|\delta_m \rangle_m, \\
B(u) |\epsilon_1 \rangle_1 \otimes \cdots \otimes
|\epsilon_m \rangle_m
&=\sum_{\delta_1,\dots,\delta_m=0,1}
[B(u)]_{\epsilon_1,\dots, \epsilon_m}^{\delta_1, \dots, \delta_m}
|\delta_1 \rangle_1 \otimes \cdots \otimes|\delta_m \rangle_m, \\
C(u) |\epsilon_1 \rangle_1 \otimes \cdots \otimes
|\epsilon_m \rangle_m
&=\sum_{\delta_1,\dots,\delta_m=0,1}
[C(u)]_{\epsilon_1,\dots, \epsilon_m}^{\delta_1, \dots, \delta_m}
|\delta_1 \rangle_1 \otimes \cdots \otimes|\delta_m \rangle_m, \\
D(u) |\epsilon_1 \rangle_1 \otimes \cdots \otimes
|\epsilon_m \rangle_m
&=\sum_{\delta_1,\dots,\delta_m=0,1}
[D(u)]_{\epsilon_1,\dots, \epsilon_m}^{\delta_1, \dots, \delta_m}
|\delta_1 \rangle_1 \otimes \cdots \otimes|\delta_m \rangle_m,
\end{align}
can be obtained from the monodromy matrix elements
$[T(u)]_{\epsilon_a,\epsilon_1,\dots, \epsilon_m}^{\delta_a,\delta_1,
\dots, \delta_m}$
by specializing $\epsilon_a$ and $\delta_a$ as follows:
\begin{align}
[A(u)]_{\epsilon_1,\dots, \epsilon_m}^{\delta_1, \dots, \delta_m}
=[T(u)]_{0,\epsilon_1,\dots, \epsilon_m}^{0,\delta_1, \dots, \delta_m},
\ \ \
[B(u)]_{\epsilon_1,\dots, \epsilon_m}^{\delta_1, \dots, \delta_m}
=[T(u)]_{1,\epsilon_1,\dots, \epsilon_m}^{0,\delta_1, \dots, \delta_m}, 
\\
[C(u)]_{\epsilon_1,\dots, \epsilon_m}^{\delta_1, \dots, \delta_m}
=[T(u)]_{0,\epsilon_1,\dots, \epsilon_m}^{1,\delta_1, \dots, \delta_m},
\ \ \
[D(u)]_{\epsilon_1,\dots, \epsilon_m}^{\delta_1, \dots, \delta_m}
=[T(u)]_{1,\epsilon_1,\dots, \epsilon_m}^{1,\delta_1, \dots, \delta_m}
.
\end{align}

Using the bra-ket notation,
the $ABCD$-operators are written as
\begin{align}
A(u)&={}_a \langle 0 |T_a(u)| 0 \rangle_a,
\\
B(u)&={}_a \langle 0 |T_a(u)| 1 \rangle_a,
\label{boperator} \\
C(u)&={}_a \langle 1 |T_a(u)| 0 \rangle_a,
\\
D(u)&={}_a \langle 1 |T_a(u)| 1 \rangle_a.
\label{doperator}
\end{align}

\begin{figure}[ht]
\includegraphics[width=10cm]{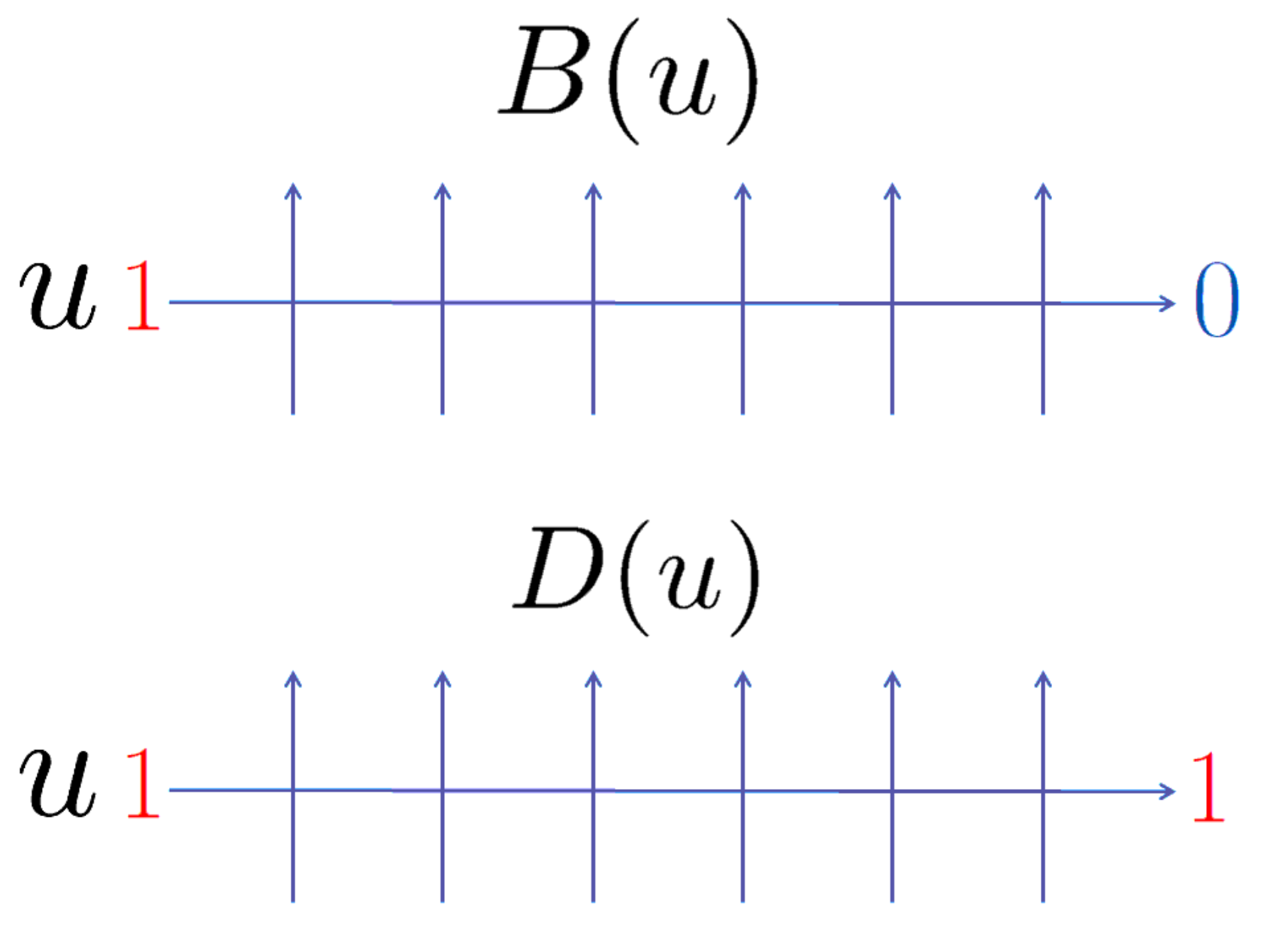}
\caption{The $B$-operator \eqref{boperator} (top)
and the $D$-operator \eqref{doperator} (bottom).
}
\label{picturebandd}
\end{figure}

In this paper, we focus on commutation relations and partition functions
constructed from the $B$-operator \eqref{boperator}
and the $D$-operator \eqref{doperator}
(Figure \ref{picturebandd}).

Let us review the commutation relations in this section.
The $ABCD$-operators satisfy commutation relations
which can be obtained from the intertwining relation
\begin{align}
R_{ab}(u_1,u_2) \widetilde{T}_{a}(u_1) \widetilde{T}_{b}(u_2)
=\widetilde{T}_{b}(u_2) \widetilde{T}_{a}(u_1)R_{ab}(u_1,u_2),
\label{rttone}
\end{align}
acting on $W_a \otimes W_b \otimes W_1 \otimes \cdots \otimes W_m$.
Here, $R_{ab}(u_1,u_2)$ acts nontrivially on $W_a \otimes W_b$
and as identity on $W_1 \otimes \cdots \otimes W_m$.
$\widetilde{T}_{a}(u_1)$ and $\widetilde{T}_{b}(u_2)$
are monodromy matrices acting on $W_a \otimes W_b \otimes W_1 \otimes \cdots \otimes W_m$
which naturally extend the monodromy matrix \eqref{monodromy}.
$\widetilde{T}_a(u_1):=R_{a,m}(u_1,1) \cdots R_{a1}(u_1,1)$
acts on $W_a,W_1,\dots,W_m$ nontrivially and as identity
on $W_b$,
and
$\widetilde{T}_b(u_2):=R_{b,m}(u_2,1) \cdots R_{b1}(u_2,1)$
acts on $W_b,W_1,\dots,W_m$ nontrivially and as identity
on $W_a$.
Note that $\widetilde{T}_a(u_1)$ can be regarded as a $2^{m+2} \times 2^{m+2}$
matrix, and since it acts on $W_b$ trivially,
the matrix elements of $\widetilde{T}_a(u_1)$
with respect to the space $W_b$ are given as
\begin{align}
{}_b \langle 0| \widetilde{T}_a(u_1)|0 \rangle_b=
{}_b \langle 1| \widetilde{T}_a(u_1)|1 \rangle_b=
T_a(u_1), \ \ \
{}_b \langle 0| \widetilde{T}_a(u_1)|1 \rangle_b=
{}_b \langle 1| \widetilde{T}_a(u_1)|0 \rangle_b=\bf{0},
\label{reductionttildeone}
\end{align}
using the bra-ket notation.
Here, $T_a(u_1)$ is the monodromy matrix \eqref{monodromy}
which acts on $W_a \otimes W_1 \otimes \cdots \otimes W_m$
which we introduced before,
and
$\bf{0}$ is the $2^{m+1} \times 2^{m+1}$ zero matrix.

Similarly, we note that $\widetilde{T}_b(u_2)$ can be regarded as a $2^{m+2} \times 2^{m+2}$ matrix which acts on $W_a$ trivially, which means
that
the matrix elements with respect to the space $W_a$ are given as
\begin{align}
{}_a \langle 0| \widetilde{T}_b(u_2)|0 \rangle_a=
{}_a \langle 1| \widetilde{T}_b(u_2)|1 \rangle_a=
T_b(u_2), \ \ \
{}_a \langle 0| \widetilde{T}_b(u_2)|1 \rangle_a=
{}_a \langle 1| \widetilde{T}_b(u_2)|0 \rangle_a=\bm{0},
\label{reductionttildetwo}
\end{align}
where $T_b(u_2)=R_{b,m}(u_2,1) \cdots R_{b1}(u_2,1)$
is the monodromy matrix
which acts on $W_b \otimes W_1 \otimes \cdots \otimes W_m$.

The intertwining relation \eqref{rttone}
follows by applying the Yang-Baxter relation \eqref{yangbaxter} repeatedly.
For example, the $m=2$ case of
\eqref{rttone} can be shown as follows:
\begin{align}
&R_{ab}(u_1,u_2) \widetilde{T}_{a}(u_1) \widetilde{T}_{b}(u_2) \nonumber \\
=&R_{ab}(u_1,u_2)R_{a2}(u_1,1)R_{a1}(u_1,1) R_{b2}(u_2,1)R_{b1}(u_2,1)
\nonumber \\
=&R_{ab}(u_1,u_2)R_{a2}(u_1,1) R_{b2}(u_2,1)
R_{a1}(u_1,1)R_{b1}(u_2,1) \nonumber \\
=&R_{b2}(u_2,1) R_{a2}(u_1,1)R_{ab}(u_1,u_2)
R_{a1}(u_1,1)R_{b1}(u_2,1) \nonumber \\
=&R_{b2}(u_2,1) R_{a2}(u_1,1)
R_{b1}(u_2,1)R_{a1}(u_1,1)R_{ab}(u_1,u_2)
\nonumber \\
=&R_{b2}(u_2,1)
R_{b1}(u_2,1)
R_{a2}(u_1,1)
R_{a1}(u_1,1)R_{ab}(u_1,u_2) \nonumber \\
=&\widetilde{T}_{b}(u_2)
\widetilde{T}_{a}(u_2)
R_{ab}(u_1,u_2). \label{exampleintertwining}
\end{align}
In the second equality in \eqref{exampleintertwining},
we used $R_{a1}(u_1,1) R_{b2}(u_2,1)=R_{b2}(u_2,1)R_{a1}(u_1,1)$.
This commutativity holds since $R_{a1}(u_1,1)$ acts nontrivially only on
$W_a$ and $W_1$ and acts as identity on the other spaces,
in particular $W_b$ and $W_2$,
and $R_{b2}(u_2,1)$ acts nontrivially only on $W_b$ and $W_2$
and acts as identity on the rest, in particular $W_a$ and $W_1$,
hence the result of acting $R_{a1}(u_1,1)$ and $R_{b2}(u_2,1)$ successively
on any vector in $W_a \otimes W_b \otimes W_1 \otimes \cdots \otimes W_m$
gives the same result and is independent of the order of the action
of the two operators.
We can also see that $R_{a2}(u_1,1)R_{b1}(u_2,1)=R_{b1}(u_2,1)R_{a2}(u_1,1)$
holds by the same arugment which is used in the fifth equality.
In the third and fourth equalities,
we used the Yang-Baxter relation \eqref{yangbaxter}.

Acting both sides of \eqref{rttone} on $|\epsilon_a \rangle_a \otimes |\epsilon_b \rangle_b \in W_a \otimes W_b$
and noting $\widetilde{T}_a(u_1)$ and $\widetilde{T}_b(u_2)$
acts trivially on $W_b$ and $W_a$ respectively, we get
\begin{align}
&\sum_{\delta_a,\delta_b,\gamma_a,\gamma_b=0,1}
|\gamma_a \rangle_a \otimes | \gamma_b \rangle_b \
[R(u_1,u_2)]_{\delta_a \delta_b}^{\gamma_a \gamma_b} \nonumber \\
&\times {}_a \langle \delta_a |
\otimes
{}_b \langle \delta_b |
\widetilde{T}_a(u_1)|\epsilon_a \rangle_a
\otimes |\delta_b \rangle_b
\
{}_a \langle \epsilon_a | \otimes
{}_b \langle \delta_b |\widetilde{T}_b(u_2)
|\epsilon_a \rangle_a \otimes
|\epsilon_b \rangle_b
\nonumber \\
=&\sum_{\delta_a,\delta_b,\gamma_a,\gamma_b=0,1}
|\gamma_a \rangle_a \otimes | \gamma_b \rangle_b \
{}_a \langle \gamma_a| \otimes
{}_b \langle \gamma_b |\widetilde{T}_b(u_2)
|\gamma_a \rangle_a \otimes |\delta_b \rangle_b \nonumber \\
&\times
{}_a \langle \gamma_a|
\otimes {}_b \langle \delta_b
|\widetilde{T}_a(u_1)|\delta_a \rangle_a
\otimes |\delta_b \rangle_b
\
[R(u_1,u_2)]_{\epsilon_a \epsilon_b}^{\delta_a \delta_b}.
\label{beforeexplicitrtt}
\end{align}

From \eqref{reductionttildeone} and \eqref{reductionttildetwo},
we have
\begin{align}
{}_a \langle \delta_a |
\otimes
{}_b \langle \delta_b |
\widetilde{T}_a(u_1)|\epsilon_a \rangle_a
\otimes |\delta_b \rangle_b
&={}_a \langle \delta_a |
T_a(u_1)|\epsilon_a \rangle_a
, \\
{}_a \langle \epsilon_a | \otimes
{}_b \langle \delta_b |\widetilde{T}_b(u_2)
|\epsilon_a \rangle_a \otimes
|\epsilon_b \rangle_b
&=
{}_b \langle \delta_b |T_b(u_2)
|\epsilon_b \rangle_b, \\
{}_a \langle \gamma_a| \otimes
{}_b \langle \gamma_b |\widetilde{T}_b(u_2)
|\gamma_a \rangle_a \otimes |\delta_b \rangle_b
&={}_b \langle \gamma_b |\widetilde{T}_b(u_2)|\delta_b \rangle_b,
\\
{}_a \langle \gamma_a|
\otimes {}_b \langle \delta_b
|\widetilde{T}_a(u_1)|\delta_a \rangle_a
\otimes |\delta_b \rangle_b
&=
{}_a \langle \gamma_a
|\widetilde{T}_a(u_1)|\delta_a \rangle_a,
\end{align}
and inserting these into \eqref{beforeexplicitrtt},
we get the following commutation relations between
the $ABCD$-operators
\begin{align}
&\sum_{\delta_a,\delta_b=0,1}
[R(u_1,u_2)]_{\delta_a \delta_b}^{\gamma_a \gamma_b} \
{}_a \langle \delta_a |T_a(u_1)|\epsilon_a \rangle_a \
{}_b \langle \delta_b |T_b(u_2)|\epsilon_b \rangle_b
\nonumber \\
=&\sum_{\delta_a,\delta_b=0,1}
{}_b \langle \gamma_b |T_b(u_2)|\delta_b \rangle_b \
{}_a \langle \gamma_a |T_a(u_1)|\delta_a \rangle_a \
[R(u_1,u_2)]_{\epsilon_a \epsilon_b}^{\delta_a \delta_b}.
\label{explicitrtt}
\end{align}
For example, the
$\epsilon_a=\epsilon_b=\gamma_a=1, \ \gamma_b=0$ case of
\eqref{explicitrtt} is
\begin{align}
\displaystyle (1-u_2/u_1)D(u_1)B(u_2)+u_2/u_1 B(u_1)D(u_2)=B(u_2)D(u_1).
\label{beforerttone1}
\end{align}
We rewrite  \eqref{beforerttone1} as
\begin{align}
D(u_1)B(u_2)&=\frac{1}{1-u_2/u_1}B(u_2)D(u_1)
+\frac{1}{1-u_1/u_2}B(u_1)D(u_2), \label{rttone1}
\end{align}
for later purpose. Some other cases of the commutation relations
\eqref{explicitrtt} which are used in this paper are
\begin{align}
D(u_1)B(u_2)&=D(u_2)B(u_1), \label{rttone2} \\
B(u_1)B(u_2)&=B(u_2)B(u_1), \label{rttone3} \\
D(u_1)D(u_2)&=D(u_2)D(u_1). \label{rttone4}
\end{align}
Note that \eqref{rttone1}, \eqref{rttone2}, \eqref{rttone3} and \eqref{rttone4}
are relations which act on $W_1 \otimes \cdots \otimes W_m$
since $B(u)$ and $D(u)$ are operators acting on
$W_1 \otimes \cdots \otimes W_m$.

Now recall the following notations for the sets
given in the introduction. $S_k^n$ is a $k$-subset of $[n]=\{1,2,\dots,n \}$,
the set of $k$-subsets of $n$ is denoted as $\binom{[n]}{k}$,
and $\overline{S_k^n}=\{1,2,\dots,n \} \backslash S_k^n$.
The following commutation relation
between the multiple operators \\
$\prod_{j=k+1}^{n} D(u_j)$
and $\prod_{j=1}^k B(u_j)$
\begin{align}
&\prod_{j=k+1}^{n} D(u_j) \prod_{j=1}^k B(u_j)
=\sum_{S_k^n \in \binom{[n]}{k}} \prod_{i \in S_k^n, j \in \overline{S_k^n}}
\frac{1}{1-u_i/u_j} \prod_{i \in S_k^n} B(u_i)
\prod_{j \in \overline{S_k^n}} D(u_j),
\label{multiplecommutation}
\end{align}
is one of the commutation relations between multiple operators
essentially derived by Shigechi and Uchiyama \cite{ShigechiUchiyama}
in their study of integrable models whose $L$-operators satisify the
$RLL$ Yang-Baxter relation with (a gauge transformed version of)
the five-vertex model $R$-matrix
\eqref{fivevertexrmatrix} intertwining the $L$-operators.
Various multiple commutation relations including this type were investigated
in \cite{ShigechiUchiyama} for the purpose of studying
generalized scalar products.
One can also use this commutation relation to give another proof
\cite{GSFNR} of an identity for the factorial Grothendieck polynomials
\cite{GS}, which is an extension of the one for the Schur polynomials
\cite{FNR}.

We record the argument given in
\cite{ShigechiUchiyama} (which appears
in Proof of Theorem 6.1) for completeness.
First, using \eqref{rttone1}, \eqref{rttone3} and \eqref{rttone4}
to move all the $B$-operators to the left of all the $D$-operators,
one notes the operator part of all the terms which appear can be expressed as
\begin{align}
\prod_{i \in S_k^n} B(u_i)
\prod_{j \in \overline{S_k^n}} D(u_j),
\label{operatorpart}
\end{align}
for $\displaystyle S_k^n \in \binom{[n]}{k}$.
To extract the coefficient of \eqref{operatorpart}
for a fixed $S_k^n$,
one uses \eqref{rttone2} repeatedly to rewrite
the left hand side of \eqref{multiplecommutation} as
\begin{align}
\prod_{j \in \overline{S_k^n}} D(u_j)
\prod_{i \in S_k^n} B(u_i).
\end{align}
Finally, we only use \eqref{rttone1} repeatedly to move
all the $B$-operators to the left of all the $D$-operators.
We only need to concentrate on the first
term of the right hand side of \eqref{rttone1}
when commuting the $B$- and $D$-operators
to extract the coefficient
of \eqref{operatorpart},
since if we once use the second term of the left hand side of \eqref{rttone1},
we get other operators.
Noting this, one finds the coefficient of the operator is given by
$\displaystyle \prod_{i \in S_k^n, j \in \overline{S_k^n}} \frac{1}{1-u_i/u_j}$
, and hence the commutation relation \eqref{multiplecommutation} follows.

\section{Partition functions}
In this section, we introduce several types of partition functions
which are used in this paper,
and give their explicit forms
using one version of skew Grothendieck polynomials.

\begin{figure}[ht]
\includegraphics[width=12cm]{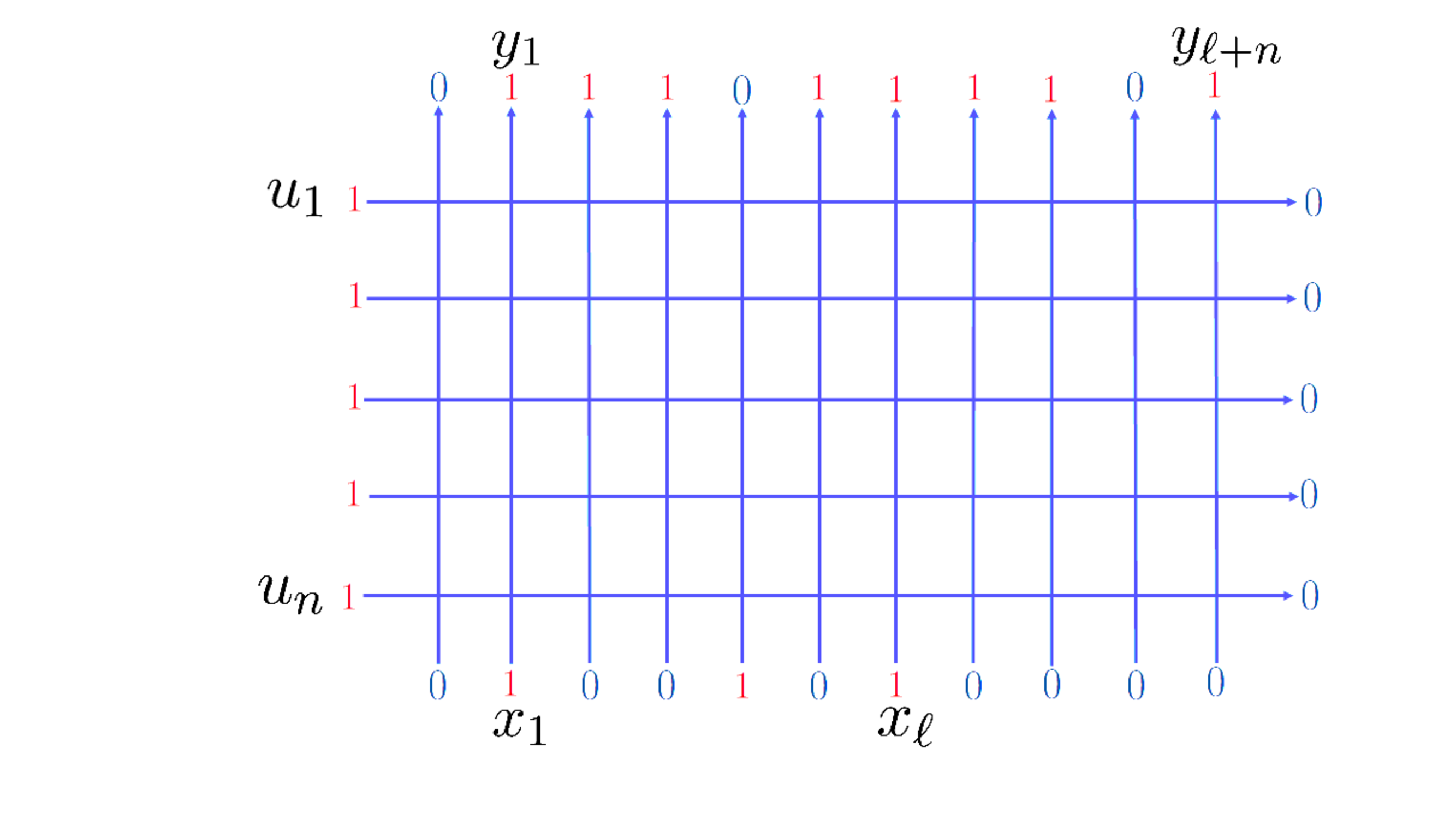}
\caption{The partition functions
$Z_{m,n}(u_1,\dots,u_n|x_1,\dots,x_\ell|y_1,\dots,y_{\ell+n})$
constructed from the $B$-operators
\eqref{fivevertexpartitionfunction}.
The figure represents the case $m=11$, $n=5$, $\ell=3$,
$(x_1,x_2,x_3)=(2,5,7)$, $(y_1,y_2,y_3,y_4,y_5,y_6,y_7,y_8)
=(2,3,4,6,7,8,9,11)$.
}
\label{picturebpartition}
\end{figure}

\begin{figure}[ht]
\includegraphics[width=12cm]{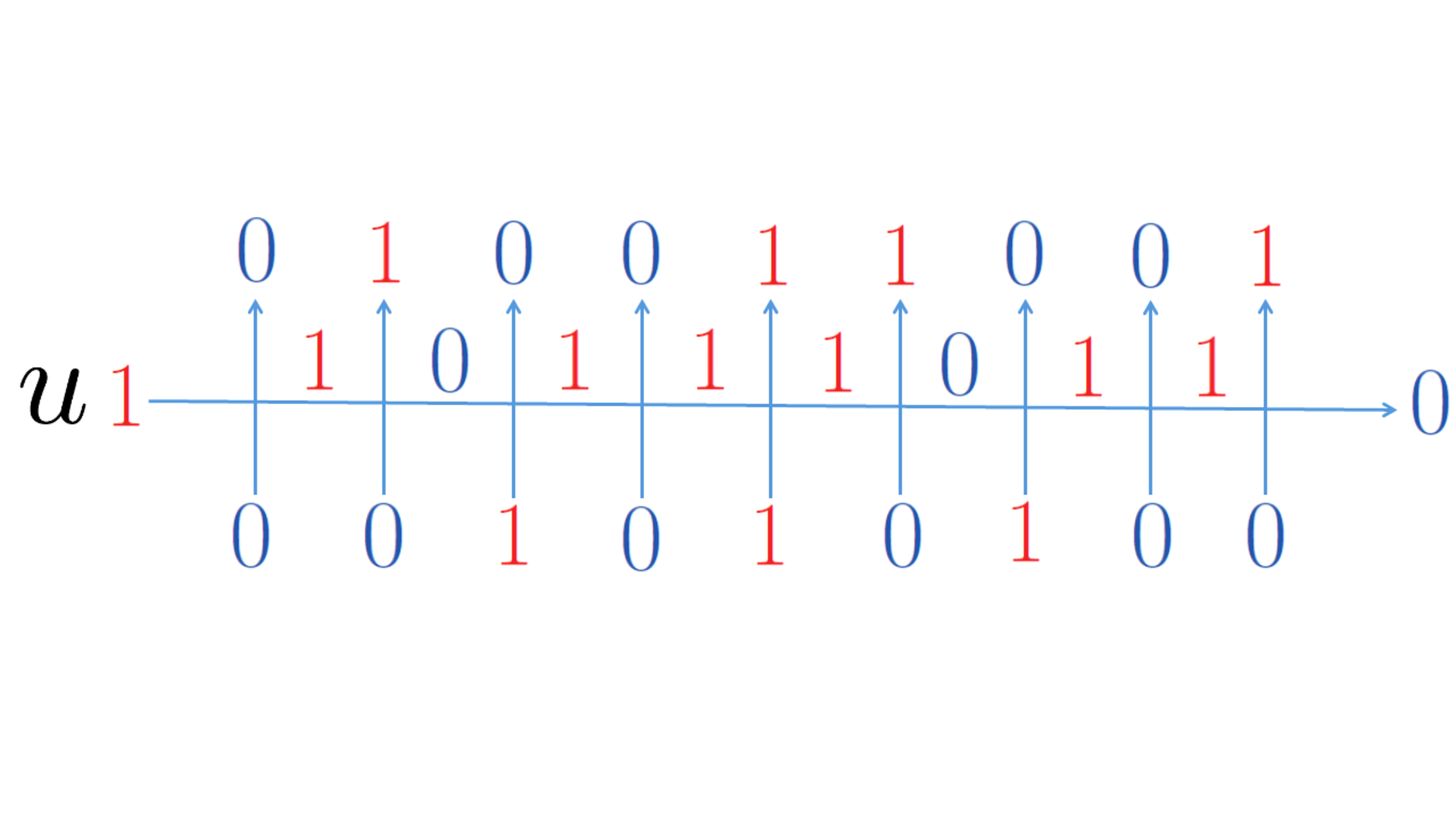}
\caption{The matrix element
${}_{m} \langle y_1 \cdots y_{\ell+1} | B(u) | x_1 \cdots x_\ell \rangle_m $
which we take as a definition of $G_{\lambda /\!/ \mu}(1-u^{-1})$.
The figure represents the case $m=9$, $\ell=3$, $(x_1,x_2,x_3)=(3,5,7)$, $(y_1,y_2,y_3,y_4)=(2,5,6,9)$.
The matrix element of the $B$-operator can be computed as
${}_{9} \langle 2,5,6,9 | B(u) | 3,5,7 \rangle_9
=
[R(u,1)]_{10}^{01}
[R(u,1)]_{10}^{10}
[R(u,1)]_{01}^{10}
[R(u,1)]_{10}^{01}
[R(u,1)]_{11}^{11}
[R(u,1)]_{10}^{10}
[R(u,1)]_{01}^{10}
[R(u,1)]_{10}^{01}
[R(u,1)]_{10}^{10}
=1 \times (1-u^{-1}) \times u^{-1} \times 1 \times 1 \times (1-u^{-1})
\times u^{-1} \times 1 \times (1-u^{-1})
=(1-u^{-1})^3 u^{-2}$.
We can compute graphically by
multiplying all nine $R$-matrix elements
$1-u^{-1}$, $1$, $u^{-1}$, $(1-u^{-1})$, $1$, $1$, $u^{-1}$, $1-u^{-1}$, $1$
in the figure (see also Figure \ref{picturermatrix}).
In terms of skew Grothendieck polynomials, $\lambda$ and $\mu$ in this example
are $\lambda=(\lambda_1,\lambda_2,\lambda_3,\lambda_4)=(y_4-4,y_3-3,y_2-2,y_3-1)=(5,3,3,1)$ and $\mu=(\mu_1,\mu_2,\mu_3)=(x_3-3,x_2-2,x_1-1)=(4,3,2)$,
so this matrix element of the $B$-operator corresponds to
$G_{(5,3,3,1) /\!/ (4,3,2)}(1-u^{-1})$.
One can check the right hand side of \eqref{singlematrixelements}
when $\lambda=(5,3,3,1)$, $\mu=(4,3,2)$ is $(1-u^{-1})^3 u^{-2}$.
}
\label{picturematrixelement}
\end{figure}

For an increasing sequence of integers $\bm{x}=(x_1,x_2,\dots,x_\ell)$ and
an integer $j$,
let us define $\delta_{j,\bm{x}}$ as $\delta_{j,\bm{x}}=1$
if there exists $x_k$ $(1 \le k \le \ell)$
such that $j=x_k$, and $\delta_{j,\bm{x}}=0$ otherwise.
For each
increasing sequence of integers
$1 \le x_1 < x_2 < \cdots < x_\ell \le m$,
we define the configuration vector and its dual as
\begin{align}
|x_1 \cdots x_\ell \rangle_m
&:=\displaystyle \otimes_{j=1}^m (
|0 \rangle_j + \delta_{j,\bm{x}} ( | 1 \rangle_j -| 0 \rangle_j ))
\in W_1 \otimes \cdots \otimes W_{m}
, \label{ordinarydualparticleconfigurationtwo}
\end{align}
and
\begin{align}
_{m} \langle x_1 \cdots x_\ell|
&:=
\displaystyle \otimes_{j=1}^m (
{}_j \langle 0 | + \delta_{j,\bm{x}} ({}_j \langle 1 |
-{}_j \langle 0|))
\in W_1^* \otimes \cdots \otimes W_{m}^*
. \label{ordinarydualparticleconfiguration}
\end{align}
For example, $|3,5 \rangle_6=|0 \rangle_1 \otimes 
|0 \rangle_2 \otimes |1 \rangle_3 \otimes 
|0 \rangle_4 \otimes |1 \rangle_5 \otimes 
|0 \rangle_6$.

By using the $B$-operators, we introduce the first type of partition functions
\\
$Z_{m,n}(u_1,\dots,u_n|x_1,\dots,x_\ell|y_1,\dots,y_{\ell+n})$
for two increasing sequences of integers
$1 \le x_1 < x_2 < \cdots < x_\ell \le m$ and
$1 \le y_1 < y_2 < \cdots < y_{\ell+n} \le m$
as (Figure \ref{picturebpartition})
\begin{align}
&Z_{m,n}(u_1,\dots,u_n|x_1,\dots,x_\ell|y_1,\dots,y_{\ell+n})
={}_{m} \langle y_1 \cdots y_{\ell+n}|
B(u_1) \cdots B(u_n)|x_1 \cdots x_\ell \rangle_{m}.
\label{fivevertexpartitionfunction}
\end{align}

For $\ell=0$, we define $Z_{m,n}(u_1,\dots,u_n|\phi|y_1,\dots,y_{n})$ as
\begin{align}
Z_{m,n}(u_1,\dots,u_n|\phi|y_1,\dots,y_{n})
:=
{}_{m} \langle y_1 \cdots y_{n}|
B(u_1) \cdots B(u_n)|0 \rangle_1 \otimes \cdots \otimes |0 \rangle_m.
\label{offshellbethedefinition}
\end{align}
The partition functions
$Z_{m,n}(u_1,\dots,u_n|\phi|y_1,\dots,y_{n})$
are called as the off-shell Bethe wavefunctions of the five-vertex model,
and can be expressed using the Grassmannian Grothendieck polynomials
\cite{LS,Buch1,Buch2,Buch3,Mc}, whose determinant representation
is given by \cite{Buch2,Lenart,IN,IS}
\begin{align}
G_\lambda(z_1,\dots,z_n)=
   \frac{\mathrm{det}(z_i^{\lambda_j+n-j}(1-z_i)^{j-1})_{i \le i,j \le n}}
        {\prod_{1 \le i < j \le n}(z_i-z_j)},
 \label{doubleGrothendieck}
\end{align}
where
$\{z_1, \dots, z_n \}$ is a set of commuting variables
and $\lambda=(\lambda_1,\lambda_2,\dots,\lambda_n)$
is a  partition, i.e. a nonincreasing sequence of nonnegative integers
whose graphical representation is naturally given by the Young diagram.

The off-shell Bethe wavefunctions \eqref{offshellbethedefinition}
of the integrable five-vertex model is a partition function representation
of the Grothendieck polynomials \cite{MS,MS2,GK,WZ}.
\begin{proposition} (\cite{MS} Lemma 5.2, \cite{MS2} Thm 2.2,  \cite{GK} Prop 4.1, \cite{WZ} Prop 1)
We have
\begin{align}
&Z_{m,n}(u_1,\dots,u_n|\phi|y_1,\dots,y_{n})
=G_\lambda(1-u_1^{-1},\dots,1-u_n^{-1}),
\label{correspondence}
\end{align}
where $\lambda$ and $y_1,\dots,y_n$ are related by
$\lambda_j=y_{n-j+1}-n+j-1 \ (j=1,\dots,n)$.
\end{proposition}

In view of the correspondence \eqref{correspondence},
we introduce a version of the skew Grothendieck polynomials
with a single variable as the matrix elements of a single
$B$-operator:
\begin{align}
G_{\lambda /\!/ \mu}(1-u^{-1})
={}_{m} \langle y_1 \cdots y_{\ell+1} | B(u)
| x_1 \cdots x_\ell \rangle_m, \label{defskew}
\end{align}
where
$\lambda_j=y_{\ell-j+2}-\ell+j-2 \ (j=1,\dots,\ell+1)$
and $\mu_j=x_{\ell-j+1}-\ell+j-1 \ (j=1,\dots,\ell)$.

One can compute the matrix elements of a single $B$-operator
${}_{m} \langle y_1 \cdots y_{\ell+1} | B(u) | x_1 \cdots x_\ell \rangle_m $
explicitly (see \cite{MS2} Prop 3.4 for example),
and through the identification \eqref{defskew},
the skew Grothendieck polynomials $G_{\lambda /\!/ \mu}(1-u^{-1})$
is explicitly given by
\begin{align}
&G_{\lambda /\!/ \mu}(1-u^{-1})
=
\left\{
\begin{array}{ll}
(1-u^{-1} )^{\sum_{j=1}^{\ell+1} \lambda_j
-\sum_{j=1}^\ell \mu_j}
\prod_{j=1}^\ell
\{
u^{-1}+
\delta_{\lambda_{j+1},\mu_j}
(
1-u^{-1}
) \},& \lambda \succ \mu, \\
0, & \mathrm{otherwise}.
\end{array}
\right.
\label{singlematrixelements}
\end{align}
Here, $\lambda \succ \mu$ is defined as the following
relation between $\lambda$ and $\mu$
\begin{align}
\lambda_1 \ge \mu_1 \ge \lambda_2 \ge \mu_2 \ge \cdots \ge \mu_\ell \ge
\lambda_{\ell+1},
\label{interlacing}
\end{align}
which is called as the interlacing relation.
See Figure \ref{picturematrixelement} for an example of
\eqref{singlematrixelements},
and \cite{MS2} Prop 3.4 for a proof of
\eqref{singlematrixelements}
(the argument in the proof of Prop 3.4
applied to the $R$-matrix \eqref{fivevertexrmatrix} gives
\eqref{singlematrixelements}.
The $R$-matrix in \cite{MS2} and
\eqref{fivevertexrmatrix} are related by gauge transformation).

Let us give a remark.
We used the notation
$G_{\lambda /\!/ \mu}(1-u^{-1})$
here instead of
$G_{\lambda / \mu}(1-u^{-1})$.
There are two versions of skew Grothendieck polynomials
$G_{\lambda/\mu}(\bm{x})$
and $G_{\lambda /\!/ \mu}(\bm{x})$
introduced by Buch in \cite{Buch2}.
$G_{\lambda/\mu}(\bm{x})$
is defined as certain types of stable Grothendieck polynomials
which are originally labeled by permutations,
and $G_{\lambda /\!/ \mu}(\bm{x})$
is introduced as a natural skew version
of the stable Grothendieck polynomials
from the coproduct structure.
A further study of this type of skew Grothendieck polynomials
was done by Yeliussizov in \cite{Yeliussizov},
in which skew Grothendieck polynomials
with finite number of variables is first defined using Schur operators,
and skew Grothendieck polynomials with infinitely many variables
is defined as a limit. 
We note that
the explicit expression for one variable case \eqref{singlematrixelements}
which we defined as matrix elements of the $B$-operators
is the same with the $\beta=1$ case of the one 
given in Proposition 4.3 in \cite{Yeliussizov},
hence we use this notation.
The skew Grothendieck polynomials with multivariables
which we later define as \eqref{skewGrothendieck}
can also be shown to be the same with the
one which is introduced using Schur operators
in \cite{Yeliussizov}, Definition 4.1. See Appendix A for details.

We can show the following relation
\begin{align}
G_\lambda(1-u_1^{-1},\dots,1-u_\ell^{-1},1-u_{\ell+1}^{-1})
=\sum_{\lambda \succ \mu} G_{\lambda /\!/ \mu}(1-u_{\ell+1}^{-1})
G_\mu(1-u_1^{-1},\dots,1-u_\ell^{-1}), \label{ktheoreticaddition}
\end{align}
using the lattice model construction as follows:
\begin{align}
&G_\lambda(1-u_1^{-1},\dots,1-u_\ell^{-1},1-u_{\ell+1}^{-1}) \nonumber \\
=&{}_m \langle y_1 \cdots y_{\ell+1}|
B(u_{1}) \cdots B(u_{\ell+1}) |0 \rangle_1 \otimes \cdots \otimes |0 \rangle_m
\nonumber \\
=&{}_m \langle y_1 \cdots y_{\ell+1}| B(u_{\ell+1}) \times \mathrm{Id} \times
B(u_{1}) \cdots B(u_\ell) |0 \rangle_1 \otimes \cdots \otimes |0 \rangle_m
\nonumber \\
=&{}_m \langle y_1 \cdots y_{\ell+1}| B(u_{\ell+1}) \times 
\Bigg(
|0 \rangle_1 \otimes \cdots \otimes |0 \rangle_m \
{}_1 \langle 0| \otimes \cdots  \otimes {}_m \langle 0|
\nonumber \\
&+
\sum_{j=1}^m
\sum_{1 \le x_1 < \cdots < x_j \le m}
|x_1 \cdots x_j \rangle_m \ {}_m \langle x_1 \cdots x_j |
\Bigg)
\times
B(u_{1}) \cdots B(u_\ell) |0 \rangle_1 \otimes \cdots \otimes |0 \rangle_m
\nonumber \\
=&
\sum_{1 \le x_1 < \cdots < x_\ell \le m}
{}_m \langle y_1 \cdots y_{\ell+1}| B(u_{\ell+1})
|x_1 \cdots x_\ell \rangle_m \ {}_m \langle x_1 \cdots x_\ell |
B(u_{1}) \cdots B(u_\ell) |0 \rangle_1 \otimes \cdots \otimes |0 \rangle_m
\nonumber \\
=&\sum_{\lambda \succ \mu} G_{\lambda /\!/ \mu}(1-u_{\ell+1}^{-1})
G_\mu(1-u_1^{-1},\dots,1-u_\ell^{-1}).
\end{align}
Here we used the commutativity of $B$-operators
\eqref{rttone3} in the second equality.
We also inserted the identity operator on $W_1 \otimes \cdots \otimes W_m$
\begin{align}
\mathrm{Id}=|0 \rangle_1 \otimes \cdots \otimes |0 \rangle_m \
{}_1 \langle 0| \otimes \cdots  \otimes {}_m \langle 0|+
\sum_{j=1}^m
\sum_{1 \le x_1 < \cdots < x_j \le m}
|x_1 \cdots x_j \rangle_m \ {}_m \langle x_1 \cdots x_j |,
\label{identityoperator}
\end{align}
which is one way of rewriting the standard expression
for the identity operator
\begin{align}
\mathrm{Id}=\sum_{\epsilon_1,\epsilon_2,\dots,\epsilon_m=0,1}
|\epsilon_1 \rangle_1 \otimes |\epsilon_2 \rangle_2 \otimes
\cdots \otimes |\epsilon_m \rangle_m \
{}_1 \langle \epsilon_1 | \otimes {}_2 \langle \epsilon_2 |
\otimes \cdots \otimes {}_m \langle \epsilon_m|.
\end{align}
The expression above follows from the bra-ket expression for the identity operator
$\displaystyle |0 \rangle {} \langle 0|+|1 \rangle {} \langle 1|=\sum_{\epsilon=0,1}
|\epsilon \rangle \langle \epsilon |$ on the two-dimensional vector space $W$.

From the ice-rule for the $R$-matrix, we note that
$
B(u_{1}) \cdots B(u_\ell) |0 \rangle_1 \otimes \cdots \otimes |0 \rangle_m$
is written as a linear combination of
$|x_1 \cdots x_\ell \rangle_m$, $1 \le x_1 < \cdots < x_\ell \le m$,
which means that
${}_m \langle x_1 \cdots x_j|
B(u_{1}) \cdots B(u_\ell) |0 \rangle_1 \otimes \cdots \otimes |0 \rangle_m=0$
unless $j=\ell$.
Hence, we can restrict the inserted identity operator
to
\begin{align}
\sum_{1 \le x_1 < \cdots < x_\ell \le m}
|x_1 \cdots x_\ell \rangle_m \ {}_m \langle x_1 \cdots x_\ell |,
\label{restrictedidentityoperator}
\end{align}
which we used in the fourth equality.
Let us call the argument of restricting
the inserted identity operator
\eqref{identityoperator} to 
\eqref{restrictedidentityoperator} as the restriction argument.

The relation \eqref{ktheoreticaddition}
can be regarded as a $K$-theoretic version
of the following relation for the Schur and skew Schur functions
\begin{align}
s_\lambda(x_1,\dots,x_\ell,x_{\ell+1})
=\sum_{\lambda \succ \mu} s_{\lambda/\mu}(x_{\ell+1})
s_\mu(x_1,\dots,x_{\ell}),
\end{align}
and as a finite variable truncation of the definition
of the skew stable Grothendieck polynomials using coproduct
$\Delta:\Gamma \rightarrow \Gamma \otimes \Gamma$ with
$(\Delta f)(x,y)=f(x,y)$
(Buch \cite{Buch2}, (6.3))
\begin{align}
\Delta(G_\lambda)=\sum_{\mu \subset \lambda} G_\mu \otimes G_{\lambda /\!/\mu}.
\label{coproductdefinition}
\end{align}
Here, $\Gamma=\oplus_\lambda \mathbb{Z} \cdot
G_\lambda$
where $G_\lambda$ are stable Grothendieck polynomials.

From \eqref{coproductdefinition}, we have
(\cite{Buch2}, section 9, \cite{Yeliussizov}, Remark 4.2)
\begin{align}
G_{\lambda /\!/ \lambda}(1-u_1^{-1},1-u_2^{-1},\dots)=\prod_{k \ge 1} u_k^{-i(\lambda)},
\end{align}
where $i(\lambda)$ is the number of indices $i$ such that $\lambda_i > \lambda_{i+1}$.
This is different from $G_{\lambda/\lambda}=G_\phi=1$ in general.
Truncating to the one variable case gives
$G_{\lambda /\!/ \lambda}(1-u_1^{-1})=u_1^{-i(\lambda)}$,
which matches the definition of the skew Grothendieck polynomials
for one variable $G_{\lambda /\!/}(1-u^{-1})$
as matrix elements
of the $B$-operator \eqref{defskew}.
For example,
$G_{(1) /\!/ (1)}(1-u_1^{-1})=
{}_{3} \langle 1,3 | B(u_1)|2 \rangle_3
=[R(u_1,1)]_{10}^{01} [R(u_1,1)]_{01}^{10} [R(u_1,1)]_{10}^{01}=u_1^{-1}
=u_1^{-i((1))}
$.

We introduce the skew Grothendieck polynomials
with multi variables by composing skew Grothendieck
polynomials of a single variable as
\begin{align}
&G_{\lambda /\!/  \mu}(1-u_1^{-1},\dots,1-u_n^{-1})
=\sum_{\lambda=\lambda^{(0)} 
\succ \lambda^{(1)} \succ \cdots \succ \lambda^{(n)}=\mu}
\prod_{j=1}^n G_{\lambda^{(j-1)} /\!/ \lambda^{(j)}}(1-u_j^{-1}).
\label{skewGrothendieck}
\end{align}

The following determinant formula for the skew Grothendieck polynomials
is shown by Iwao \cite{Iwaoskew} by the boson-fermion correspondence.
\begin{proposition} (Iwao \cite{Iwaoskew}, $\beta=-1$ case of Proposition 4.8)
For $n \le r-\ell(\mu)$, we have
\begin{align}
\displaystyle
G_{\lambda /\!/ \mu}(z_1,\dots,z_n)
=\mathrm{det} \Bigg( \sum_{k=0}^{\mu_j-j+r} \binom{1-j}{k}
H_{\lambda_i-\mu_j-i+j+k}^{(i-1)}(z_1,\dots,z_n)
\Bigg)_{1 \le i,j \le r},
\end{align}
where
\begin{align}
\displaystyle
H_p^{(i)}(z_1,\dots,z_n):=\sum_{\ell=0}^\infty \binom{i}{\ell} (-1)^\ell
h_{p+\ell}(z_1,\dots,z_n).
\end{align}
Here, binomials are defined as
$\displaystyle \binom{p}{q}=\frac{\prod_{j=0}^{q-1}(p-j)}{q!}$
for $p,q$ integers, and $h_i(z_1,\dots,z_n)$ is the $i$-th
complete symmetric polynomial with commuting variables $z_1,\dots,z_n$.
\end{proposition}

\begin{figure}[ht]
\includegraphics[width=12cm]{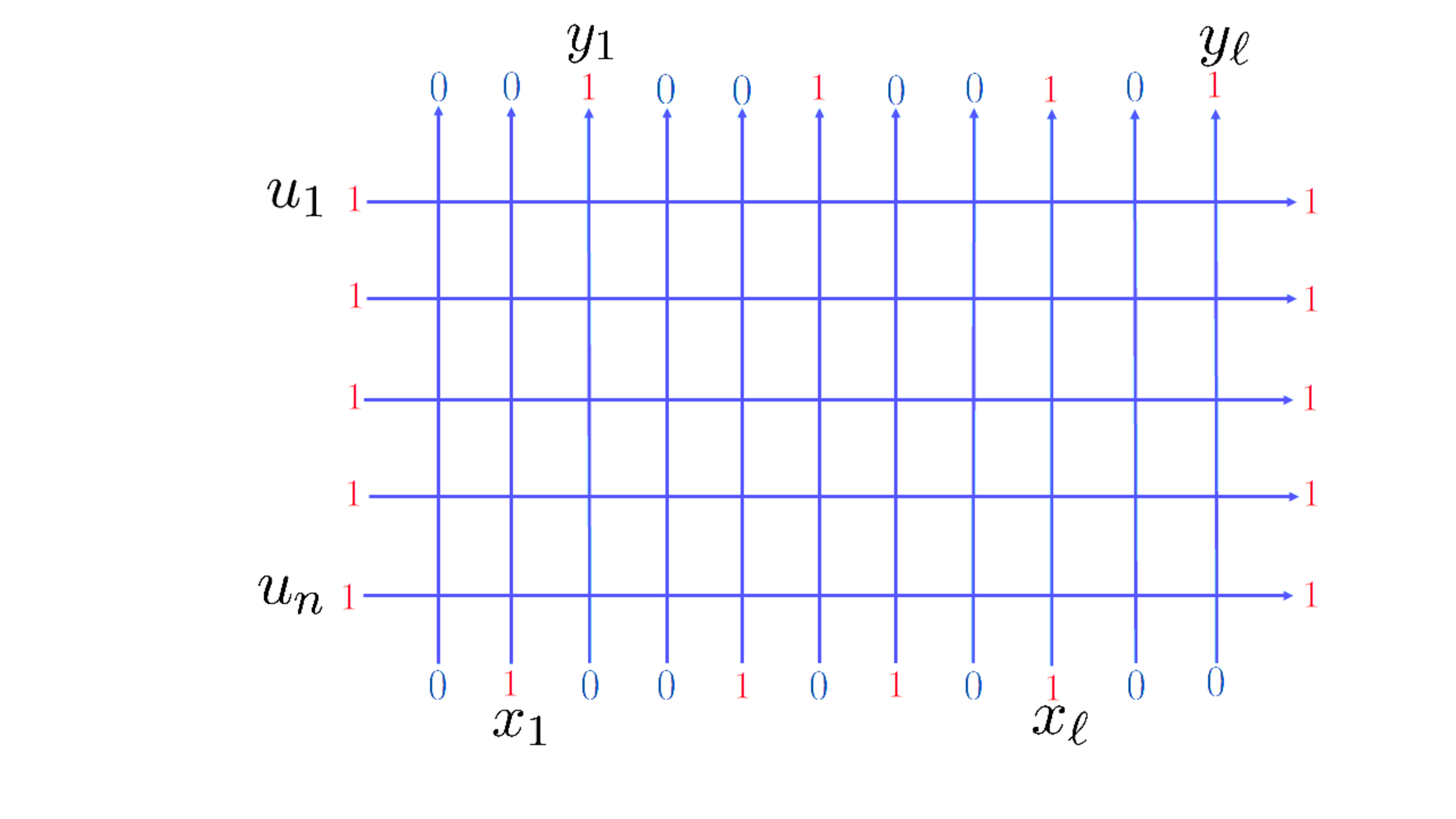}
\caption{The partition functions
$U_{m,n}(u_1,\dots,u_n|x_1,\dots,x_\ell|y_1,\dots,y_{\ell})$
constructed fom the $D$-operators
\eqref{Doperatorpartitionfunction}.
The figure represents the case $m=11$, $n=5$, $\ell=4$,
$(x_1,x_2,x_3,x_4)=(2,5,7,9)$, $(y_1,y_2,y_3,y_4)
=(3,6,9,11)$.
}
\label{picturedpartition}
\end{figure}

\begin{figure}[ht]
\includegraphics[width=12cm]{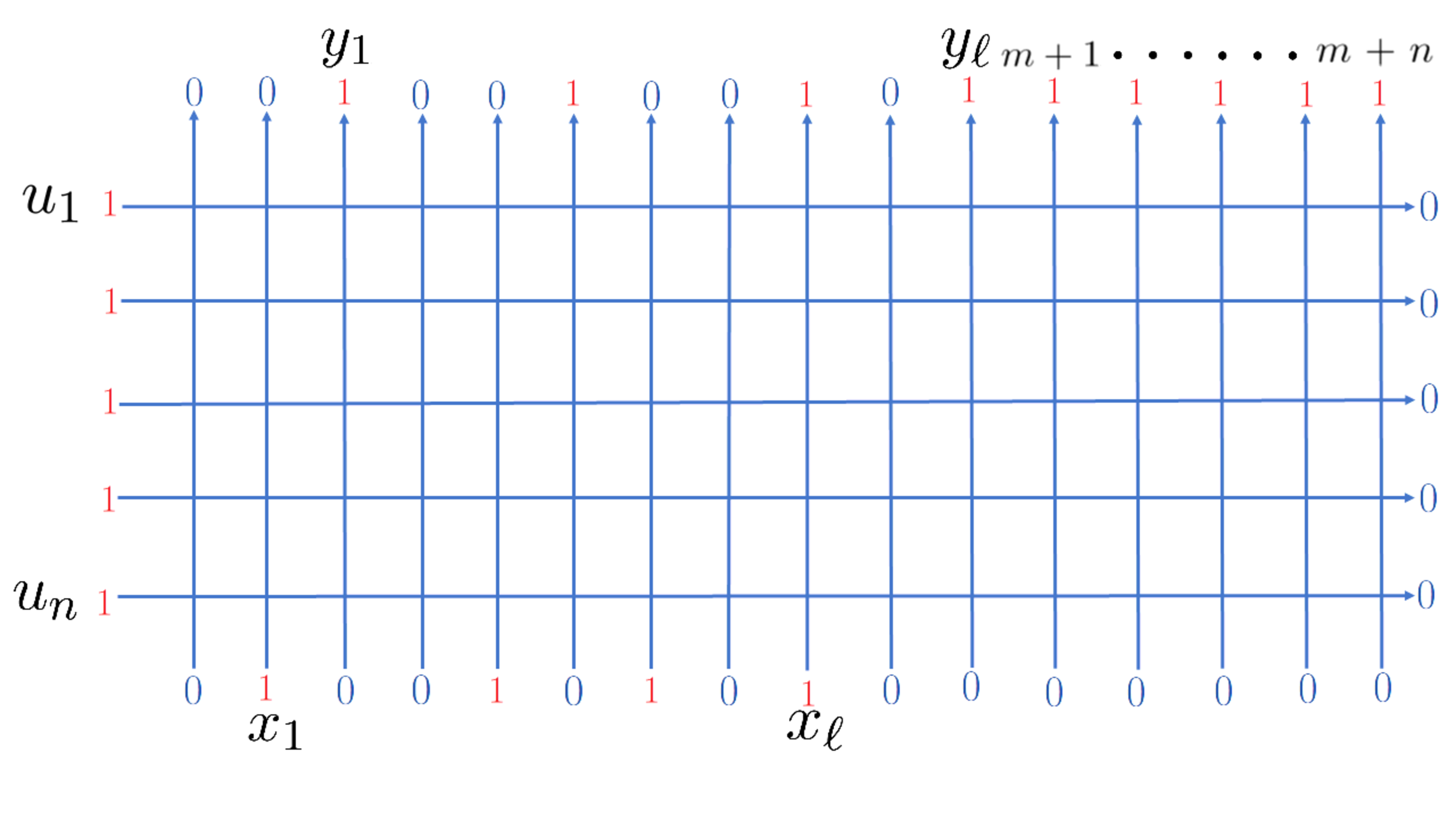}
\caption{The partition functions
$Z_{m+n,n}(u_1,\dots,u_n|x_1,\dots,x_\ell|y_1,\dots,y_{\ell},m+1,\dots,m+n)$ \eqref{anotherlarger}.
The figure represents the case $m=11$, $n=5$, $\ell=4$,
$(x_1,x_2,x_3,x_4)=(2,5,7,9)$, $(y_1,y_2,y_3,y_4)
=(3,6,9,11)$.
}
\label{picturebpartitiondefinition}
\end{figure}

The partition functions
$Z_{m,n}(u_1,\dots,u_n|x_1,\dots,x_\ell|y_1,\dots,y_{\ell+n})$
\eqref{fivevertexpartitionfunction}
are constructed by taking matrix elements
of multiple $B$-operators.
By inserting the identity operator \eqref{identityoperator}
between the $B$-operators
and using the restriction argument and \eqref{defskew},
the partition functions can be
rewritten as the right hand side of \eqref{skewGrothendieck},
hence the following correspondence holds:
\begin{align}
&Z_{m,n}(u_1,\dots,u_n|x_1,\dots,x_\ell|y_1,\dots,y_{\ell+n})
=G_{\lambda /\!/ \mu}(1-u_1^{-1},\dots,1-u_n^{-1}),
\label{skewcorrespondence}
\end{align}
where
$\lambda_j=y_{\ell-j+n+1}-\ell+j-n-1 \ (j=1,\dots,\ell+n)$
and $\mu_j=x_{\ell-j+1}-\ell+j-1 \ (j=1,\dots,\ell)$.

Let us illustrate the $n=2$ case of showing \eqref{skewcorrespondence}
(the general case can be shown in the same way).
This can be shown in the following way
\begin{align}
&Z_{m,2}(u_1,u_2|x_1,\dots,x_\ell|y_1,\dots,y_{\ell+2}) \nonumber \\
=&{}_m \langle y_1 \cdots y_{\ell+2}|B(u_1) B(u_2) |x_1 \cdots x_\ell
\rangle_m \nonumber \\
=&{}_m \langle y_1 \cdots y_{\ell+2}|B(u_1)
\times \mathrm{Id} \times
B(u_2) |x_1 \cdots x_\ell
\rangle_m \nonumber \\
=&
{}_m \langle y_1 \cdots y_{\ell+2}|B(u_1)
\times
\Bigg( |0 \rangle_1 \otimes \cdots \otimes |0 \rangle_m \
{}_1 \langle 0| \otimes \cdots \otimes {}_m \langle 0|
\nonumber \\
+&
\sum_{j=1}^m
\sum_{1 \le z_1 < \cdots < z_j \le m}
|z_1 \cdots z_j \rangle_m \ {}_m \langle z_1 \cdots z_j |
\Bigg) \times B(u_2) |x_1 \cdots x_\ell
\rangle_m \nonumber \\
=&\sum_{1 \le z_1 < \cdots < z_{\ell+1} \le m}
{}_m \langle y_1 \cdots y_{\ell+2}|B(u_1)
|z_1 \cdots z_{\ell+1} \rangle_m \ {}_m \langle z_1 \cdots z_{\ell+1} |
B(u_2) |x_1 \cdots x_\ell
\rangle_m \nonumber \\
=&\sum_{\lambda \succ \lambda^{(1)} \succ \mu}
G_{\lambda /\!/ \lambda^{(1)}}(1-u_1^{-1})
G_{\lambda^{(1)} /\!/ \mu}(1-u_2^{-1}) \nonumber \\
=&G_{\lambda /\!/ \mu}(1-u_1^{-1},1-u_2^{-1}), \label{n=2proof}
\end{align}
where $\lambda^{(1)}=(\lambda_1^{(1)},\dots,\lambda_{\ell+1}^{(1)})$
is defined as $\lambda_j^{(1)}=z_{\ell-j+2}-\ell+j-2$ $(j=1,\dots,\ell+1)$
.
Note that we used the property
${}_m \langle z_1 \cdots z_{j} |
B(u_2) |x_1 \cdots x_\ell
\rangle_m=0$ unless $j=\ell+1$ in the fourth equality
in \eqref{n=2proof}
which comes from the ice-rule of the $R$-matrix,
and used \eqref{defskew} and \eqref{singlematrixelements}
in the fifth equality.

Next, we introduce the second type of partition functions
constructed from the $D$-operators.
Define
$U_{m,n}(u_1,\dots,u_n|x_1,\dots,x_\ell|y_1,\dots,y_{\ell})$
for two increasing sequences of integers
$1 \le x_1 < x_2 < \cdots < x_\ell \le m$ and
$1 \le y_1 < y_2 < \cdots < y_{\ell} \le m$
as (Figure \ref{picturedpartition})
\begin{align}
&U_{m,n}(u_1,\dots,u_n|x_1,\dots,x_\ell|y_1,\dots,y_{\ell})
={}_{m} \langle y_1 \cdots y_{\ell}|
D(u_1) \cdots
D(u_n)|x_1 \cdots x_\ell \rangle_{m}.
\label{Doperatorpartitionfunction}
\end{align}
The partition functions \eqref{Doperatorpartitionfunction} introduced
as matrix elements of multiple $D$-operators,
can be connected with partition functions of the first type
which are constructed from the $B$-operators,
and hence with the skew Grothendieck polynomials.
Consider the following partition functions
constructed from the $B$-operators (Figure \ref{picturebpartitiondefinition})
\begin{align}
&Z_{m+n,n}(u_1,\dots,u_n|x_1,\dots,x_\ell|y_1,\dots,y_{\ell},m+1,\dots,m+n)
\nonumber \\
=&_{m+n} \langle y_1 \cdots y_{\ell},m+1,\dots,m+n|
B^{(m+n)}(u_1) \cdots B^{(m+n)}(u_n)|x_1 \cdots x_\ell \rangle_{m+n}, \label{anotherlarger}
\end{align}
where
\begin{align}
B^{(m+n)}(u)={}_a \langle 0|R_{a,m+n}(u) \cdots R_{a1}(u)|1 \rangle_a \in \mathrm{End}(W_1 \otimes \cdots \otimes W_{m+n}).
\end{align}

We can show that the matrix elements of multiple $D$-operators
\eqref{Doperatorpartitionfunction}
are the same with
those of multiple $B$-operators
\eqref{anotherlarger}.
\begin{lemma} \label{firstlemmaproofinappendix}
The following relation holds:
\begin{align}
&Z_{m+n,n}(u_1,\dots,u_n|x_1,\dots,x_\ell|y_1,\dots,y_{\ell},m+1,\dots,m+n) \nonumber \\
=&U_{m,n}(u_1,\dots,u_n|x_1,\dots,x_\ell|y_1,\dots,y_{\ell}).
\label{relationBandD}
\end{align}
\end{lemma}
In Appendix B, a full proof
of \eqref{relationBandD} using equations only is given.
We also give a graphical derivation of \eqref{relationBandD} in Appendix C.

Applying the correspondence between partition functions
and skew Grothendieck polynomials \eqref{skewcorrespondence}
to $Z_{m+n,n}(u_1,\dots,u_n|x_1,\dots,x_\ell|y_1,\dots,y_{\ell},m+1,\dots,m+n)$ and
combining with \eqref{relationBandD},
one finds that the partition functions of the $D$-operators \\
$U_{m,n}(u_1,\dots,u_n|x_1,\dots,x_\ell|y_1,\dots,y_{\ell})$
are explicitly given by the skew Grothendieck polynomials as
\begin{align}
&U_{m,n}(u_1,\dots,u_n|x_1,\dots,x_\ell|y_1,\dots,y_{\ell})
=G_{((m-\ell)^n,\lambda) /\!/ \mu}(1-u_1^{-1},\dots,1-u_n^{-1}),
\label{skewcorrespondenceforD}
\end{align}
where
$\lambda_j=y_{\ell-j+1}-\ell+j-1 \ (j=1,\dots,\ell)$
and $\mu_j=x_{\ell-j+1}-\ell+j-1 \ (j=1,\dots,\ell)$.

\begin{figure}[ht]
\includegraphics[width=12cm]{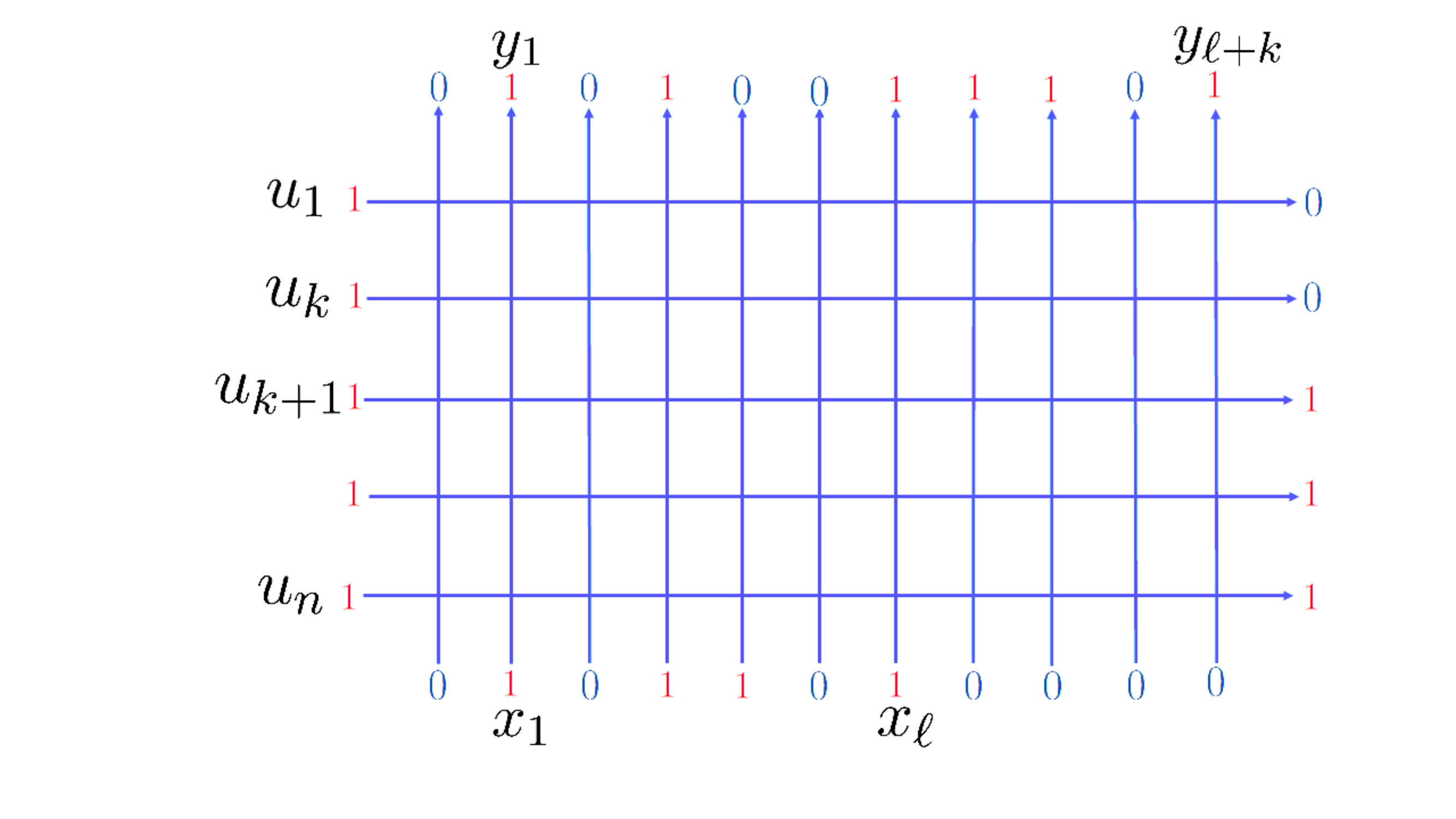}
\caption{The mixed partition functions
$ZBD_{m,n,k}(u_1,\dots,u_k|u_{k+1},\dots,u_n|x_1,\dots,x_\ell|y_1,\dots,y_{\ell+k})$ constructed from both the
$B$- and $D$-operators \eqref{bdpartitionfunction}.
The figure represents the case $m=11$, $n=5$, $\ell=4$, $k=2$
$(x_1,x_2,x_3,x_4)=(2,4,5,7)$, $(y_1,y_2,y_3,y_4,y_5,y_6)
=(2,4,7,8,9,11)$.
}
\label{picturebdpartition}
\end{figure}

In this section, we also introduce
two types of partition functions
which are constructed from both the $B$-operators and the $D$-operators,
and describe their explicit forms using skew Grothendieck polynoimals.

The first type of mixed partition functions is
(Figure \ref{picturebdpartition})
\begin{align}
&ZBD_{m,n,k}(u_1,\dots,u_k|u_{k+1},\dots,u_n|x_1,\dots,x_\ell|y_1,\dots,y_{\ell+k})
\nonumber
\\
=&
_{m} \langle y_1 \cdots y_{\ell+k}|
B(u_1) \cdots B(u_k)
D(u_{k+1}) \cdots D(u_n)
|x_1 \cdots x_\ell \rangle_{m}.
\label{bdpartitionfunction}
\end{align}
By inserting the identity operator \eqref{identityoperator}
between
$B(u_k)$ and
$D(u_{k+1})$ and using
the restriction argument and the correspondences
\eqref{skewcorrespondence} and
\eqref{skewcorrespondenceforD},
\eqref{bdpartitionfunction} can be described
as a sum of products of the skew Grothendieck polynomials
as
\begin{align}
&ZBD_{m,n,k}(u_1,\dots,u_k|u_{k+1},\dots,u_n|x_1,\dots,x_\ell|y_1,\dots,y_{\ell+k})
\nonumber
\\
=&\sum_{\nu \subseteq (m-\ell)^\ell}
G_{\lambda /\!/ \nu}(1-u_1^{-1},\dots,1-u_k^{-1})
G_{((m-\ell)^{n-k},\nu) /\!/ \mu}(1-u_{k+1}^{-1},\dots,1-u_n^{-1}),
\label{bdpartitionfunctionexplicit}
\end{align}
where
$\lambda_j=y_{\ell-j+k+1}-\ell+j-k-1 \ (j=1,\dots,\ell+k)$
and
$\mu_j=x_{\ell-j+1}-\ell+j-1 \ (j=1,\dots,\ell)$.
\eqref{bdpartitionfunctionexplicit} can be shown as follows
\begin{align}
&ZBD_{m,n,k}(u_1,\dots,u_k|u_{k+1},\dots,u_n|x_1,\dots,x_\ell|y_1,\dots,y_{\ell+k})
\nonumber
\\
=&
_{m} \langle y_1 \cdots y_{\ell+k}|
B(u_1) \cdots B(u_k)
D(u_{k+1}) \cdots D(u_n)
|x_1 \cdots x_\ell \rangle_{m} \nonumber \\
=&
_{m} \langle y_1 \cdots y_{\ell+k}|
B(u_1) \cdots B(u_k) \times \mathrm{Id} \times
D(u_{k+1}) \cdots D(u_n)
|x_1 \cdots x_\ell \rangle_{m} \nonumber \\
=&
_{m} \langle y_1 \cdots y_{\ell+k}|
B(u_1) \cdots B(u_k) \times
\Bigg( |0 \rangle_1 \otimes \cdots \otimes |0 \rangle_m \
{}_1 \langle 0| \otimes \cdots \otimes {}_m \langle 0|
\nonumber \\
+&
\sum_{j=1}^m
\sum_{1 \le z_1 < \cdots < z_j \le m}
|z_1 \cdots z_j \rangle_m \ {}_m \langle z_1 \cdots z_j |
\Bigg)
\times
D(u_{k+1}) \cdots D(u_n)
|x_1 \cdots x_\ell \rangle_{m} \nonumber \\
=&\sum_{1 \le z_1 < \cdots < z_\ell \le m}
{}_{m} \langle y_1 \cdots y_{\ell+k}|
B(u_1) \cdots B(u_k)
|z_1 \cdots z_\ell \rangle_m \nonumber \\
&\times {}_m \langle z_1 \cdots z_\ell |
D(u_{k+1}) \cdots D(u_n)
|x_1 \cdots x_\ell \rangle_{m}
\nonumber \\
=&\sum_{1 \le z_1 < \cdots < z_\ell \le m}
Z_{m,k}(u_{1},\dots,u_{k}|z_1,\dots,z_\ell|y_1,\dots,y_{\ell+k})
\nonumber \\
&\times U_{m,n-k}(u_{k+1},\dots,u_{n}|x_1,\dots,x_\ell|z_1,\dots,z_{\ell})
\nonumber \\
=&\sum_{\nu \subseteq (m-\ell)^\ell}
G_{\lambda /\!/ \nu}(1-u_1^{-1},\dots,1-u_k^{-1})
G_{((m-\ell)^{n-k},\nu) /\!/ \mu}(1-u_{k+1}^{-1},\dots,1-u_n^{-1}),
\label{bdpartitionfunctionexplicitproof}
\end{align}
where $\nu=(\nu_1,\dots,\nu_\ell)$ is defined as
$\nu_j=z_{\ell-j+1}-\ell+j-1 \ (j=1,\dots,\ell)$.
Note that we used
${}_{m} \langle y_1 \cdots y_{\ell+k}|
B(u_1) \cdots B(u_k)
|z_1 \cdots z_j \rangle_m=0$ unless
$j=\ell$ in the fourth equality in
\eqref{bdpartitionfunctionexplicitproof}
which follows from the ice-rule of the $R$-matrix,
and \eqref{skewcorrespondence}, \eqref{skewcorrespondenceforD}
in the last equality.

\begin{figure}[ht]
\includegraphics[width=12cm]{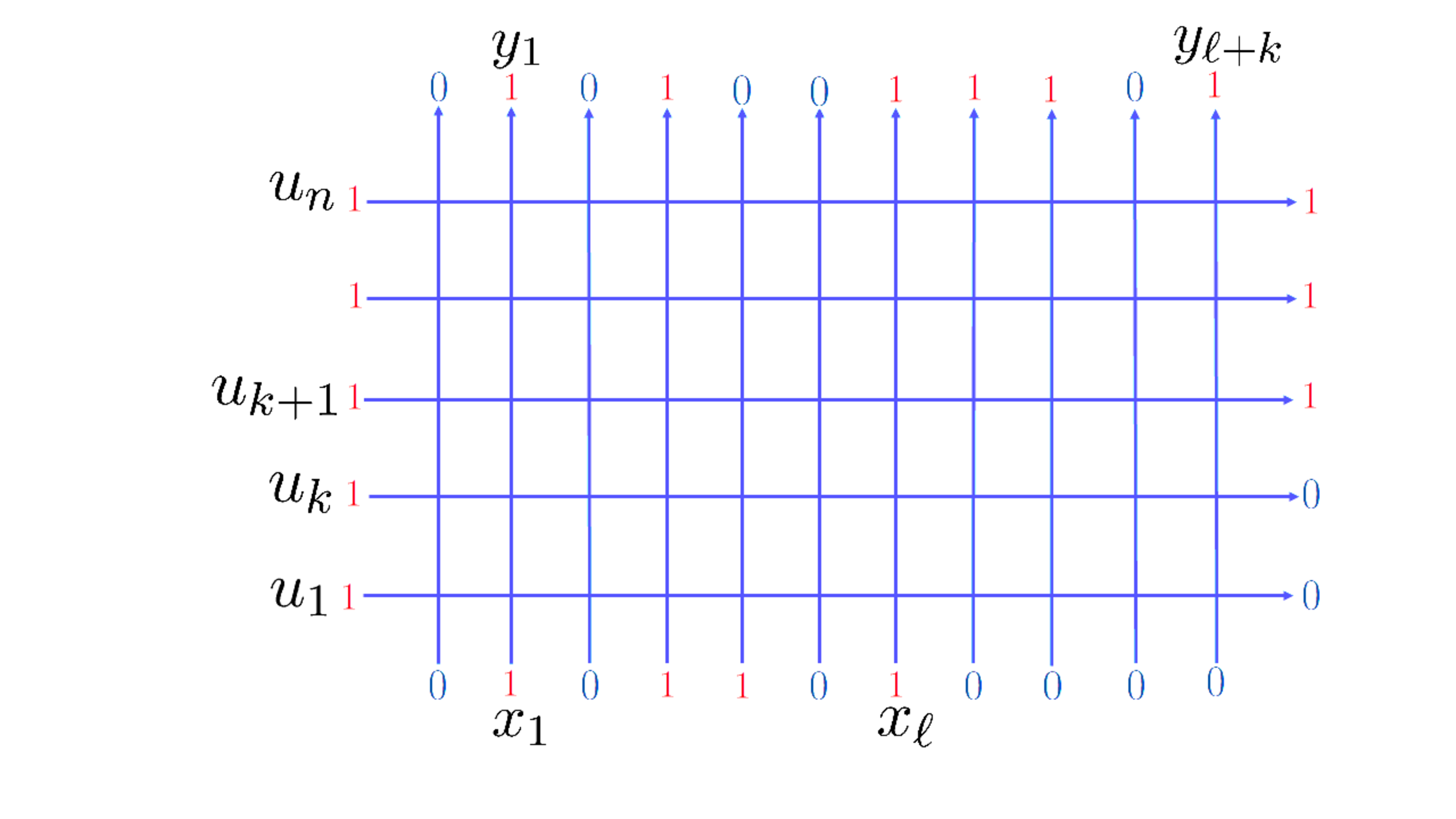}
\caption{The mixed partition functions
$ZDB_{m,n,k}(u_{k+1},\dots,u_n|u_1,\dots,u_k|x_1,\dots,x_\ell|y_1,\dots,y_{\ell+k})$
constructed from both the $B$- and $D$-operators
\eqref{dbpartitionfunction}.
Note the order of the layers of the $B$-operators and
those of the $D$-operators is reversed from
\eqref{bdpartitionfunction}.
The figure represents the case $m=11$, $n=5$, $\ell=4$, $k=2$
$(x_1,x_2,x_3,x_4)=(2,4,5,7)$, $(y_1,y_2,y_3,y_4,y_5,y_6)
=(2,4,7,8,9,11)$.
}
\label{picturedbpartition}
\end{figure}

\begin{figure}[ht]
\includegraphics[width=12cm]{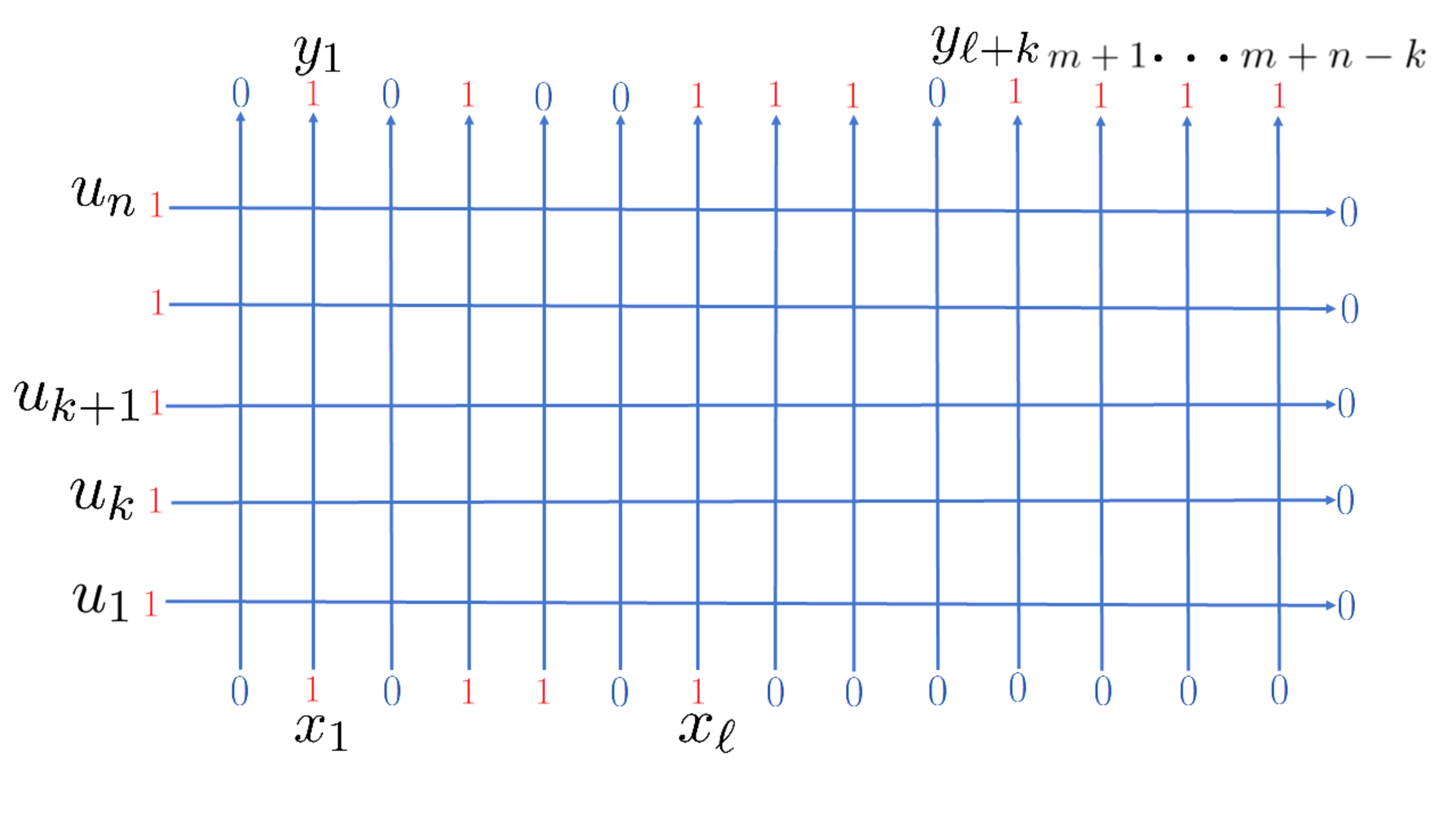}
\caption{The partition functions
$Z_{m+n-k,n}
(u_1,\dots,u_n|x_1,\dots,x_\ell|y_1,\dots,y_{\ell+k},
m+1,\dots,m+n-k)$ \eqref{larger}.
The figure represents the case $m=11$, $n=5$, $\ell=4$, $k=2$
$(x_1,x_2,x_3,x_4)=(2,4,5,7)$, $(y_1,y_2,y_3,y_4,y_5,y_6)
=(2,4,7,8,9,11)$.
}
\label{picturedbpartitiondefinition}
\end{figure}

The second type of mixed partition functions
we use in this paper is (Figure \ref{picturedbpartition})
\begin{align}
&ZDB_{m,n,k}(u_{k+1},\dots,u_n|u_1,\dots,u_k|x_1,\dots,x_\ell|y_1,\dots,y_{\ell+k})
\nonumber
\\
=&
_{m} \langle y_1 \cdots y_{\ell+k}|
D(u_n) \cdots D(u_{k+1}) B(u_k) \cdots B(u_1)
|x_1 \cdots x_\ell \rangle_{m},
\label{dbpartitionfunction}
\end{align}
where the order of the layers of the $B$-operators and
those of the $D$-operators is reversed from
the first type of mixed partition functions \eqref{bdpartitionfunction}.
One can write the explicit form of \eqref{dbpartitionfunction}
exactly as the skew Grothendieck polynomials.
Consider the following partition functions
constructed only from the $B$-operators \eqref{fivevertexpartitionfunction}
(Figure \ref{picturedbpartitiondefinition})
\begin{align}
&Z_{m+n-k,n}
(u_1,\dots,u_n|x_1,\dots,x_\ell|y_1,\dots,y_{\ell+k},
m+1,\dots,m+n-k)
\nonumber \\
=&_{m} \langle y_1 \cdots y_{\ell+k},m+1 \cdots m+n-k|
B^{(m+n-k)}(u_n) \cdots B^{(m+n-k)}(u_1)|x_1 \cdots x_\ell \rangle_{m},
\label{larger}
\end{align}
where
\begin{align}
B^{(m+n-k)}(u)={}_a \langle 0|R_{a,m+n-k}(u) \cdots R_{a1}(u)|1 \rangle_a \in \mathrm{End}(W_1 \otimes \cdots \otimes W_{m+n-k}).
\end{align}
We can also show that the mixed partition functions \eqref{dbpartitionfunction}
are the same with the matrix elements \eqref{larger}.
\begin{lemma} \label{secondlemmaproofinappendix}
The following relation holds:
\begin{align}
&Z_{m+n-k,n}
(u_1,\dots,u_n|x_1,\dots,x_\ell|y_1,\dots,y_{\ell+k},
m+1,\dots,m+n-k)
\nonumber \\
=&ZDB_{m,n,k}(u_{k+1},\dots,u_n|u_1,\dots,u_k|x_1,\dots,x_\ell|y_1,\dots,y_{\ell+k}). \label{largeanddb}
\end{align}
\end{lemma}
In Appendix B,
we show a proof of \eqref{largeanddb} using equations only.
A graphical derivation is given in Appendix C.

Applying the correspondence \eqref{skewcorrespondence}
to the partition functions \\
$Z_{m+n-k,n}
(u_1,\dots,u_n|x_1,\dots,x_\ell|y_1,\dots,y_{\ell+k},
m+1,\dots,m+n-k)$
and combining with \eqref{largeanddb}, we get
\begin{align}
&ZDB_{m,n,k}(u_{k+1},\dots,u_n|u_1,\dots,u_k|x_1,\dots,x_\ell|y_1,\dots,y_{\ell+k})
\nonumber
\\
=&
G_{((m-\ell-k)^{n-k},\lambda) /\!/ \mu}(1-u_{1}^{-1},\dots,1-u_n^{-1}),
\label{dbpartitionfunctionexplicit}
\end{align}
where
$\lambda_j=y_{\ell-j+k+1}-\ell+j-k-1 \ (j=1,\dots,\ell+k)$
and $\mu_j=x_{\ell-j+1}-\ell+j-1 \ (j=1,\dots,\ell)$.

\section{Grothendieck classes and $K$-theoretic pushforward}

In this section, we use two types of partition functions
of the integrable five-vertex model introduced in the last section
and the Shigechi-Uchiyama commutation relation
to show that two Grothendieck classes
are directly connected via the $K$-theoretic pushforward.
Let us first recall the algebraic geometry side 
\cite{Buch1,Buch2,AllPhD,All} again.

Let $X$ be a smooth projective complex variety
and $K(X)$ the Grothendieck ring of isomorphism classes
of algebraic vector bundles over $X$.
For a subvariety $Y$ of $X$, the Grothendieck class of the
structure sheaf of $Y$ is denoted $[\mathcal{O}_Y]$, or $[Y]$ for short.
Let $\mathcal{A} \rightarrow X$ be a rank $n$ vector bundle,
and $\pi: \mathrm{Gr}_k(\mathcal{A}) \rightarrow X$ the Grassmann bundle
of $k$-planes in $\mathcal{A}$ with universal tautological exact sequence
$0 \rightarrow \mathcal{S} \rightarrow \pi^* \mathcal{A} \rightarrow \mathcal{Q} \rightarrow 0$ of vector bundles over $\mathrm{Gr}_k(\mathcal{A})$.
Let $\sigma=\{ \sigma_1, \dots, \sigma_k \}$ and $\omega=\{ \omega_1,\dots,\omega_{n-k} \}$
be the set of Grothendieck roots
for the subbundle $\mathcal{S}$ and the 
quotient bundle $\mathcal{Q}$ respectively, i.e., 
defined by the exterior powers of
$\mathcal{S}$ and $\mathcal{Q}$ by the relation
${[} \bigwedge^i \mathcal{S} {]}=e_i(\sigma)$
and ${[} \bigwedge^j \mathcal{Q} {]}=e_j(\omega)$,
where $e_i(\sigma)$ and $e_j(\omega)$ are the $i$th and $j$th
elementary symmetric functions with the set of variables $\sigma$ and $\omega$,
respectively.

Let $f(\sigma;\omega)$ be a Laurent polynomial
in $K(X)[\sigma_i^{\pm};\omega_j^{\pm}]$ which are symmetric in
$\sigma$ variables and also symmetric in $\omega$ variables,
which makes it possible for the polynomial to be viewed as 
a certain Grothendieck class in $K(\mathrm{Gr}_k(\mathcal{A}))$.
We consider the following $K$-theoretic pushforward
$\pi_*:K(\mathrm{Gr}_k(\mathcal{A})) \rightarrow K(X)$
induced from $\pi$.
By using quantum integrability, we show that the following holds
for the $K$-theoretic pushforward:
\begin{theorem} \label{maintheorem}
Let $\mathcal{A}$ be a vector bundle on $X$
of rank $n$, and $\pi: \mathrm{Gr}_k(\mathcal{A}) \rightarrow X$ the Grassmann bundle of $k$-planes in $\mathcal{A}$ with universal tautological exact sequence $0 \rightarrow \mathcal{S} \rightarrow \pi^* \mathcal{A} \rightarrow \mathcal{Q} \rightarrow 0$ of vector bundles over $\mathrm{Gr}_k(\mathcal{A})$.
Let $\alpha=\{\alpha_1,\dots,\alpha_n \}$,
$\sigma=\{ \sigma_1, \dots, \sigma_k \}$
and $\omega=\{ \omega_1,\dots,\omega_{n-k} \}$
be the sets of Grothendieck roots
for bundles $\mathcal{A}$, $\mathcal{S}$ and
$\mathcal{Q}$ respectively. Then we have
\begin{align}
&\pi_*\Bigg( \sum_{\nu \subseteq (m-\ell)^\ell}
G_{\lambda /\!/ \nu}(1-\sigma_1^{-1},\dots,1-\sigma_k^{-1})
G_{((m-\ell)^{n-k},\nu) /\!/ \mu}(1-\omega_{1}^{-1},\dots,1-\omega_{n-k}^{-1}) \Bigg)
\nonumber
\\
=&G_{((m-\ell-k)^{n-k},\lambda) /\!/ \mu}(1-\alpha_{1}^{-1},\dots,1-\alpha_n^{-1}),
\label{mapofgrothendieckclasses}
\end{align}
where $\ell$ is an integer satisfying $0 \le \ell \le m-k$
and $\lambda$, $\mu$ and $\nu$ are partitions with
$\lambda \subseteq (m-k-\ell)^k$.
\end{theorem}
\begin{proof}
The following sum of products of the skew Grothendieck polynomials
\begin{align}
g(\sigma;\omega)=&\sum_{\nu \subseteq (m-\ell)^\ell}
G_{\lambda /\!/ \nu}(1-\sigma_1^{-1},\dots,1-\sigma_k^{-1})
G_{((m-\ell)^{n-k}, \nu) /\!/ \mu}(1-\omega_{1}^{-1},\dots,1-\omega_{n-k}^{-1}), \label{shorthand}
\end{align}
is the right hand side of \eqref{bdpartitionfunctionexplicit}
under the following renaming of variables
$u_j=\sigma_j$, $(j=1,\dots,k)$, $u_j=\omega_{j-k}$, $(j=k+1,\dots,n)$.
By \eqref{bdpartitionfunctionexplicit}, we get
the partition function representation of $g(\sigma;\omega)$
\begin{align}
g(\sigma;\omega)
=&ZBD_{m,n,k}(\sigma_1,\dots,\sigma_k|\omega_1,\dots,\omega_{n-k}|x_1,\dots,x_\ell|y_1,\dots,y_{\ell+k})
\nonumber
\\
=&
_{m} \langle y_1 \cdots y_{\ell+k}|
B(\sigma_1) \cdots B(\sigma_k) D(\omega_{1}) \cdots D(\omega_{n-k})
|x_1 \cdots x_\ell \rangle_{m}, \label{partfuncrep}
\end{align}
where
$\lambda_j=y_{\ell-j+k+1}-\ell+j-k-1 \ (j=1,\dots,\ell+k)$
and $\mu_j=x_{\ell-j+1}-\ell+j-1 \ (j=1,\dots,\ell)$.
From the partition function representation \eqref{partfuncrep},
one can see for example that from the commutativity relation
\eqref{rttone3} and \eqref{rttone4},
$g(\sigma;\omega)$ is symmetric in $\sigma$ variables
and also in $\omega$ variables.
We also note from the expressions of the matrix elements of the $R$-matrix
for the integrable five-vertex model \eqref{fivevertexrmatrix}
that if one expands the partition function representation
of $g(\sigma;\omega)$
in the $\sigma$ and $\omega$ variables,
the coefficient of each term is some integer which can be viewed as 
$1=G_\phi \in K(X)$ multiplied by the integer,
hence $g(\sigma;\omega)$ is a Laurent polynomial
in $K(X)[\sigma_i^{\pm};\omega_j^{\pm}]$ which are symmetric in
$\sigma$ and in $\omega$,
and can be viewed as a certain Grothendieck class in $K(\mathrm{Gr}_k(\mathcal{A}))$.

To prove \eqref{mapofgrothendieckclasses},
one first applies the formula \eqref{KGysin} for the $K$-theoretic pushforward
to $g(\sigma;\omega)$ and then use
the partition function representation of
$g(\sigma;\omega)$ \eqref{partfuncrep} to get
\begin{align}
&\pi_*(g(\sigma;\omega)) \nonumber \\
&=\sum_{S_k^n \in \binom{[n]}{k}}
\prod_{i \in S_k^n, j \in \overline{S_k^n}}
\frac{1}{1-\alpha_i/\alpha_j}
g(\alpha_{S_k^n};\alpha_{\overline{{S_k^n}}}) \nonumber \\
&=\sum_{S_k^n \in \binom{[n]}{k}}
\prod_{i \in S_k^n, j \in \overline{S_k^n}}
\frac{1}{1-\alpha_i/\alpha_j}
{}_{m} \langle y_1 \cdots y_{\ell+k}|
\prod_{i \in S_k^n} B(\alpha_i)
\prod_{j \in \overline{S_k^n}} D(\alpha_j)
|x_1 \cdots x_\ell \rangle_{m} \nonumber \\
&={}_{m} \langle y_1 \cdots y_{\ell+k}|
\sum_{S_k^n \in \binom{[n]}{k}}
\prod_{i \in S_k^n, j \in \overline{S_k^n}}
\frac{1}{1-\alpha_i/\alpha_j}
\prod_{i \in S_k^n} B(\alpha_i)
\prod_{j \in \overline{S_k^n}} D(\alpha_j)
|x_1 \cdots x_\ell \rangle_{m}.
\label{tochutoformula}
\end{align}
Recall that $S_k^n$ is a $k$-subset of $[n]=\{1,2,\dots,n \}$,
the set of $k$-subsets of $n$ is denoted as $\binom{[n]}{k}$,
and $\overline{S_k^n}=\{1,2,\dots,n \} \backslash S_k^n$.
Next, we use the commutation relation \eqref{multiplecommutation}
to \eqref{tochutoformula}. Note that
$|x_1 \cdots x_\ell \rangle_{m}$ and $_{m} \langle y_1 \cdots y_{\ell+k}|$
which are vectors on the tensor product of (dual) quantum spaces
$W_1 \otimes \cdots \otimes W_m$ ($W_1^* \otimes \cdots \otimes W_m^*$)
 do not disturb one to use the commutation relation.
Applying the commutation relation \eqref{multiplecommutation}
to \eqref{tochutoformula},
we get
\begin{align}
&\pi_*(g(\sigma;\omega))
={}_{m} \langle y_1 \cdots y_{\ell+k}|
\prod_{j=k+1}^{n} D(\alpha_j) \prod_{j=1}^k B(\alpha_j)
|x_1 \cdots x_\ell \rangle_{m}. \label{tochuformulatwo}
\end{align}
The right hand side of \eqref{tochuformulatwo}
is nothing but the second type of the
mixed partition functions
$ZDB_{m,n,k}(\alpha_{k+1},\dots,\alpha_n|\alpha_1,\dots,\alpha_k
|x_1,\dots,x_\ell|y_1,\dots,y_{\ell+k})$.
Combining \eqref{tochuformulatwo} with
the correspondence formula
\eqref{dbpartitionfunctionexplicit}, we find the pushforward
of the Grothendieck class $g(\sigma;\omega)$
is given by the skew Grothendieck polynomials
\begin{align}
&\pi_*(g(\sigma;\omega))=
G_{((m-\ell-k)^{n-k},\lambda) /\!/ \mu}(1-\alpha_{1}^{-1},\dots,1-\alpha_n^{-1}).
\end{align}
\end{proof}
Note that in the proof above,
what played the crucial role are
matrix elements of the commutation relation
\eqref{SUcommutation}
\begin{align}
&ZDB_{m,n,k}(u_{k+1},\dots,u_n|u_1,\dots,u_k|x_1,\dots,x_\ell|y_1,\dots,y_{\ell+k}) \nonumber \\
=&\sum_{S_k^n \in \binom{[n]}{k}} \prod_{i \in S_k^n, j \in \overline{S_k^n}}
\frac{1}{1-u_i/u_j}
ZBD_{m,n,k}(u_{S_k^n}|u_{\overline{S_k^n}}
|x_1,\dots,x_\ell|y_1,\dots,y_{\ell+k}),
\end{align}
where $u_{S_k^n}=\{u_i \ | \ i \in S_k^n \}$
and $u_{\overline{S_k^n}}=\{ u_j \ | \ j \in \overline{S_k^n} \}$.
Using
\eqref{bdpartitionfunctionexplicit} and
\eqref{dbpartitionfunctionexplicit}, this can be written as the following
identity for the skew Grothendieck polynomials.
\begin{theorem} \label{skewfnrgs}
We have the following identity.
\begin{align}
&G_{((m-\ell-k)^{n-k},\lambda) /\!/ \mu}(1-u_{1}^{-1},\dots,1-u_n^{-1})
\nonumber \\
=&\sum_{S_k^n \in \binom{[n]}{k}}
\prod_{i \in S_k^n, j \in \overline{S_k^n}}
\frac{1}{1-u_i/u_j}
 \sum_{\nu \subseteq (m-\ell)^\ell}
G_{\lambda /\!/ \nu}(1-u_{S_k^n}^{-1})
G_{((m-\ell)^{n-k},\nu) /\!/ \mu}(1-u_{\overline{S_k^n}}^{-1}),
\label{skewtnrgsidentity}
\end{align}
where $\ell$ is an integer satisfying $0 \le \ell \le m-k$
and $\lambda$, $\mu$ and $\nu$ are partitions with
$\lambda \subseteq (m-k-\ell)^k$,
and
$1-u_{S_k^n}^{-1}=\{1-u_i^{-1} \ | \ i \in S_k^n \}$
and
$1-u_{\overline{S_k^n}}^{-1}=
\{1-u_j^{-1} \ | \ j \in \overline{S_k^n} \}$.
\end{theorem}

\section{Special cases and application}
Let us see in this section that certain special cases
of Theorem \ref{maintheorem}
reproduce special cases of the formulas derived by Buch \cite{Buch1}.

Let us consider $\ell=0$, $\mu=\phi$
of \eqref{mapofgrothendieckclasses} in Theorem \ref{maintheorem}. 
In this case, only the summand correseponding to $\nu=\phi$
in $g(\sigma;\omega)$ \eqref{shorthand} survives,
and $g(\sigma;\omega)$ becomes
the following product of two Grothendieck polynomials
\begin{align}
g(\sigma;\omega)
&=
G_{\lambda}(1-\sigma_1^{-1},\dots,1-\sigma_k^{-1})
G_{(m)^{n-k}}(1-\omega_{1}^{-1},\dots,1-\omega_{n-k}^{-1}),
\end{align}
where $\lambda$ satisfies $\lambda \subseteq (m-k)^k$.
Specializing the pushforward formula \eqref{mapofgrothendieckclasses}
to this case, we get the following result:
\begin{align}
&\pi_*(
G_{\lambda}(1-\sigma_1^{-1},\dots,1-\sigma_k^{-1})
G_{(m)^{n-k}}(1-\omega_{1}^{-1},\dots,1-\omega_{n-k}^{-1})
) \nonumber \\
&=G_{((m-k)^{n-k},\lambda)}(1-\alpha_{1}^{-1},\dots,1-\alpha_n^{-1}).
\label{special}
\end{align}
This is also special cases of the following $K$-theoretic pushforward formulas
by Buch \cite{Buch1}, Lemma 7.1 and Theorem 7.3
(see also \cite{Pragacz,FP} for cohomological versions).
\begin{lemma} (Buch \cite{Buch1}, Lemma 7.1) \label{lemmabyBuch}
Let $\mathcal{A}$ be a vector bundle of rank $n$ over a variety $X$.
Let $\pi: \mathbb {P}^*(\mathcal{A}) \rightarrow X$
be the dual projective bundle
of $\mathcal{A}$ and $\mathcal{Q}$ be the tautological quotient
of $\pi^* \mathcal{A}$.
Then we have
\begin{align}
\pi_* (G_m (\mathcal{Q}))=G_{m-n+1}(\mathcal{A}), \label{Buchlemma}
\end{align}
for any integer $m$.
\end{lemma}

\begin{theorem} (Buch \cite{Buch1}, Theorem 7.3) \label{theorembyBuch}
Let $\mathcal{A}$ and $\mathcal{B}$ be vector bundles on $X$
of rank $n$ and $\ell$ respectively.
Let $\pi: \mathrm{Gr}_k(\mathcal{A}) \rightarrow X$ the Grassmann bundle of $k$-planes in $\mathcal{A}$ with universal tautological exact sequence $0 \rightarrow \mathcal{S} \rightarrow \pi^* \mathcal{A} \rightarrow \mathcal{Q} \rightarrow 0$ of vector bundles over $\mathrm{Gr}_k(\mathcal{A})$.
Let $\alpha=\{\alpha_1,\dots,\alpha_n \}$,
$\beta=\{\beta_1,\dots,\beta_\ell \}$,
$\sigma=\{ \sigma_1, \dots, \sigma_k \}$
and $\omega=\{ \omega_1,\dots,\omega_{n-k} \}$
be the sets of Grothendieck roots
for bundles $\mathcal{A}$, $\mathcal{B}$, $\mathcal{S}$ and
$\mathcal{Q}$ respectively.
Let $I=(I_1,\dots,I_{n-k})$ and $J=(J_1,J_2,\dots)$
be sequences of integers satisfying $I_j \geq \ell$ for all $j$.
Then we have
\begin{align}
\pi_*(G_I(\mathcal{Q}-\mathcal{B}) G_J(\mathcal{S}-\mathcal{B}))
=G_{I-(k)^{n-k},J}(\mathcal{A}-\mathcal{B}). \label{Buchformula}
\end{align}
\end{theorem}
Here, $G_I(\mathcal{A}-\mathcal{B})$ in \eqref{Buchlemma}, \eqref{Buchformula}
are certain Grothendieck classes of $K(X)$
extended to sequences of integers \cite{Buch1}
from the the case when $I$ are partitions
$I=\lambda=(\lambda_1,\dots,\lambda_i)$
satisfying $n \ge i$, in which case
$G_\lambda(\mathcal{A}-\mathcal{B})$
are the Grassmannian double Grothendieck polynomials
\begin{align}
G_\lambda(\mathcal{A}-\mathcal{B})
=G_\lambda(1-\alpha_1^{-1},\dots,1-\alpha_n^{-1}|1-\beta_1,\dots,1-\beta_\ell).
\end{align}
The Grassmannian double Grothendieck polynomials
\cite{LS,FK,Buch1,Buch2,Buch3,IN} has the following determinant form
\begin{align}
G_\lambda(z_1,\dots,z_n|v_1,\dots,v_\ell)=
   \frac{\mathrm{det}([z_i|\bs{v}]^{\lambda_j+n-j}(1-z_i)^{j-1})_{1 \le i,j \le n}}
        {\prod_{1 \le i < j \le n}(z_i-z_j)}.
\end{align}
where $\lambda=(\lambda_1,\lambda_2,\dots,\lambda_n)$
is a  partition, $\{z_1, \dots, z_n \}$ and $\bs{v}=\{v_1,v_2,\dots,v_\ell \}$ are sets of variables and
\begin{align}
[z_i|\bs{v}]^j=(z_i \oplus v_1) (z_i \oplus v_2)
\cdots (z_i \oplus v_j),
\end{align}
where $z \oplus v:=z+v- z v$.

The case $\lambda=\phi$, $k=n-1$ of \eqref{special}
\begin{align}
&\pi_*(
G_{m}(1-\omega_{1}^{-1})
) =G_{m-n+1}(1-\alpha_{1}^{-1},\dots,1-\alpha_n^{-1}),
\end{align}
which corresponds to the case when the rank of the quotient bundle
$\mathcal{Q}$ is 1, is nothing but \eqref{Buchlemma} in Lemma \ref{lemmabyBuch}
(Buch \cite{Buch1}, Lemma 7.1) when $m$ is positive.

We can also see that \eqref{special} is the special case
$\ell=0$, $I=(m)^{n-k}$, $J=\lambda$ of \eqref{Buchformula}
in Theorem \ref{theorembyBuch} (Buch \cite{Buch1}, Theorem 7.3),
which is explicitly
\begin{align}
\pi_*(G_{(m)^{n-k}}(\mathcal{Q}) G_\lambda(\mathcal{S}))
=G_{(m)^{n-k}-(k)^{n-k},\lambda}(\mathcal{A})
=G_{((m-k)^{n-k},\lambda)}(\mathcal{A}).
\end{align}
Let us also see that
Theorem \ref{skewfnrgs} is a skew generalization of an identity
for the Grothendieck polynomials
recently found by Guo and Sun \cite{GS},
which is a generalization of the one for Schur polynomials
by Feh\'er, N\'emethi and Rim\'anyi \cite{FNR}.
In fact, $\ell=0$, $\mu=\phi$ of
\eqref{skewtnrgsidentity} is
\begin{align}
&G_{(m^{n-k},\lambda)}(1-u_{1}^{-1},\dots,1-u_n^{-1})
\nonumber \\
=&\sum_{S_k^n \in \binom{[n]}{k}}
\prod_{i \in S_k^n, j \in \overline{S_k^n}}
\frac{1}{1-u_i/u_j}
G_{\lambda}(1-u_{S_k^n}^{-1})
G_{m^{n-k}}(1-u_{\overline{S_k^n}}^{-1}).
\label{specialcaseskewtnrgsidentity}
\end{align}
It is easy to see from the determinant representation
\eqref{doubleGrothendieck} that
$G_{m^{n-k}}(1-u_{\overline{S_k^n}}^{-1})$
has the factorized form
\begin{align}
\displaystyle G_{m^{n-k}}(1-u_{\overline{S_k^n}}^{-1})
=\prod_{j \in \overline{S_k^n}}(1-u_j^{-1})^m.
\label{factorizedform}
\end{align}
Inserting \eqref{factorizedform} into
\eqref{specialcaseskewtnrgsidentity}, we get
\begin{align}
\displaystyle
G_{(m^{n-k},\lambda)}(1-u_{1}^{-1},\dots,1-u_n^{-1})
=\sum_{S_k^n \in \binom{[n]}{k}}
G_{\lambda}(1-u_{S_k^n}^{-1})
\frac{\displaystyle
\prod_{j \in \overline{S_k^n}}(1-u_j^{-1})^m
}{\displaystyle
\prod_{i \in S_k^n, j \in \overline{S_k^n}}
(1-u_i/u_j)}. \label{fnrgszvariable}
\end{align}
Using the $z$-variables $z_j=1-u_j^{-1}$, $j=1,\dots,n$,
\eqref{fnrgszvariable} can be rewritten as
\begin{align}
\displaystyle
G_{(m^{n-k},\lambda)}(z_1,\dots,z_n)
=\sum_{S_k^n \in \binom{[n]}{k}}
G_{\lambda}(z_{S_k^n})
\frac{\displaystyle
\prod_{i \in S_k^n} (1-z_i)^{n-k}
\prod_{j \in \overline{S_k^n}}z_j^m
}{\displaystyle
\prod_{i \in S_k^n, j \in \overline{S_k^n}}
(z_j-z_i)}.
\end{align}
This is the $\beta=-1$ and nonfactorial case of Theorem 3.1 in \cite{GS} by Guo and Sun.

Finally, let us discuss an application of the pushforward formula.
There are studies on deriving algebraic formulas and identities
by investigating the pushforward map in various ways and combining them.
One example is the geometric derivation of the
Jacobi-Trudi type formulas \cite{HIMN,NN3}.

Let us derive an integration formula (residue formula)
by combining the pushforward formula which we derived
with another expression which is obtained by applying the following
formula.
\begin{proposition} (Allman \cite{All}, Prop 6.3)
Let $\mathcal{A}$ be a vector bundle on $X$
of rank $n$, and $\pi: \mathrm{Gr}_k(\mathcal{A}) \rightarrow X$ the Grassmann bundle of $k$-planes in $\mathcal{A}$ with universal tautological exact sequence $0 \rightarrow \mathcal{S} \rightarrow \pi^* \mathcal{A} \rightarrow \mathcal{Q} \rightarrow 0$ of vector bundles over $\mathrm{Gr}_k(\mathcal{A})$.
Let $\alpha=\{\alpha_1,\dots,\alpha_n \}$,
$\sigma=\{ \sigma_1, \dots, \sigma_k \}$
and $\omega=\{ \omega_1,\dots,\omega_{n-k} \}$
be the sets of Grothendieck roots
for bundles $\mathcal{A}$, $\mathcal{S}$ and
$\mathcal{Q}$ respectively.
Let $f(\sigma;\omega)$ be a Grothendieck class in 
$K(\mathrm{Gr}_k(\mathcal{A}))$,
and ${\bf z}=\{z_1,\dots,z_n \}$  an alphabet of ordered,
commuting variables. If $f$ does not have poles in $K(X)$
aside from zero and the point at infinity, we have
\begin{align}
\displaystyle
\pi_*(f(\sigma;\omega))=
\frac{1}{(2 \pi i)^n} \oint_C \cdots \oint_C
f({\bf z}) \frac{\displaystyle \prod_{1 \le i < j \le n}(1-z_j/z_i)}
{\displaystyle \prod_{i,j=1}^n (z_i/\alpha_j-1)} \prod_{i=1}^n
\frac{d z_i}{z_i}, \label{integrationformula}
\end{align}
where $f({\bf z})=f(z_1,\dots,z_k;z_{k+1},\dots,z_n)$
and the integration contour $C$ surrounds $\alpha_j$, $j=1,\dots,n$.

\end{proposition}
Note that \eqref{integrationformula}
is equivalent to the expression
in \cite{All}, which is written in the form of
taking residues at zero and  the point at infinity.
Applying the integration formula \eqref{integrationformula}
to $g(\sigma;\omega)$ \eqref{shorthand}, we get
\begin{align}
&\pi_* \Bigg( \sum_{\nu \subseteq (m-\ell)^\ell}
G_{\lambda /\!/ \nu}(1-\sigma_1^{-1},\dots,1-\sigma_k^{-1})
G_{((m-\ell)^{n-k},\nu) /\!/ \mu}(1-\omega_{1}^{-1},\dots,1-\omega_{n-k}^{-1})
\Bigg)
\nonumber \\
=&\frac{1}{(2 \pi i)^n}
\sum_{\nu \subseteq (m-\ell)^\ell}
 \oint_C \cdots \oint_C
G_{\lambda /\!/ \nu}(1-z_1^{-1},\dots,1-z_k^{-1})
\nonumber \\
&\times G_{((m-\ell)^{n-k},\nu) /\!/ \mu}(1-z_{k+1}^{-1},\dots,1-z_{n}^{-1}) \frac{\displaystyle \prod_{1 \le i < j \le n}(1-z_j/z_i)}
{\displaystyle \prod_{i,j=1}^n (z_i/\alpha_j-1)} \prod_{i=1}^n
\frac{d z_i}{z_i} \label{applyingintegrationformula}.
\end{align}
Combining
\eqref{mapofgrothendieckclasses} and \eqref{applyingintegrationformula},
we get the following.
\begin{corollary}
We have the following identity.
\begin{align}
&G_{((m-\ell-k)^{n-k},\lambda) /\!/ \mu}(1-\alpha_{1}^{-1},\dots,1-\alpha_n^{-1})
\nonumber \\
=&\frac{1}{(2 \pi i)^n}
\sum_{\nu \subseteq (m-\ell)^\ell}
 \oint_C \cdots \oint_C
G_{\lambda /\!/ \nu}(1-z_1^{-1},\dots,1-z_k^{-1}) \nonumber \\
&\times G_{((m-\ell)^{n-k},\nu) /\!/ \mu}(1-z_{k+1}^{-1},\dots,1-z_{n}^{-1}) \frac{\displaystyle \prod_{1 \le i < j \le n}(1-z_j/z_i)}
{\displaystyle \prod_{i,j=1}^n (z_i/\alpha_j-1)} \prod_{i=1}^n
\frac{d z_i}{z_i}, \label{corollaryintegration}
\end{align}
where $\ell$ is an integer satisfying $0 \le \ell \le m-k$
and $\lambda$, $\mu$ and $\nu$ are partitions with
$\lambda \subseteq (m-k-\ell)^k$.
\end{corollary}
Taking $\ell=0$, $\mu=\phi$ in
\eqref{corollaryintegration} gives
\begin{align}
&G_{((m-k)^{n-k},\lambda)}(1-\alpha_{1}^{-1},\dots,1-\alpha_n^{-1})
\nonumber \\
=&\frac{1}{(2 \pi i)^n}
 \oint_C \cdots \oint_C
G_{\lambda}(1-z_1^{-1},\dots,1-z_k^{-1})
\frac{\displaystyle
\prod_{j=k+1}^n (1-z_j^{-1})^m
\prod_{1 \le i < j \le n}(1-z_j/z_i)}
{\displaystyle \prod_{i,j=1}^n (z_i/\alpha_j-1)} \prod_{i=1}^n
\frac{d z_i}{z_i}, \label{corollaryintegrationtwo}
\end{align}
by using
$\displaystyle G_{m^{n-k}}(1-z_{k+1}^{-1},\dots,1-z_{n}^{-1})
=\prod_{j=k+1}^n (1-z_j^{-1})^{m}
$.
Further, setting $m=k$, $\lambda=\phi$ and using $G_\phi=1$,
\eqref{corollaryintegrationtwo} becomes
\begin{align}
1=&\frac{1}{(2 \pi i)^n}
 \oint_C \cdots \oint_C
\frac{\displaystyle
\prod_{j=k+1}^n (1-z_j^{-1})^k
\prod_{1 \le i < j \le n}(1-z_j/z_i)}
{\displaystyle \prod_{i,j=1}^n (z_i/\alpha_j-1)} \prod_{i=1}^n
\frac{d z_i}{z_i}.
\end{align}

\section*{Acknowledgments}
We appreciate the referee for careful reading and
numerous invaluable comments and suggestions to improve the paper.
This work was partially supported by grant-in-Aid
for Scientific Research (C) No. 18K03205 and No. 16K05468.

\renewcommand{\theequation}{A.\arabic{equation} }
\setcounter{equation}{0}

\section*{Appendix A. On skew Grothendieck polynomials}

\begin{figure}[ht]
\includegraphics[width=10cm]{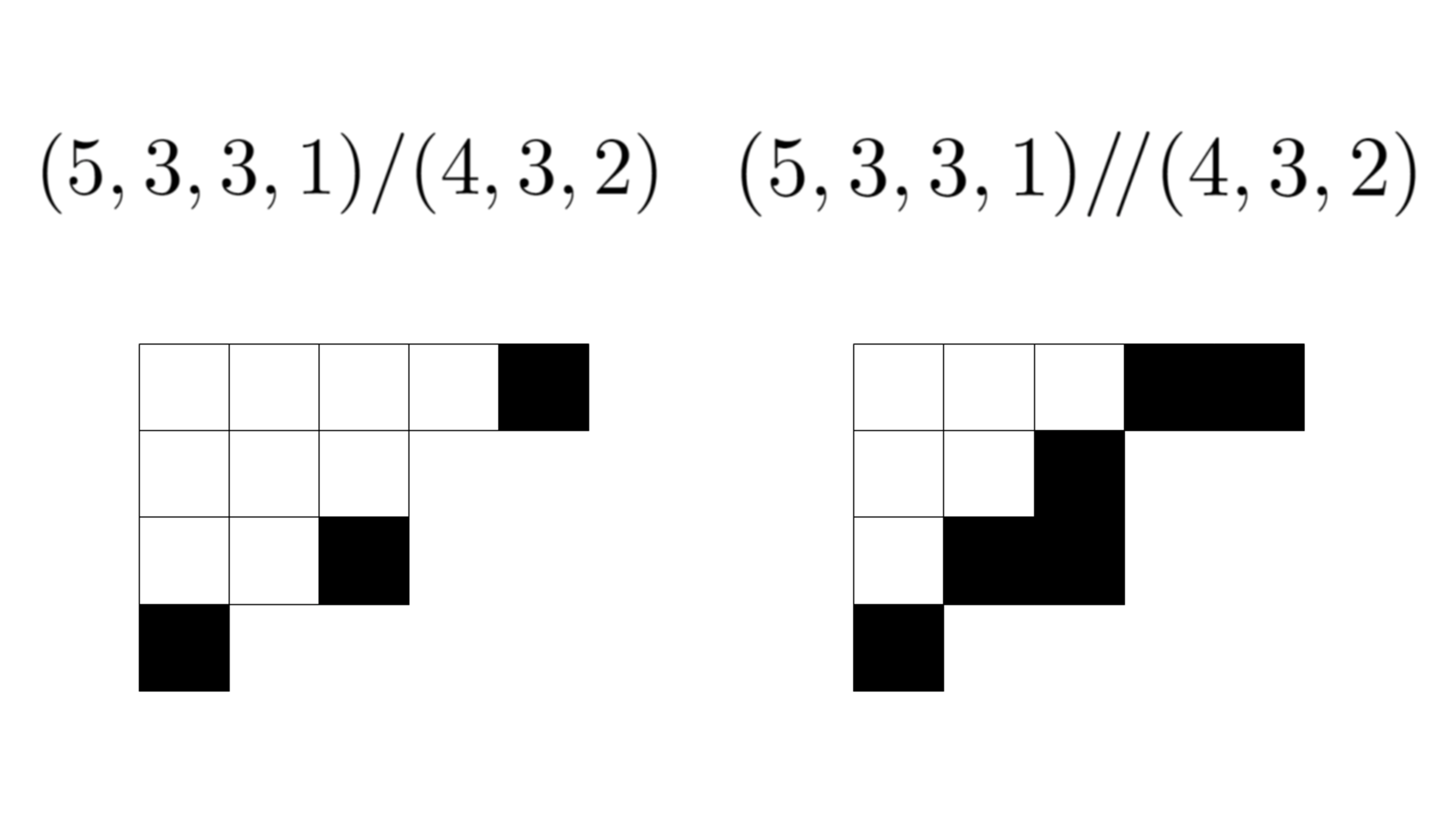}
\caption{The skew shape $(5,3,3,1) / (4,3,2)$ 
are boxes which are shaded in the left figure.
Removing a box whose coordinate is
$(1,4)$, $(2,3)$ or $(3,2)$ from the Young diagram
$(4,3,2)$ still yields a Young diagram again.
Adding these boxes to $(5,3,3,1) / (4,3,2)$
gives the extended skew shape $(5,3,3,1) /\!/ (4,3,2)$,
which are boxes shaded in the right figure.
}
\label{pictureyoungdiagram}
\end{figure}

In this appendix, we record a formulation
of the skew Grothendieck polynomials
$G_{\lambda /\!/ \mu}$ by
Yeliussizov \cite{Yeliussizov}, which was originally introduced by
Buch \cite{Buch2}. Note that in this Appendix,
we respect the notations in \cite{Yeliussizov}
as possible as we can, so the meaning of the
$A$-operator, $u$- and $x$-variables are different from the ones
used in the main part of this paper.

We first prepare notations.
A partition is a nonincreasing sequence of nonnegative integers
$\lambda=(\lambda_1,\lambda_2,\dots,\lambda_{\ell(\lambda)},\lambda_{\ell(\lambda)+1}=0,\dots
)$, $\lambda_1 \geq \lambda_2 \geq \cdots \geq \lambda_{\ell(\lambda)} >0$
with only finitely many nonzero terms. $\ell(\lambda)$ is called the length
of the partition.
We define $|\lambda|$ for a partition $\lambda$ as
$|\lambda|=\lambda_1+\lambda_2+\cdots+\lambda_{\ell(\lambda)}$.
We identify a partition $\lambda$ with a Young diagram
which is the set $\{(i,j) \ | \ 1 \leq i \leq \ell(\lambda),
1 \le j \leq \lambda_i \}$.

For a Young diagram $\mu$, we say that a box in $\mu$ is removable
if removing it yields a Young diagram of a partition.
Let $I(\mu)$ be the set of removable boxes of $\mu$.
For two partitions $\lambda$ and $\mu$ satisfying $\mu \subset \lambda$,
we define $\lambda /\!/ \mu$ as $\lambda /\!/ \mu:=\lambda / \mu \cup I(\mu)$.
This is an extension of a skew shape $\lambda / \mu$.
We define $a(\lambda /\!/ \mu)$ as the number of columns of
$\lambda /\!/ \mu$ that are not columns of $\lambda/\mu$.
We say that a skew shape $\lambda/\mu$ is a horizontal strip
if $|\lambda / \mu|$ is equal to the number of columns of $\lambda/\mu$.

For example, let us take $\lambda$ and $\mu$ as
$\lambda=(5,3,3,1)$, $\mu=(4,3,2)$ (Figure \ref{pictureyoungdiagram}).
$I((4,3,2))$ consists of boxes whose coordinates are $(1,4)$, $(2,3)$, $(3,2)$.
There are boxes which are
in the second and fourth column of $(5,3,3,1) /\!/ (4,3,2)$
but not in $(5,3,3,1) / (4,3,2)$, hence $a((5,3,3,1) /\!/ (4,3,2))=2$.

Consider the free
$\mathbb{Z}$-module $\mathbb{Z} P=\oplus_{\lambda} \
\mathbb{Z} \cdot \lambda$ with a basis of all partitions.
Let $u=(u_1,u_2,\dots)$ and $d=(d_1,d_2,\dots)$ be sets of linear operators
on $\mathbb{Z} P$, which act on bases as
\begin{align}
u_i \cdot \lambda=
\left\{
\begin{array}{ll}
\lambda+\Box \ \mathrm{in} \ \mathrm{column} \ i, & \mathrm{if} \ \mathrm{possible}, \\
0, & \mathrm{otherwise},
\end{array}
\right.
\end{align}
\begin{align}
d_i \cdot \lambda=
\left\{
\begin{array}{ll}
\lambda-\Box \ \mathrm{in} \ \mathrm{column} \ i, & \mathrm{if} \ \mathrm{possible}, \\
0, & \mathrm{otherwise}.
\end{array}
\right.
\end{align}

Let $\beta$ be a scalar parameter and we consider the free
$\mathbb{Z}[\beta]$-module $\mathbb{Z}[\beta]P=\oplus_{\lambda}
\mathbb{Z}[\beta] \cdot \lambda$ with a basis of all partitions.
We define $\widetilde{u}=(\widetilde{u}_1,\widetilde{u}_2,\dots)$
as linear operators which act on $\mathbb{Z}[\beta]P$
as $\widetilde{u}_i:=u_i-\beta u_i d_i$.

Let $x$ be an indeterminate commuting with $u$ and $d$,
and define $A(x)$ as
\begin{align}
A(x)=\cdots (1+x \widetilde{u}_2) (1+x \widetilde{u}_1).
\end{align}

Let $\langle \cdot , \cdot \rangle$ be a bilinear pairing
on $\mathbb{Z}[\beta]P$ given by $\langle \lambda, \mu \rangle=
\delta_{\lambda \mu}$.
For partitions $\lambda$ and $\mu$,
the skew Grothendieck polynomials
$G_{\lambda /\!/ \mu}^\beta(x_1,\dots,x_n)$
is defined in \cite{Yeliussizov}, Definition 4.1 as
\begin{align}
G_{\lambda /\!/ \mu}^\beta(x_1,\dots,x_n):=
\langle A(x_n) \cdots A(x_1) \cdot \mu, \lambda \rangle.
\label{anotherskewdefinition}
\end{align}
The following is shown in \cite{Yeliussizov}. \\
\\
{\bf Proposition A.1.} (Yeliussizov \cite{Yeliussizov}, Proposition 4.3) \\
{\it For a single variable $x$, we have}
\begin{align}
&G_{\lambda /\!/ \mu}^{\beta}(x)
=
\left\{
\begin{array}{ll}
(1-\beta x)^{a(\lambda /\!/ \mu)} x^{|\lambda/\mu|} ,& \lambda/\mu \ 
is \  a \ horizontal \ strip, \\
0, & otherwise.
\end{array}
\right.
\label{singlematrixelementsYe}
\end{align}
Noting that
$(1-u^{-1} )^{\sum_{j=1}^{\ell+1} \lambda_j
-\sum_{j=1}^\ell \mu_j}
\prod_{j=1}^\ell
\{
u^{-1}+
\delta_{\lambda_{j+1},\mu_j}
(
1-u^{-1}
) \}$ in \eqref{singlematrixelements}
can be rewritten as
\begin{align}
&(1-u^{-1} )^{\sum_{j=1}^{\ell+1} \lambda_j
-\sum_{j=1}^\ell \mu_j}
\prod_{j=1}^\ell
\{
u^{-1}+
\delta_{\lambda_{j+1},\mu_j}
(
1-u^{-1}
) \} \nonumber \\
=&(1-u^{-1} )^{|\lambda|-|\mu|}
(1-(1-u^{-1}))^{\# \{j | \lambda_{j+1} \neq \mu_j \}}
\nonumber \\
=&(1-u^{-1})^{|\lambda/\mu|}
(1-(1-u^{-1}))^{a(\lambda /\!/ \mu)},
\end{align}
and the interlacing relation $\lambda \succ \mu$
is another way of saying that $\lambda/\mu$
is a horizontal strip, we find that
the expression of the skew Grothendieck polynomials
\eqref{singlematrixelements} which we defined as matrix elements
of the $B$-operator
\eqref{defskew}, is the same with the $\beta=1$
case of \eqref{singlematrixelementsYe} under the substitution
$x=1-u^{-1}$.

\eqref{singlematrixelementsYe} and the $n=1$ case of
\eqref{anotherskewdefinition} means
\begin{align}
A(x) \cdot \mu
=\sum_{
\lambda \succ \mu
}G_{\lambda /\!/ \mu}^{\beta}(x) \cdot \lambda,
\end{align}
and applying this relation repeatedly
to the right hand side of \eqref{anotherskewdefinition} for generic $n$,
we get
\begin{align}
&G_{\lambda /\!/  \mu}^{\beta}(x_1,\dots,x_n)
=\sum_{\lambda=\lambda^{(0)} 
\succ \lambda^{(1)} \succ \cdots \succ \lambda^{(n)}=\mu}
\prod_{j=1}^n G_{\lambda^{(j-1)} /\!/ \lambda^{(j)}}^{\beta}(x_j).
\label{skewGrothendieckYe}
\end{align}
Comparing with
\eqref{skewGrothendieck} in section 3,
we note that the skew Grothendieck polynomials
$G_{\lambda /\!/ \mu}(1-u_1^{-1},\dots,1-u_n^{-1})$ which we defined
and has partition function expressions using $B$-operators
\eqref{skewcorrespondence} is the same with the $\beta=1$ case of
the one \eqref{anotherskewdefinition} which are introduced
using Schur operators, where we substitute
the $u$-variables in the main part of this paper
into the $x$-variables in this Appendix
by $x_j=1-u_j^{-1}$, $j=1,\dots,n$.

\renewcommand{\theequation}{B.\arabic{equation} }
\setcounter{equation}{0}

\section*{Appendix B. Details of proofs of Lemma
\ref{firstlemmaproofinappendix} and Lemma
\ref{secondlemmaproofinappendix}
}
In this appendix,
we show Lemma
\ref{firstlemmaproofinappendix} and Lemma
\ref{secondlemmaproofinappendix} using equations only.
We abbreviate $R(u,1)$ as $R(u)$ in this Appendix.

Let us first show \eqref{relationBandD}.
For $j=1,\dots,n-1$, let $\bm{ \epsilon}_{m+n}^j$
be a sequence of 0s and 1s
$\bm{ \epsilon}_{m+n}^j=\{\epsilon_1^j,\epsilon_2^j,\dots,\epsilon_{m+n}^j \}$,
$\epsilon_1^j,\epsilon_2^j,\dots,\epsilon_{m+n}^j=0,1$,
and define $| \bm{\epsilon}_{m+n}^j \rangle_{m+n}$ and ${}_{m+n} \langle \bm{\epsilon}_{m+n}^j|$ as
\begin{align}
| \bm{\epsilon}_{m+n}^j \rangle_{m+n}
&:=| \epsilon_1^j \rangle_1 \otimes
| \epsilon_2^j \rangle_2 \otimes \cdots \otimes
| \epsilon_{m+n}^j \rangle_{m+n}, \\
{}_{m+n} \langle \bm{\epsilon}_{m+n}^j|
&:={}_1 \langle  \epsilon_1^j | \otimes
{}_2 \langle  \epsilon_2^j | \otimes \cdots \otimes
{}_{m+n} \langle  \epsilon_{m+n}^j |.
\end{align}
We also define $| \bm{\epsilon}_{m}^j \rangle_{m}$ and ${}_{m} \langle \bm{\epsilon}_{m}^j|$ similarly.

We write the identity operator on $W_1 \otimes \cdots \otimes W_{m+n}$ as
\begin{align}
\mathrm{Id}=\sum_{\bm{\epsilon}_{m+n}^j}
| \bm{\epsilon}_{m+n}^j \rangle_{m+n} \ {}_{m+n} \langle 
\bm{\epsilon}_{m+n}^j|, \label{appendixidentityoperatorj}
\end{align}
where $\displaystyle \sum_{\bm{\epsilon}_{m+n}^j}$ means
$\displaystyle \sum_{\epsilon_1^j,\dots,\epsilon_{m+n}^j=0,1}$.
We insert \eqref{appendixidentityoperatorj}
between
${}_{m+n} \langle y_1 \cdots y_{\ell},m+1,\dots,m+n|$,
$B^{(m+n)}(u_j)$, $j=1,\dots,n$
and $|x_1 \cdots x_\ell \rangle_{m+n}$,
and decompose the partition functions
$Z_{m+n,n}(u_1,\dots,u_n|x_1,\dots,x_\ell|y_1,\dots,y_{\ell},m+1,\dots,m+n)$
\eqref{anotherlarger}
as
\begin{align}
&Z_{m+n,n}(u_1,\dots,u_n|x_1,\dots,x_\ell|y_1,\dots,y_{\ell},m+1,\dots,m+n)
\nonumber \\
=&{}_{m+n} \langle y_1 \cdots y_{\ell},m+1,\dots,m+n|
\prod_{j=1}^{n}
B^{(m+n)}(u_j) |x_1 \cdots x_\ell \rangle_{m+n}
\nonumber \\
=&\sum_{\bm{\epsilon}_{m+n}^1, \bm{\epsilon}_{m+n}^2,\dots,\bm{\epsilon}_{m+n}^{n-1}}
{}_{m+n} \langle y_1 \cdots y_{\ell},m+1,\dots,m+n|
B^{(m+n)}(u_1) | \bm{\epsilon}_{m+n}^1 \rangle_{m+n} \nonumber \\
\times&\prod_{j=2}^{n-1}
{}_{m+n} \langle \bm{\epsilon}_{m+n}^{j-1}|
B^{(m+n)}(u_j) | \bm{\epsilon}_{m+n}^j \rangle_{m+n}
\times
{}_{m+n} \langle \bm{\epsilon}_{m+n}^{n-1} |
B^{(m+n)}(u_n)|x_1 \cdots x_\ell \rangle_{m+n}.
\label{decompositionforproof}
\end{align}
First, we examine the factor
${}_{m+n} \langle y_1 \cdots y_{\ell},m+1,\dots,m+n|
B^{(m+n)}(u_1) | \bm{\epsilon}_{m+n}^1 \rangle_{m+n}$
in the right hand side of \eqref{decompositionforproof}.
We use the explicit expression for the $B$-operator using $R$-matrices
$B^{(m+n)}(u_1)={}_a \langle 0| R_{a,m+n}(u_1) \cdots R_{a1}(u_1)|1 \rangle_a$
and rewrite the factor as
\begin{align}
&{}_{m+n} \langle y_1 \cdots y_{\ell},m+1,\dots,m+n|
B^{(m+n)}(u_1) | \bm{\epsilon}_{m+n}^1 \rangle_{m+n}
\nonumber \\
=&
{}_a \langle 0| \otimes {}_{m} \langle y_1 \cdots y_{\ell}|
\otimes {}_{m+1} \langle 1| \otimes \cdots \otimes
{}_{m+n} \langle 1|
R_{a,m+n}(u_1) \cdots R_{a1}(u_1) \nonumber \\
&\times |1 \rangle_a \otimes
| \bm{\epsilon}_{m}^1 \rangle_{m}
\otimes | \epsilon_{m+1}^1 \rangle_{m+1}
\otimes \cdots \otimes | \epsilon_{m+n}^1 \rangle_{m+n}.
\label{appendixtsuika}
\end{align}
We then write
the identity operator on $W_a \otimes W_1 \otimes \cdots \otimes W_{m+n}$
as
\begin{align}
\mathrm{Id}
=\sum_{\gamma_1,\epsilon_1,\dots,\epsilon_{m+n}=0,1}
|\gamma_1 \rangle_a \otimes |\epsilon_1 \rangle_1 \otimes
\cdots \otimes |\epsilon_{m+n} \rangle_{m+n} \
{}_a \langle \gamma_1| \otimes {}_1 \langle \epsilon_1 |
\otimes \cdots \otimes {}_{m+n} \langle \epsilon_{m+n}|,
\end{align}
and insert between $R_{a,m+1}(u_1)$ and $R_{am}(u_1)$
in \eqref{appendixtsuika} as
\begin{align}
&{}_{m+n} \langle y_1 \cdots y_{\ell},m+1,\dots,m+n|
B^{(m+n)}(u_1) | \bm{\epsilon}_{m+n}^1 \rangle_{m+n}
\nonumber \\
=&
\sum_{\gamma_1,\epsilon_1,\dots,\epsilon_{m+n}=0,1}
{}_a \langle 0| \otimes {}_{m} \langle y_1 \cdots y_{\ell}|
\otimes {}_{m+1} \langle 1| \otimes \cdots \otimes
{}_{m+n} \langle 1|
R_{a,m+n}(u_1) \cdots R_{a,m+1}(u_1) \nonumber \\
&\times
|\gamma_1 \rangle_a \otimes |\epsilon_1 \rangle_1 \otimes
\cdots \otimes |\epsilon_{m+n} \rangle_{m+n} \
{}_a \langle \gamma_1| \otimes {}_1 \langle \epsilon_1 |
\otimes \cdots \otimes {}_{m+n} \langle \epsilon_{m+n}|
R_{am}(u_1) \cdots R_{a1}(u_1) \nonumber \\
&\times |1 \rangle_a \otimes
| \bm{\epsilon}_{m}^1 \rangle_{m}
\otimes | \epsilon_{m+1}^1 \rangle_{m+1}
\otimes \cdots \otimes | \epsilon_{m+n}^1 \rangle_{m+n}.
\label{appendixtsuikatwo}
\end{align}
Since the product of $R$-matrices $R_{am}(u_1) \cdots R_{a1}(u_1)$
 act nontrivially
on $W_a$, $W_1 \dots,W_m$ and as identity
on $W_{m+1} \dots,W_{m+n}$, we have
\begin{align}
&{}_a \langle \gamma_1| \otimes {}_1 \langle \epsilon_1 |
\otimes \cdots \otimes {}_{m+n} \langle \epsilon_{m+n}|
R_{a,m}(u_1) \cdots R_{a1}(u_1)
|1 \rangle_a \otimes
| \bm{\epsilon}_{m}^1 \rangle_{m}
\otimes | \epsilon_{m+1}^1 \rangle_{m+1}
\otimes \cdots \otimes | \epsilon_{m+n}^1 \rangle_{m+n} \nonumber \\
&
=\prod_{j=m+1}^{m+n} \delta_{\epsilon_j \epsilon_j^1} \
{}_a \langle \gamma_1| \otimes
{}_1 \langle \epsilon_1 |
\otimes \cdots \otimes {}_{m} \langle \epsilon_{m}|
R_{am}(u_1) \cdots R_{a1}(u_1)
|1 \rangle_a \otimes
| \bm{\epsilon}_{m}^1 \rangle_{m}, \label{appendixtsuikathree}
\end{align}
where $R_{aj}(u_1)$, $j=1,\dots,m$ on the right hand side
are operators acting on $W_a \otimes W_1 \otimes \cdots \otimes W_m$.
Inserting \eqref{appendixtsuikathree} into \eqref{appendixtsuikatwo},
one gets
\begin{align}
&{}_{m+n} \langle y_1 \cdots y_{\ell},m+1,\dots,m+n|
B^{(m+n)}(u_1) | \bm{\epsilon}_{m+n}^1 \rangle_{m+n}
\nonumber \\
=&
\sum_{\gamma_1,\epsilon_1,\dots,\epsilon_m=0,1}
{}_a \langle 0| \otimes {}_{m} \langle y_1 \cdots y_{\ell}|
\otimes {}_{m+1} \langle 1| \otimes \cdots \otimes
{}_{m+n} \langle 1|
R_{a,m+n}(u_1) \cdots R_{a,m+1}(u_1) \nonumber \\
&\times
|\gamma_1 \rangle_a \otimes |\epsilon_1 \rangle_1 \otimes
\cdots \otimes |\epsilon_{m} \rangle_{m} \otimes
|\epsilon_{m+1}^1 \rangle_{m+1} \otimes \cdots \otimes
|\epsilon_{m+n}^1 \rangle_{m+n} \nonumber \\
&\times
{}_a \langle \gamma_1| \otimes
{}_1 \langle \epsilon_1 |
\otimes \cdots \otimes {}_{m} \langle \epsilon_{m}|
R_{am}(u_1) \cdots R_{a1}(u_1)
|1 \rangle_a \otimes
| \bm{\epsilon}_{m}^1 \rangle_{m}
.
\label{appendixtsuikafour}
\end{align}
Since the product $R_{a,m+n}(u_1) \cdots R_{a,m+1}(u_1)$
acts nontrivially on $W_a$, $W_{m+1}, \dots, W_{m+n}$
and as identity on $W_1, \dots, W_m$, we get
\begin{align}
&
{}_a \langle 0| \otimes {}_{m} \langle y_1 \cdots y_{\ell}|
\otimes {}_{m+1} \langle 1| \otimes \cdots \otimes
{}_{m+n} \langle 1|
R_{a,m+n}(u_1) \cdots R_{a,m+1}(u_1) \nonumber \\
&\times
|\gamma_1 \rangle_a \otimes |\epsilon_1 \rangle_1 \otimes
\cdots \otimes |\epsilon_{m} \rangle_{m} \otimes
|\epsilon_{m+1}^1 \rangle_{m+1} \otimes \cdots \otimes
|\epsilon_{m+n}^1 \rangle_{m+n} \nonumber \\
=&
{}_a \langle 0|
\otimes {}_{m+1} \langle 1| \otimes \cdots \otimes
{}_{m+n} \langle 1|
R_{a,m+n}(u_1) \cdots R_{a,m+1}(u_1) 
|\gamma_1 \rangle_a \otimes
|\epsilon_{m+1}^1 \rangle_{m+1} \otimes \cdots \otimes
|\epsilon_{m+n}^1 \rangle_{m+n},
\end{align}
if $\{\epsilon_1,\dots,\epsilon_m \}$ satisfies
$|\epsilon_1 \rangle_1 \otimes \cdots \otimes |\epsilon_m \rangle_m
=|y_1 \cdots y_\ell \rangle_m$
(${}_1 \langle \epsilon_1| \otimes \cdots \otimes {}_m \langle \epsilon_m|
={}_m \langle y_1 \cdots y_\ell|$),
and
\begin{align}
&
{}_a \langle 0| \otimes {}_{m} \langle y_1 \cdots y_{\ell}|
\otimes {}_{m+1} \langle 1| \otimes \cdots \otimes
{}_{m+n} \langle 1|
R_{a,m+n}(u_1) \cdots R_{a,m+1}(u_1) \nonumber \\
&\times
|\gamma_1 \rangle_a \otimes |\epsilon_1 \rangle_1 \otimes
\cdots \otimes |\epsilon_{m} \rangle_{m} \otimes
|\epsilon_{m+1}^1 \rangle_{m+1} \otimes \cdots \otimes
|\epsilon_{m+n}^1 \rangle_{m+n}=0,
\end{align}
otherwise.
Then we find \eqref{appendixtsuikafour} can be rewritten as
\begin{align}
&{}_{m+n} \langle y_1 \cdots y_{\ell},m+1,\dots,m+n|
B^{(m+n)}(u_1) | \bm{\epsilon}_{m+n}^1 \rangle_{m+n}
\nonumber \\
=&
\sum_{\gamma_1=0,1}
{}_a \langle 0|
\otimes {}_{m+1} \langle 1| \otimes \cdots \otimes
{}_{m+n} \langle 1|
R_{a,m+n}(u_1) \cdots R_{a,m+1}(u_1) \nonumber \\
&\times
|\gamma_1 \rangle_a \otimes
|\epsilon_{m+1}^1 \rangle_{m+1} \otimes \cdots \otimes
|\epsilon_{m+n}^1 \rangle_{m+n} \nonumber \\
&\times
{}_a \langle \gamma_1| \otimes
{}_m \langle y_1 \cdots y_{\ell}|
R_{am}(u_1) \cdots R_{a1}(u_1)
|1 \rangle_a \otimes
| \bm{\epsilon}_{m}^1 \rangle_{m}
.
\label{appendixtsuikafive}
\end{align}
The factor ${}_a \langle 0|
\otimes {}_{m+1} \langle 1| \otimes \cdots \otimes
{}_{m+n} \langle 1|
R_{a,m+n}(u_1) \cdots R_{a,m+1}(u_1)
|\gamma_1 \rangle_a \otimes
|\epsilon_{m+1}^1 \rangle_{m+1} \otimes \cdots \otimes
|\epsilon_{m+n}^1 \rangle_{m+n}$
can be explicitly expressed using $R$-matrix elements as
\begin{align}
&{}_a \langle 0|
\otimes {}_{m+1} \langle 1| \otimes \cdots \otimes
{}_{m+n} \langle 1|
R_{a,m+n}(u_1) \cdots R_{a,m+1}(u_1)
|\gamma_1 \rangle_a \otimes
|\epsilon_{m+1}^1 \rangle_{m+1} \otimes \cdots \otimes
|\epsilon_{m+n}^1 \rangle_{m+n} \nonumber \\
=&\sum_{\gamma_2,\dots,\gamma_n=0,1}
[R(u_1)]_{\gamma_n \epsilon_{m+n}^1}^{01}
\prod_{j=1}^{n-1}
[R(u_1)]_{\gamma_{j}, \epsilon_{m+j}^1}^{\gamma_{j+1}, 1}.
\label{appendixtsuikasix}
\end{align}
Inserting \eqref{appendixtsuikasix} into
\eqref{appendixtsuikafive} gives
\begin{align}
&{}_{m+n} \langle y_1 \cdots y_{\ell},m+1,\dots,m+n|
B^{(m+n)}(u_1) | \bm{\epsilon}_{m+n}^1 \rangle_{m+n} \nonumber \\
=&\sum_{\gamma_1,\dots,\gamma_n=0,1}
[R(u_1)]_{\gamma_n \epsilon_{m+n}^1}^{01}
\prod_{j=1}^{n-1}
[R(u_1)]_{\gamma_{j}, \epsilon_{m+j}^1}^{\gamma_{j+1}, 1} \nonumber \\
\times&{}_a \langle \gamma_1 | \otimes {}_m \langle y_1 \cdots y_\ell|
R_{am}(u_1) \cdots R_{a1}(u_1)|1 \rangle_a \otimes |\bm{\epsilon}_{m}^1 \rangle_m.
\label{explicitbelement}
\end{align}
Since $[R(u_1)]_{\gamma_n \epsilon_{m+n}^1}^{01}=1$
if $\gamma_n=1,\epsilon_{m+n}^1=0$ and
$[R(u_1)]_{\gamma_n \epsilon_{m+n}^1}^{01}=0$
otherwise,
we can restrict the sum over $\gamma_n$ in the right hand side of \eqref{explicitbelement} to $\gamma_n=1$.
Then we find the factor
$[R(u_1)]_{\gamma_{n-1} \epsilon_{m+n-1}^1}^{\gamma_n 1}$
becomes $[R(u_1)]_{\gamma_{n-1} \epsilon_{m+n-1}^1}^{1 1}$,
which is equal to
1 if $\gamma_{n-1}=\epsilon_{m+n-1}^1=1$ and vanishes otherwise.
Continuing this argument, we find that 
we only need to consider the case which
$\epsilon_{m+1}^1=\cdots=\epsilon_{m+n-1}^1=1$, $\epsilon_{m+n}^1=0$
is satisfied since otherwise 
${}_{m+n} \langle y_1 \cdots y_{\ell},m+1,\dots,m+n|
B^{(m+n)}(u_1) | \bm{\epsilon}_{m+n}^1 \rangle_{m+n}=0$,
and we can restrict the sum in
the right hand side of \eqref{explicitbelement} to
the summand corresponding to $\gamma_1=\cdots=\gamma_n=1$.
Then we find the right hand side of
\eqref{explicitbelement} becomes
\begin{align}
&[R(u_1)]_{10}^{01}
([R(u_1)]_{11}^{11})^{n-1}
{}_a \langle 1 | \otimes {}_m \langle y_1 \cdots y_\ell|
R_{am}(u_1) \cdots R_{a1}(u_1)|1 \rangle_a \otimes |\bm{\epsilon}_m^1 \rangle_m
\nonumber \\
=&{}_m \langle y_1 \cdots y_\ell| D^{(m)}(u_1) |\bm{\epsilon}_m^1 \rangle_m,
\end{align}
hence we conclude
\begin{align}
&{}_{m+n} \langle y_1 \cdots y_{\ell},m+1,\dots,m+n|
B^{(m+n)}(u_1) | \bm{\epsilon}_{m+n}^1 \rangle_{m+n} 
={}_m \langle y_1 \cdots y_\ell| D^{(m)}(u_1) |\bm{\epsilon}_m^1 \rangle_m,
\label{detailedfactorizationone}
\end{align}
if $\bm{\epsilon}_{m+n}^1$ satisfies
$\epsilon_{m+1}^1=\cdots=\epsilon_{m+n-1}^1=1$, $\epsilon_{m+n}^1=0$,
and
${}_{m+n} \langle y_1 \cdots y_{\ell},m+1,\dots,m+n|
B^{(m+n)}(u_1) | \bm{\epsilon}_{m+n}^1 \rangle_{m+n} =0$ otherwise.

Next, we examine
${}_{m+n} \langle \bm{\epsilon}_{m+n}^{1}|
B^{(m+n)}(u_2) | \bm{\epsilon}_{m+n}^2 \rangle_{m+n}$
under $\epsilon_{m+1}^1=\cdots=\epsilon_{m+n-1}^1=1$, $\epsilon_{m+n}^1=0$.
We first write this explicitly as
\begin{align}
{}_{m+n} \langle \bm{\epsilon}_{m+n}^{1}|
B^{(m+n)}(u_2) | \bm{\epsilon}_{m+n}^2 \rangle_{m+n}
=&\sum_{\gamma_1,\dots,\gamma_n=0,1}
[R(u_2)]_{\gamma_n \epsilon_{m+n}^2}^{00}
\prod_{j=1}^{n-1}
[R(u_2)]_{\gamma_{j}, \epsilon_{m+j}^2}^{\gamma_{j+1}, 1} \nonumber \\
\times&{}_a \langle \gamma_1 | \otimes {}_m \langle \bm{\epsilon}_m^1 |
R_{am}(u_2) \cdots R_{a1}(u_2)|1 \rangle_a \otimes |\bm{\epsilon}_m^2 \rangle_m,
\label{explicitbelementtwo}
\end{align}
in the same way we did as above.
Then we observe that the factor $[R(u_2)]_{\gamma_n \epsilon_{m+n}^2}^{00}$
in the right hand side of \eqref{explicitbelementtwo}
is equal to 1 if $\gamma_n=\epsilon_{m+n}^2=0$, and vanishes otherwise.
Then we note the matrix element $[R(u_2)]_{\gamma_{n-1} \epsilon_{m+n-1}^2}^{\gamma_n 1}$ becomes $[R(u_2)]_{\gamma_{n-1} \epsilon_{m+n-1}^2}^{0 1}
$ and find that we only need to consider $\gamma_{n-1}=1, \epsilon_{m+n-1}^2=0$,
since otherwise the matrix element becomes 0.
Continuing this argument,
we find ${}_{m+n} \langle \bm{\epsilon}_{m+n}^{1}|
B^{(m+n)}(u_2) | \bm{\epsilon}_{m+n}^2 \rangle_{m+n}=0$
unless $\epsilon_{m+1}^2=\cdots=\epsilon_{m+n-2}^2=1$, $\epsilon_{m+n-1}^2=\epsilon_{m+n}^2=0$,
and if $\bm{\epsilon}_{m+n}^{2}$
satisfies
$\epsilon_{m+1}^2=\cdots=\epsilon_{m+n-2}^2=1$, $\epsilon_{m+n-1}^2=\epsilon_{m+n}^2=0$,
we can restrict the sum in the right hand side of
\eqref{explicitbelementtwo} to the summand corresponding to
$\gamma_1=\cdots=\gamma_{n-1}=1$, $\gamma_n=0$
and we have
\begin{align}
{}_{m+n} \langle \bm{\epsilon}_{m+n}^{1}|
B^{(m+n)}(u_2) | \bm{\epsilon}_{m+n}^2 \rangle_{m+n}
=&
[R(u_2)]_{00}^{00}
[R(u_2)]_{10}^{01}
([R(u_2)]_{11}^{11})^{n-2} \nonumber \\
\times&{}_a \langle 1 | \otimes {}_m \langle \bm{\epsilon}_m^1 |
R_{am}(u_2) \cdots R_{a1}(u_2)|1 \rangle_a \otimes |\bm{\epsilon}_m^2 \rangle_m
\nonumber \\
=&{}_m \langle \bm{\epsilon}_m^1 | D^{(m)}(u_2) |\bm{\epsilon}_m^2 \rangle_m.
\end{align}

We continue this argument and show the following for $j=2,\dots,n-1$:
for $\bm{\epsilon}_{m+n}^{j-1}$ such that
$\epsilon_{m+1}^{j-1}=\cdots=\epsilon_{m+n-j+1}^{j-1}=1$,
$\epsilon_{m+n-j+2}^{j-1}=\cdots=\epsilon_{m+n}^{j-1}=0$,
the following relation holds
\begin{align}
{}_{m+n} \langle \bm{\epsilon}_{m+n}^{j-1}|
B^{(m+n)}(u_j) | \bm{\epsilon}_{m+n}^j \rangle_{m+n}
={}_{m} \langle \bm{\epsilon}_m^{j-1}|
D^{(m)}(u_j) | \bm{\epsilon}_m^j \rangle_{m}, \label{detailedfactorizationtwo}
\end{align}
if $\bm{\epsilon}_{m+n}^{j}$ satisfies
$\epsilon_{m+1}^{j}=\cdots=\epsilon_{m+n-j}^{j}=1$,
$\epsilon_{m+n-j+1}^{j}=\cdots=\epsilon_{m+n}^{j}=0$.
Otherwise,
${}_{m+n} \langle \bm{\epsilon}_{m+n}^{j-1}|
B^{(m+n)}(u_j) | \bm{\epsilon}_{m+n}^j \rangle_{m+n}
=0$.

One can also show in a similar way that
for $\bm{\epsilon}_{m+n}^{n-1}$ such that
$\epsilon_{m+1}^{n-1}=1$,
$\epsilon_{m+2}^{n-1}=\cdots=\epsilon_{m+n}^{n-1}=0$,
the following relation holds
\begin{align}
{}_{m+n} \langle \bm{\epsilon}_{m+n}^{n-1}|
B^{(m+n)}(u_n) | x_1 \cdots x_\ell \rangle_{m+n}
={}_{m} \langle \bm{\epsilon}_m^{n-1}|
D^{(m)}(u_n) | x_1 \cdots x_\ell \rangle_{m}.
\label{detailedfactorizationthree}
\end{align}

From the arguments above and
using
\eqref{detailedfactorizationone},
\eqref{detailedfactorizationtwo} and
\eqref{detailedfactorizationthree},
the right hand side of \eqref{decompositionforproof}
can be rewritten in the following way:
\begin{align}
&\sum_{\bm{\epsilon}_{m+n}^1, \bm{\epsilon}_{m+n}^2,\dots,\bm{\epsilon}_{m+n}^{n-1}}
{}_{m+n} \langle y_1 \cdots y_{\ell},m+1,\dots,m+n|
B^{(m+n)}(u_1) | \bm{\epsilon}_{m+n}^1 \rangle_{m+n} \nonumber \\
\times&\prod_{j=2}^{n-1}
{}_{m+n} \langle \bm{\epsilon}_{m+n}^{j-1}|
B^{(m+n)}(u_j) | \bm{\epsilon}_{m+n}^j \rangle_{m+n}
\times
{}_{m+n} \langle \bm{\epsilon}_{m+n}^{n-1} |
B^{(m+n)}(u_n)|x_1 \cdots x_\ell \rangle_{m+n} \nonumber \\
=&\sum_{\bm{\epsilon}_m^1, \bm{\epsilon}_m^2,\dots,\bm{\epsilon}_m^{n-1}}
{}_{m} \langle y_1 \cdots y_{\ell}|
D^{(m)}(u_1) | \bm{\epsilon}_m^1 \rangle_{m} \nonumber \\
\times&\prod_{j=2}^{n-1}
{}_{m} \langle \bm{\epsilon}_m^{j-1}|
D^{(m)}(u_j) | \bm{\epsilon}_m^j \rangle_{m}
\times
{}_{m} \langle \bm{\epsilon}_m^{n-1} |
D^{(m)}(u_n)|x_1 \cdots x_\ell \rangle_{m} \nonumber \\
=&
{}_{m} \langle y_1 \cdots y_{\ell}|
\prod_{j=1}^n D^{(m)}(u_j)|x_1 \cdots x_\ell \rangle_{m} \nonumber \\
=&U_{m,n}(u_1,\dots,u_n|x_1,\dots,x_\ell|y_1,\dots,y_{\ell}),
\label{appendixequality}
\end{align}
hence we conclude \eqref{relationBandD} holds.
Note that we used the expression for the
identity operator acting on $W_1 \otimes \cdots \otimes W_m$
\begin{align}
\mathrm{Id}=\sum_{\bm{\epsilon}_{m}^j}
| \bm{\epsilon}_{m}^j \rangle_{m} \ {}_{m} \langle 
\bm{\epsilon}_{m}^j|, \label{identityoperatorappendix}
\end{align}
in the second equality in \eqref{appendixequality}.
We also remark that from the argument above,
$\epsilon_k^j$, $j=1,\dots,n-1$, $k=m+1,\dots,m+n$
are fixed to
$\epsilon_{m+1}^{j}=\cdots=\epsilon_{m+n-j}^{j}=1$,
$\epsilon_{m+n-j+1}^{j}=\cdots=\epsilon_{m+n}^{j}=0$
and the sum
$\displaystyle \sum_{\bm{\epsilon}_{m+n}^1, \bm{\epsilon}_{m+n}^2,\dots,\bm{\epsilon}_{m+n}^{n-1}}$ is reduced to
$\displaystyle \sum_{\bm{\epsilon}_{m}^1, \bm{\epsilon}_{m}^2,\dots,\bm{\epsilon}_{m}^{n-1}}$ which is used in the first equality in \eqref{appendixequality}. \\
\\

We can show \eqref{largeanddb}
in the same way. 
For $j=1,\dots,n-1$,
we define
$| \bm{\epsilon}_{m+n-k}^j \rangle_{m+n-k}:=|\epsilon_1^j \rangle_1 \otimes \cdots \otimes |\epsilon_{m+n-k}^j \rangle_{m+n-k} $
and ${}_{m+n-k} \langle \bm{\epsilon}_{m+n-k}^j|
:={}_{1} \langle \epsilon_1^j | \otimes \cdots \otimes {}_{m+n-k} \langle \epsilon_{m+n-k}^j |$
for a sequence of 0s and 1s $\bm{\epsilon}_{m+n-k}^j
=\{\epsilon_1^j, \epsilon_2^j,\dots,\epsilon_{m+n-k}^j \}$ ($\epsilon_1^j,\epsilon_2^j,\dots,\epsilon_{m+n-k}^j=0,1$),
write the identity operator
on $W_1 \otimes \cdots \otimes W_{m+n-k}$ as
\begin{align}
\mathrm{Id}=\sum_{\bm{\epsilon}_{m+n-k}^j}
| \bm{\epsilon}_{m+n-k}^j \rangle_{m+n-k} \ {}_{m+n-k} \langle 
\bm{\epsilon}_{m+n-k}^j|,
\end{align}
and insert into
$Z_{m+n-k,n}
(u_1,\dots,u_n|x_1,\dots,x_\ell|y_1,\dots,y_{\ell+k},
m+1,\dots,m+n-k)$ \eqref{larger} and decompose as
\begin{align}
&Z_{m+n-k,n}
(u_1,\dots,u_n|x_1,\dots,x_\ell|y_1,\dots,y_{\ell+k},
m+1,\dots,m+n-k) \nonumber \\
=&_{m} \langle y_1 \cdots y_{\ell+k},m+1 \cdots m+n-k|
B^{(m+n-k)}(u_n) \cdots B^{(m+n-k)}(u_1)|x_1 \cdots x_\ell \rangle_{m}
\nonumber \\
=&\sum_{\bm{\epsilon}_{m+n-k}^1,\dots,\bm{\epsilon}_{m+n-k}^{n-1}} {}_{m+n-k} \langle y_1 \cdots y_{\ell+k},m+1,\dots,m+n-k|
B^{(m+n-k)}(u_n)| \bm{\epsilon}_{m+n-k}^1 \rangle_{m+n-k} \nonumber \\
&\times\prod_{j=2}^{n-1} {}_{m+n-k} \langle \bm{\epsilon}_{m+n-k}^{j-1}| B^{(m+n-k)}(u_{n-j+1})|\bm{\epsilon}_{m+n-k}^j \rangle_{m+n-k} \nonumber \\
&\times
{}_{m+n-k} \langle \bm{\epsilon}_{m+n-k}^{n-1}|B^{(m+n-k)}(u_1)|x_1 \cdots x_\ell \rangle_{m+n-k}.
\label{decompositionintomatrixelements}
\end{align}
We analyze the factors in
\eqref{decompositionintomatrixelements} in the following order.

One first shows that if
$\bm{\epsilon}_{m+n-k}^1$ satisfies
$\epsilon_{m+1}^1=\cdots=\epsilon_{m+n-k-1}^1=1, \epsilon_{m+n-k}^1=0$,
the following relation holds
\begin{align}
&{}_{m+n-k} \langle y_1 \cdots y_{\ell+k},m+1,\dots,m+n-k|
B^{(m+n-k)}(u_n)| \bm{\epsilon}_{m+n-k}^1 \rangle_{m+n-k} \nonumber \\
=&
{}_{m} \langle y_1 \cdots y_{\ell+k}|
D^{(m)}(u_n)| \bm{\epsilon_m}^1 \rangle_{m},
\label{reductioncompone}
\end{align}
and
${}_{m+n-k} \langle y_1 \cdots y_{\ell+k},m+1,\dots,m+n-k|
B^{(m+n-k)}(u_n)| \bm{\epsilon}_{m+n-k}^1 \rangle_{m+n-k}=0$
otherwise.

Next, we show the following for $j=2,\dots,n-k$:
for $\bm{\epsilon}_{m+n-k}^{j-1}$ such that
$\epsilon_{m+1}^{j-1}=\cdots=\epsilon_{m+n-k-j+1}^{j-1}=1, \epsilon_{m+n-k-j+2}^{j-1}=\cdots=\epsilon_{m+n-k}^{j-1}=0$,
the following relation holds
\begin{align}
&{}_{m+n-k} \langle \bm{\epsilon}_{m+n-k}^{j-1}| B^{(m+n-k)}(u_{n-j+1})|\bm{\epsilon}_{m+n-k}^j \rangle_{m+n-k}
={}_{m} \langle \bm{\epsilon}_m^{j-1}| D^{(m)}(u_{n-j+1})|\bm{\epsilon}_m^j \rangle_{m}, \label{reductioncomptwo}
\end{align}
if $\bm{\epsilon}_{m+n-k}^j$ satisfies
$\epsilon_{m+1}^j=\cdots=\epsilon_{m+n-k-j}^j=1, \epsilon_{m+n-k-j+1}^j=\cdots=\epsilon_{m+n-k}^j=0$.
Otherwise,
${}_{m+n-k} \langle \bm{\epsilon}_{m+n-k}^{j-1}| B^{(m+n-k)}(u_{n-j+1})|\bm{\epsilon}_{m+n-k}^j \rangle_{m+n-k}=0$.

The third step is to show the following for $j=n-k+1,\dots,n-1$:
for $\bm{\epsilon}_{m+n-k}^{j-1}$ such that
$\epsilon_{m+1}^{j-1}=\cdots=\epsilon_{m+n-k}^{j-1}=0$,
the following relation holds
\begin{align}
&{}_{m+n-k} \langle \bm{\epsilon}_{m+n-k}^{j-1}| B^{(m+n-k)}(u_{n-j+1})
|\bm{\epsilon}_{m+n-k}^j \rangle_{m+n-k}
={}_{m} \langle \bm{\epsilon}_m^{j-1}| B^{(m)}(u_{n-j+1})|\bm{\epsilon}_m^j \rangle_{m}, \label{reductioncompthree}
\end{align}
if $\bm{\epsilon}_{m+n-k}^j$ satisfies
$\epsilon_{m+1}^j=\cdots=\epsilon_{m+n-k}^j=0$.
Otherwise, we have \\
${}_{m+n-k} \langle \bm{\epsilon}_{m+n-k}^{j-1}| B^{(m+n-k)}(u_{n-j+1})
|\bm{\epsilon}_{m+n-k}^j \rangle_{m+n-k}=0$.

Finally, we show that for $\bm{\epsilon}_{m+n-k}^{n-1}$ such that
$\epsilon_{m+1}^{n-1}=\cdots=\epsilon_{m+n-k}^{n-1}=0$,
the following relation holds
\begin{align}
&{}_{m+n-k} \langle \bm{\epsilon}_{m+n-k}^{n-1}|B^{(m+n-k)}(u_1)|x_1 \cdots x_\ell \rangle_{m+n-k}
=
{}_{m} \langle \bm{\epsilon}_m^{n-1}|B^{(m)}(u_1)|x_1 \cdots x_\ell \rangle_{m}.
\label{reductioncompfour}
\end{align}
From the arguments above and
inserting \eqref{reductioncompone}, \eqref{reductioncomptwo},
\eqref{reductioncompthree} and \eqref{reductioncompfour}
into the right hand side of
\eqref{decompositionintomatrixelements}
and using \eqref{identityoperatorappendix},
we can show the right hand side of
\eqref{decompositionintomatrixelements}
can be rewritten in the following way:
\begin{align}
&\sum_{\bm{\epsilon}_{m+n-k}^1,\dots,\bm{\epsilon}_{m+n-k}^{n-1}} {}_{m+n-k} \langle y_1 \cdots y_{\ell+k},m+1,\dots,m+n-k|
B^{(m+n-k)}(u_n)| \bm{\epsilon}_{m+n-k}^1 \rangle_{m+n-k} \nonumber \\
&\times\prod_{j=2}^{n-1} {}_{m+n-k} \langle \bm{\epsilon}_{m+n-k}^{j-1}| B^{(m+n-k)}(u_{n-j+1})|\bm{\epsilon}_{m+n-k}^j \rangle_{m+n-k} 
\nonumber \\
&\times
{}_{m+n-k} \langle \bm{\epsilon}_{m+n-k}^{n-1}|B^{(m+n-k)}(u_1)|x_1 \cdots x_\ell \rangle_{m+n-k} \nonumber \\
=&\sum_{\bm{\epsilon}_m^1,\dots,\bm{\epsilon}_m^{n-1}} {}_{m+n-k} \langle y_1 \cdots y_{\ell+k}|
D^{(m)}(u_n)| \bm{\epsilon}_m^1 \rangle_{m}
\prod_{j=2}^{n-k} {}_{m} \langle \bm{\epsilon}_m^{j-1}| D^{(m)}(u_{n-j+1})|
\bm{\epsilon}_m^j \rangle_{m}
\nonumber \\
&\times \prod_{j=n-k+1}^{n-1} {}_{m} \langle \bm{\epsilon}_m^{j-1}| B^{(m)}(u_{n-j+1})|\bm{\epsilon}_m^j \rangle_{m} \times
{}_{m} \langle \bm{\epsilon}_m^{n-1}|B^{(m)}(u_1)|x_1 \cdots x_\ell \rangle_{m} \nonumber \\
=&
\langle y_1 \cdots y_{\ell+k}|
D^{(m)}(u_{n}) \cdots D^{(m)}(u_{k+1}) B^{(m)}(u_k) \cdots B^{(m)}(u_1)
|x_1 \cdots x_\ell \rangle_{m} \nonumber \\
=&ZDB_{m,n,k}(u_{k+1},\dots,u_n|u_1,\dots,u_k|x_1,\dots,x_\ell|y_1,\dots,y_{\ell+k}), \label{appendixtsuikasaigo}
\end{align}
hence we conclude \eqref{largeanddb} holds.
Note that $\epsilon_k^j$, $j=1,\dots,n-1$, $k=m+1,\dots,m+n$
are fixed to
$\epsilon_{m+1}^j=\cdots=\epsilon_{m+n-k-j}^j=1, \epsilon_{m+n-k-j+1}^j=\cdots=\epsilon_{m+n-k}^j=0$ for $j=1,\dots,n-k-1$,
and $\epsilon_{m+1}^j=\cdots=\epsilon_{m+n-k}^j=0$
for $j=n-k,\dots,n-1$ by the argument above,
and the sum
$\displaystyle \sum_{\bm{\epsilon}_{m+n-k}^1,\dots,\bm{\epsilon}_{m+n-k}^{n-1}}$ is reduced to
$\displaystyle \sum_{\bm{\epsilon}_{m}^1,\dots,\bm{\epsilon}_{m}^{n-1}}$ which is used in the first equality in \eqref{appendixtsuikasaigo}.

\renewcommand{\theequation}{C.\arabic{equation} }
\setcounter{equation}{0}

\section*{Appendix C.
Graphical derivations of
Lemma
\ref{firstlemmaproofinappendix} and Lemma
\ref{secondlemmaproofinappendix}
}
In this appendix, we give graphical derivations
of Lemma
\ref{firstlemmaproofinappendix} and Lemma
\ref{secondlemmaproofinappendix}.

\begin{figure}[h]
\includegraphics[width=12cm]{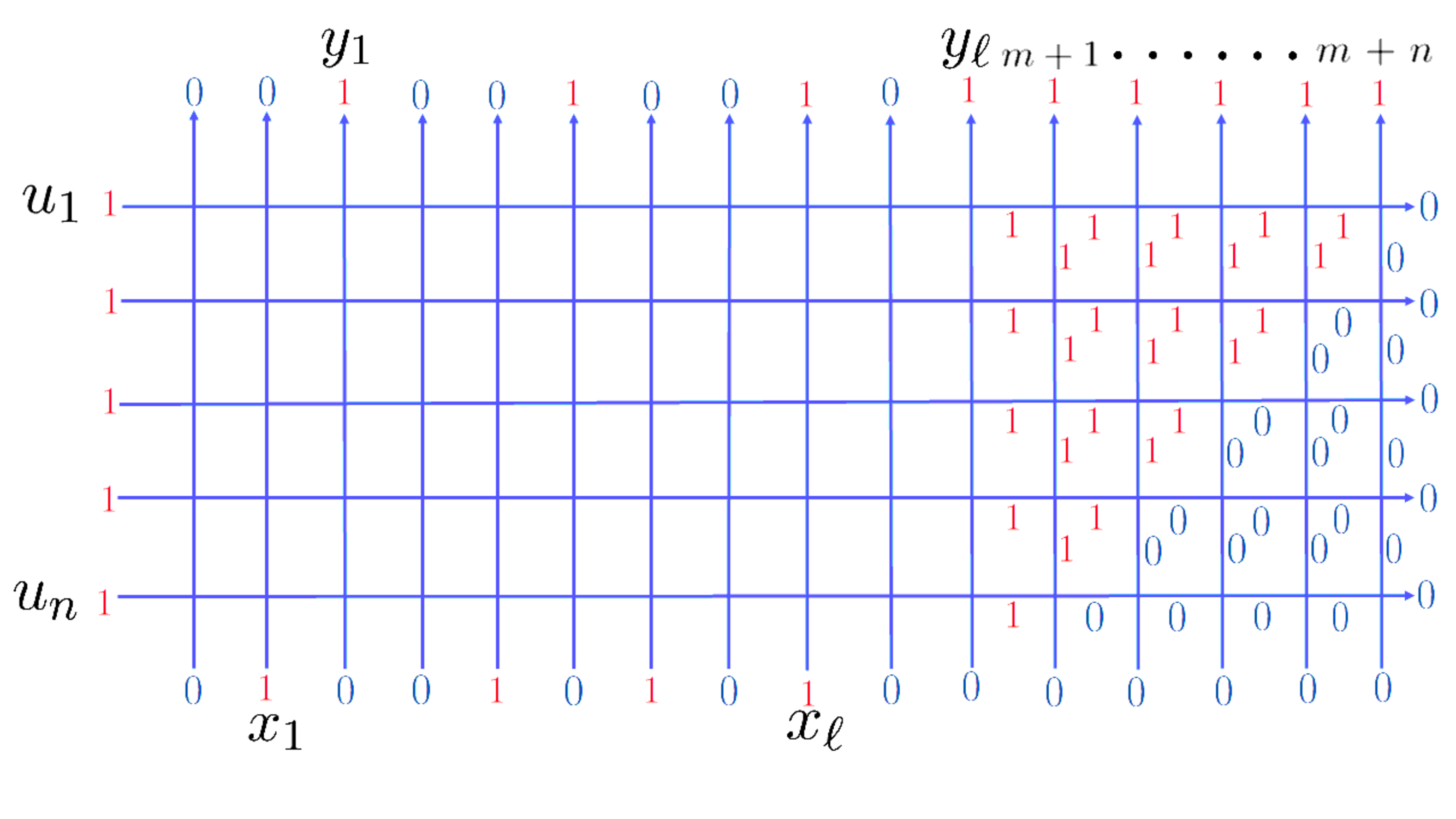}
\caption{
A graphical description which shows
$
Z_{m+n,n}(u_1,\dots,u_n|x_1,\dots,x_\ell|y_1,\dots,y_{\ell},m+1,\dots,m+n)
=U_{m,n}(u_1,\dots,u_n|x_1,\dots,x_\ell|y_1,\dots,y_{\ell})
$ \eqref{relationBandD}.
}
\label{picturebpartitionreduction}
\end{figure}

First, let us show \eqref{relationBandD}.
Figure \ref{picturebpartitiondefinition}
in section 3 represents the partition function \\
$Z_{m+n,n}(u_1,\dots,u_n|x_1,\dots,x_\ell|y_1,\dots,y_{\ell},m+1,\dots,m+n)$ \eqref{anotherlarger}, and by graphical observation,
one can see that only one configuration is allowed in
the rightmost $n$ columns, which is graphically
Figure \ref{picturebpartitionreduction}.
Let us explain this observation in more detail.
First, we look the $R$-matrix at the upper right corner in
Figure \ref{picturebpartitiondefinition}
and note that
the output of the auxiliary space and the quantum space $W_{m+n}$
is 0 and 1, respectively.
From this, we note that we can restrict
the input of the auxiliary space and the quantum space $W_{m+n}$
to 1 and 0 respectively,
since $[R_{a,m+n}(u_1,1)]_{\gamma_n \epsilon_{m+n}^1}^{01} \not\equiv 0$
if and only if $\gamma_n=1$ and $\epsilon_{m+n}^1=0$
(see Figure \ref{picturermatrix}).
$[R_{a,m+n}(u_1,1)]_{\gamma_n \epsilon_{m+n}^1}^{01}$
corresponds to the factor
$[R(u_1)]_{\gamma_n \epsilon_{m+n}^1}^{01}$ in \eqref{explicitbelement}
in the previous appendix.
Next, we move to the second $R$-matrix on the top row, counted from right.
The outputs of the auxiliary space and the quantum space $W_{m+n-1}$ are
both 1, hence we find the inputs of both spaces can be restricted to 1,
since $[R_{a,m+n-1}(u_1,1)]_{\gamma_{n-1} \epsilon_{m+n-1}^1}^{11}$
is not identically zero if and only if $\gamma_{n-1} =\epsilon_{m+n-1}^1=1$
(see Figure \ref{picturermatrix}).
Continuing this observation,
one can see that only one configuration is allowed in
the rightmost $n$ columns,
and the part constructed from the remaining $m$ columns
is nothing but the partition functions
$U_{m,n}(u_1,\dots,u_n|x_1,\dots,x_\ell|y_1,\dots,y_{\ell})$
\eqref{Doperatorpartitionfunction}
(Figure \ref{picturedpartition}),
which is the graphical meaning of
Figure \ref{picturebpartitionreduction}.
The product of the $R$-matrix elements
on the rightmost $n$ columns is 1
since all $R$-matrix elements which appear in that part
are either $[R(u_j,1)]_{00}^{00}=1$, $[R(u_j,1)]_{11}^{11}=1$ or
$[R(u_j,1)]_{10}^{01}=1$.
Hence we conclude
$Z_{m+n,n}(u_1,\dots,u_n|x_1,\dots,x_\ell|y_1,\dots,y_{\ell},m+1,\dots,m+n)$
is equal to $U_{m,n}(u_1,\dots,u_n|x_1,\dots,x_\ell|y_1,\dots,y_{\ell})$
multiplied by 1, and we get \eqref{relationBandD}. \\
\\

\begin{figure}[h]
\includegraphics[width=12cm]{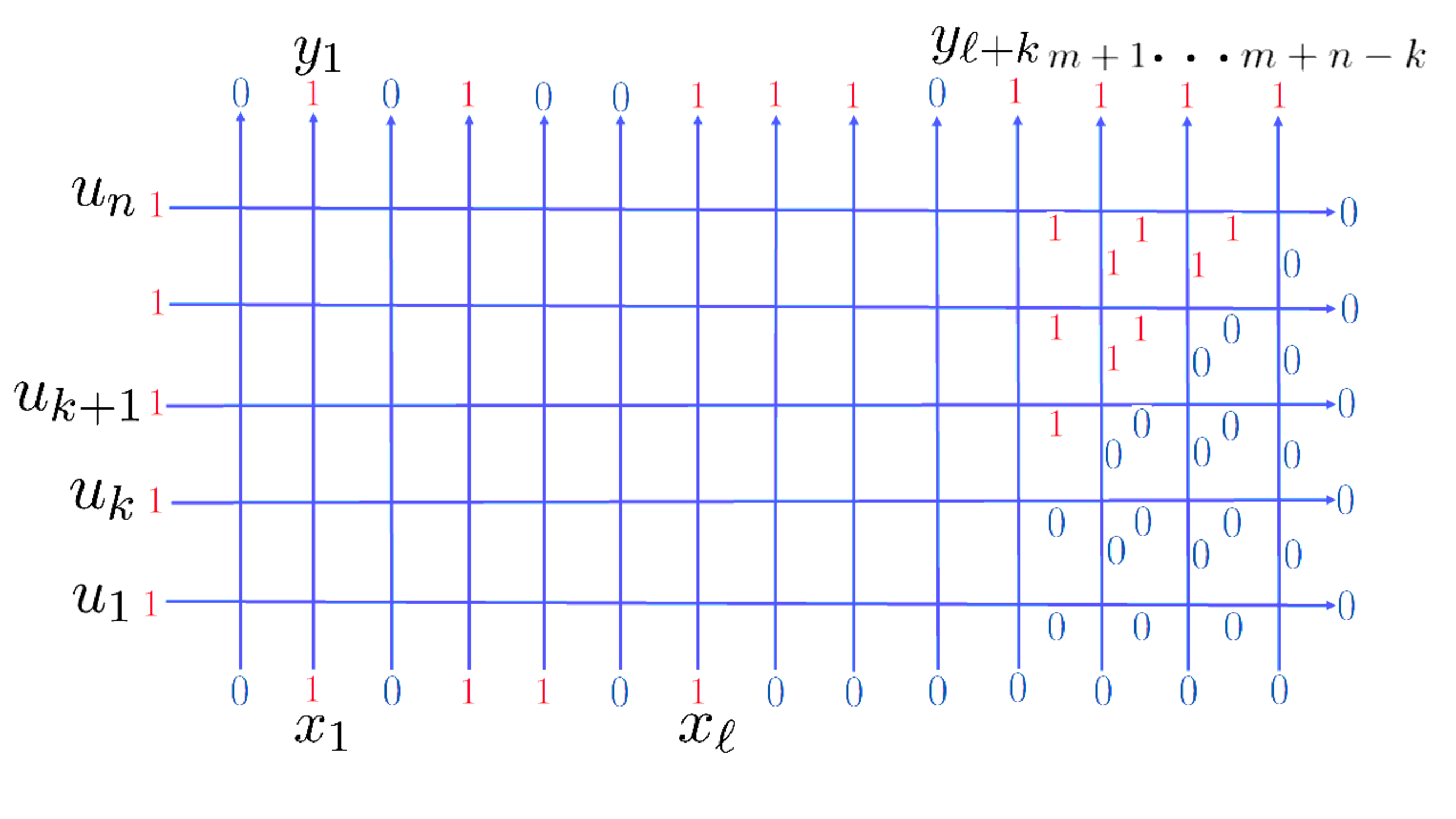}
\caption{
A graphical description which shows
$Z_{m+n-k,n}
(u_1,\dots,u_n|x_1,\dots,x_\ell|y_1,\dots,y_{\ell+k},
m+1,\dots,m+n-k)
=ZDB_{m,n,k}(u_{k+1},\dots,u_n|u_1,\dots,u_k|x_1,\dots,x_\ell|y_1,\dots,y_{\ell+k})$ \eqref{largeanddb}.
}
\label{picturedbpartitionreduction}
\end{figure}

\eqref{largeanddb} can be graphically derived in a similar way.
Figure \ref{picturedbpartitiondefinition} in section 3
represents the parition function
$Z_{m+n-k,n}
(u_1,\dots,u_n|x_1,\dots,x_\ell|y_1,\dots,y_{\ell+k},
m+1,\dots,m+n-k)$
\eqref{larger},
and by graphical observation, we find that
only one configuration is allowed in the rightmost $n-k$ columns
( Figure \ref{picturedbpartitionreduction}).
In more detail, we first perform graphical observations starting from the
$R$-matrix at the upper right corner,
and note that we can restrict the configuration of the
rightmost $n-k$ columns of the top $n-k$ rows in the same way as
we have seen in the observation when going from Figure \ref{picturebpartitiondefinition}
to Figure \ref{picturebpartitionreduction}.
Next, one looks the $(n-k+1)$-th row counted from top.
Since the outputs of the auxiliary space and the quantum spaces
$W_{m+1},\dots,W_{m+n-k}$ are all 0, we can conclude
that we can restrict all the inputs of those spaces to 0,
since
$[R(u_k,1)]_{j_1 j_2}^{0 0} \not\equiv 0$ if and only if $j_1=j_2=0$,
and $[R(u_k,1)]_{j_1 j_2}^{0 0}= 0$ otherwise
(see Figure \ref{picturermatrix}).
This observation applies to the remaining bottom $k-1$ rows as well,
and one can see that
only one configuration is allowed in the rightmost $n-k$ columns,
and the part constructed from the remaining
$m$ columns is nothing but
the mixed partition functions
$ZDB_{m,n,k}(u_{k+1},\dots,u_n|u_1,\dots,u_k|x_1,\dots,x_\ell|y_1,\dots,y_{\ell+k})$  \eqref{dbpartitionfunction} (Figure \ref{picturedbpartition}),
which is the graphical meaning of Figure \ref{picturedbpartitionreduction}.
Since the $R$-matrix elements which appear in
the rightmost $n-k$ columns
are either $[R(u_j,1)]_{00}^{00}=1$, $[R(u_j,1)]_{11}^{11}=1$
or $[R(u_j,1)]_{10}^{01}=1$,
the product of all $R$-matrix elements in
the rightmost $n-k$ columns is 1.
Multiplying this 1 by $ZDB_{m,n,k}(u_{k+1},\dots,u_n|u_1,\dots,u_k|x_1,\dots,x_\ell|y_1,\dots,y_{\ell+k})$ gives
$Z_{m+n-k,n}
(u_1,\dots,u_n|x_1,\dots,x_\ell|y_1,\dots,y_{\ell+k},
m+1,\dots,m+n-k)$, hence we conclude \eqref{largeanddb}.

\end{document}